\documentclass[a4paper,oneside,submitp]{pioArxiv}

\definecolor{refcolor}{rgb}{0,0,0.4}    
\definecolor{anchorcolor}{rgb}{0,0,0}   
\definecolor{citecolor}{rgb}{0,0.4,0.4} 
\definecolor{urlcolor}{rgb}{0,0,0.690}   
\definecolor{filecolor}{rgb}{0,0,0.690}  
\definecolor{menucolor}{rgb}{0.75,0,0}  
\definecolor{pagecolor}{rgb}{0.75,0,0}  

\usepackage[square]{natbib}
\usepackage{afterpage}
\usepackage{here}
\usepackage{tabularx,pdflscape}
\usepackage{array,longtable}
\newcolumntype{Y}{>{\raggedright\arraybackslash}X}
\newcounter{Fig}

\newcommand\leqt[1]{\protect\label{eq:#1}}
\newcommand\reqtn[1]{\ref{eq:#1}}
\newcommand\reqt[1]{(\reqtn{#1})}

\newcommand\lsect[1]{\protect\label{sect:#1}}
\newcommand\rsect[1]{\ref{sect:#1}}

\newcommand\REMOVE[1]{}

\DeclareMathOperator*{\sech}{sech}

\begin{document}
\sloppy

\title{Optical Vortices and Vortex Solitons}
\shorttitle{Optical Vortices}

\author{Anton S. Desyatnikov$^1$, Lluis Torner$^2$, and Yuri S. Kivshar$^1$\\
\begin{small}
$^1$ Nonlinear Physics Center and Center for Ultra-high bandwidth Devices for Optical Systems, Research School of Physical Sciences and Engineering, Australian National University, Canberra ACT 0200, Australia\\
HomePage:~\href{http://wwwrsphysse.anu.edu.au/nonlinear}{wwwrsphysse.anu.edu.au/nonlinear}\\
$^2$ ICFO-Institut de Ciencies Fotoniques, and Department of Signal Theory and Communications, Universitat Politecnica de Catalunya, Barcelona ES 08034, Spain\\
HomePage:~\href{http://www.icfo.es}{www.icfo.es}\\
\end{small}
}

\maketitle

\begin{abstract}
Optical vortices are phase singularities nested in electromagnetic waves that constitute a fascinating source of phenomena in the physics of light and display deep similarities to their close relatives, quantized vortices in superfluids and Bose-Einstein condensates. We present a brief overview of the major advances in
the study of optical vortices in different types of nonlinear media, with emphasis on the properties of {\em vortex solitons}. Self-focusing nonlinearity leads, in general, to the azimuthal instability of a vortex-carrying beam, but it can also support novel types of stable or meta-stable self-trapped beams carrying nonzero angular momentum, such as ring-like solitons, necklace beams, and soliton clusters. We describe vortex solitons created by multi-component beams, by parametrically coupled beams in quadratic nonlinear media, and in partially incoherent light, as well as discrete vortex solitons in periodic photonic lattices.
\end{abstract}

\tableofcontents

\section{Introduction} \lsect{intro}

In physics, wave propagation is traditionally analyzed by means of
regular solutions of wave equations. However, solutions of wave
equations in two and three dimensions often possess singularities,
the points or lines in space at which mathematical quantities that
describe physical properties of waves become infinite or change
abruptly (\cite{Berry:2000-21:NAT}). For example, at the point of
phase singularity, the phase of the wave is undefined and wave
intensity vanishes. Phase singularities are recognized as
important features common to all waves. They were first discussed
in depth in a seminal paper by~\cite{Nye:1974-165:PRSA}. However,
the earliest known scientific description of phase singularity was
made in the 1830's by Whewell, as discussed
by~\cite{Berry:2000-21:NAT}. While Whewell studied the ocean
tides, he came to the extraordinary conclusion that rotary systems
of tidal waves possess a singular point at which all cotidal lines
meet and at which tide height vanishes. Waves that possess a phase
singularity and a rotational flow around the singular point are
called {\em vortices}. They can be found in physical systems of
different nature and scale, ranging from water whirlpools and
atmospheric tornadoes to quantized vortices in superfluids and
quantized lines of magnetic flux in superconductors
(\cite{Pismen:1999:VorticesNonlinear}).

In a light wave, the phase singularity is known to form {\em an
optical vortex}: The energy flow rotates around the vortex core in
a given direction; at the center, the velocity of this rotation
would be infinite and thus the light intensity must vanish. The
study of optical vortices and associated localized objects is
important from the viewpoint of both fundamental and applied
physics. The unique nature of vortex fields is expected to lead to
applications in many areas that include optical data storage,
distribution, and processing. Optical vortices propagating, e.g.,
in air, have been suggested also for the establishment of optical
interconnects between electronic chips and boards
(\cite{Scheuer:1999-230:SCI}), as well as free-space communication
links (\cite{Gibson:2004-5448:OE, Bouchal:2004-131:NJP}), based on
the multidimensional alphabets afforded by the corresponding
angular momentum states (\cite{Molina-Terriza:2002-13601:PRL}).
The ability to use light vortices to create reconfigurable
patterns of complex intensity in an optical medium could aid
optical trapping of particles in a vortex field
(\cite{Gahagan:1999-533:JOSB}), and could enable light to be
guided by the light itself, or in other words by the waveguides
created by optical vortices (\cite{Truscott:1999-1438:PRL,
Law:2000-55:OL, Carlsson:2000-660:OL, Salgueiro:2004-593:OL}).
Thus, {\em singular optics}, the study of wave singularities in
optics (\cite{Nye:1974-165:PRSA, Vasnetsov:1999:OpticalVortices,
Soskin:2001-219:ProgressOptics}), is now emerging as {\em a new
discipline} (for an extended list of references, see http://www.u.arizona.edu/~grovers/SO/so.html).

In a broad perspective, the study of optical vortices brings
inspiring similarities between different and seemingly disparate
fields of physics; the comparison of singularities of optical and
other origins leads to theories that transcend the confines of
specific fields. Vortices play an important role in many branches
of physics, even those not directly related to wave propagation.
An example is the Kosterlitz-Thouless phase transition
(\cite{Kosterlitz:1973-1181:JPC}) in solid-state physics models,
characterized by creation of tightly bound pairs of point-like
vortices that restore the quasi-long-range order of a
two-dimensional model at low temperatures. Such vortex-induced
phase transitions can be observed in superfluid helium films, thin
superconducting films, and surfaces of solids, as well as in
models of interest to particle physicists and cosmologists.

The Bose-Einstein condensate (BEC), a state of matter in which a
macroscopic number of particles share the same quantum state,
constitutes a well-researched example of a superfluid in which
topological defects with a circulating persistent current are
observed. Nearly 75 years ago, Bose and Einstein introduced the
idea of condensate of a dilute gas at temperatures close to
absolute zero. The BEC was experimentally created in 1995 by the
JILA group (\cite{Anderson:1995-198:SCI}), who trapped thousands
(later, millions) of alkali $^{87}$Rb atoms in a 10-$\mu$m cloud
and then cooled them to a millionth of a degree above absolute
zero. The extensive study of vortices in BEC
(\cite{Williams:1999-568:NAT, Matthews:1999-2498:PRL,
Madison:2000-806:PRL, Raman:2001-210402:PRL}) promises a deeper
understanding of deep links between the physics of superfluidity,
condensation, and nonlinear singular optics.

To introduce the notion of optical vortices, we recall that a
light wave can be represented by a complex scalar function $\psi$
(e.g., an envelope of an electric field), which varies smoothly in
space and/or time. Phase singularities of the wave function $\psi$
appear at the points (or lines in space) at which its modulus
vanishes, i.e., when ${\rm Re} \, \psi = {\rm Im} \, \psi = 0$.
Such points are referred to as wave-front screw dislocations or
optical vortices, because the surface of constant phase
structurally resembles a screw dislocation in a crystal lattice,
and because the phase gradient direction swirls around the
singular line much like fluid in a whirlpool. Optical vortices are
associated with zeros in light intensity (black spots) and can be
recognized by a specific helical wave front. If the complex wave
function is presented as, $\psi({\bf r},t) = \rho({\bf r},t)
\exp\{i \theta({\bf r},t)\}$, in terms of its real modulus $\rho
(r,t)$ and phase $\theta (r,t)$, the dislocation strength
(sometimes referred to as the vortex topological charge) is
defined by the circulation of the phase gradient around the
singularity,
\begin{equation}
\leqt{S}
S= \frac{1}{2\pi} \oint \nabla \theta d{\bf r}.
\end{equation}

The result is an integer because the phase changes by a multiple
of $2\pi$. Under appropriate conditions, it also measures an
orbital angular momentum of the vortex associated with the helical
wave-front structure. If a light wave is characterized by an extra
parameter, e.g., the wave polarization, its mathematical
representation is no longer a scalar but a vector field. In vector
fields, several types of line singularity exist; for example,
those analogous to disclinations in liquid crystals, which could
be edge type, screw type, or mixed edge-screw type, that could
move relative to background wave fronts and could interact in
several different ways (\cite{Nye:1974-165:PRSA,
Soskin:2001-219:ProgressOptics}). In {\em the linear theory of
waves}, each wave dislocation could be understood as a simple
consequence of destructive wave interference. In this review we
mostly address screw phase singularities existing in scalar wave
fields and thus we concentrate our analysis in the corresponding
vortices. However, other types of singularities whose analysis
falls beyond the scope of this review, such us polarization
singularities (\cite{Freund:2004-539:OL, Freund:2004-875:OL}), do
exist and exhibit fascinating properties.

A laser beam with a phase singularity generally has a
doughnut-like shape and diffracts when it propagates in a free
space. However, when the vortex-bearing beam propagates in {\em a
nonlinear medium}, a variety of interesting effects can be
observed. Nonlinear optical media are characterized by the
electromagnetic response that depends on the strength of the
propagating light. The polarization of such a medium can be
described as $P = \chi^{(1)}E + \chi^{(2)}E^2 + \chi^{(3)}E^3$,
where $E$ is the amplitude of the light wave's electric field, and
the coefficients characterize both the linear and the nonlinear
response of the medium (\cite{Shen:1984:PrinciplesNonlinear,
Butcher:1992:ElementsNonlinear, Boyd:1992:NonlinearOptics}). The
$\chi^{(1)}$ coefficient describes the linear refractive index of
the medium. When $\chi^{(2)}$ vanishes (as happens in the case of
centro-symmetric media), the main nonlinear effect is produced by
the third term that can be presented as an intensity-induced
change of the refractive index proportional to $\chi^{(3)}E^3$. An
important consequence of such intensity-dependent nonlinearity is
the spontaneous focusing of a beam that is due to the lensing
property of {\em a self-focusing medium} (i.e. when
$\chi^{(3)}>0$). This focusing action of a nonlinear medium can
precisely balance the diffraction of a laser beam, resulting in
the creation of optical solitons, which are self-trapped light
beams that do not change shape during propagation
(\cite{Kivshar:1998-81:PRP}).

\pict[0.708]{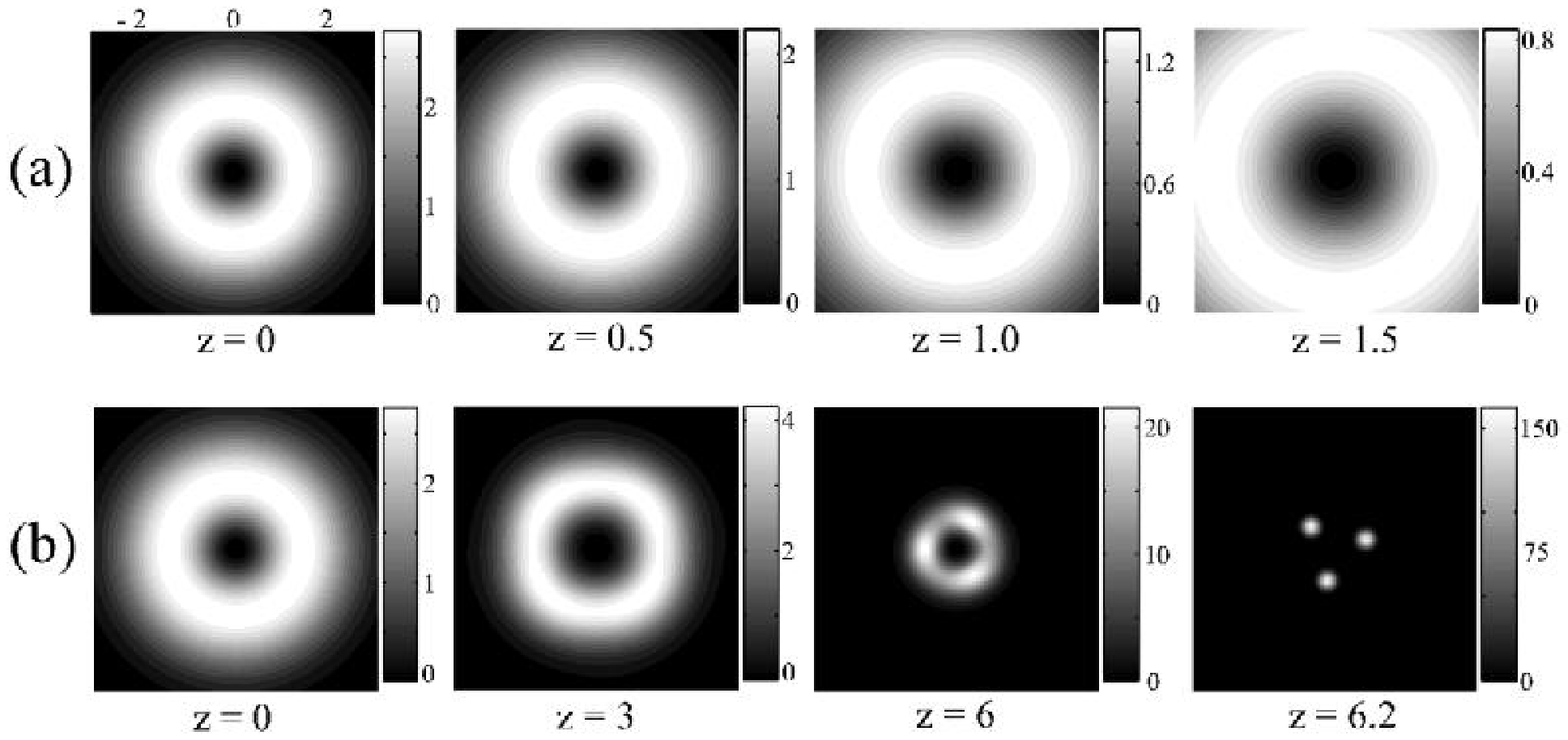}{fig1}{Propagation of the Gaussian beam with
a phase dislocation generated by the input beam
$E(r,\varphi,z=0)=2r\exp\left (-r^2/4+i\varphi\right)$, in (a)
linear medium, $\chi^{(3)}=0$, and (b) self-focusing Kerr medium,
$\chi^{(3)}>0$. Shown is the field intensity. Note the difference
in the intensity scales.}

\pict[0.708]{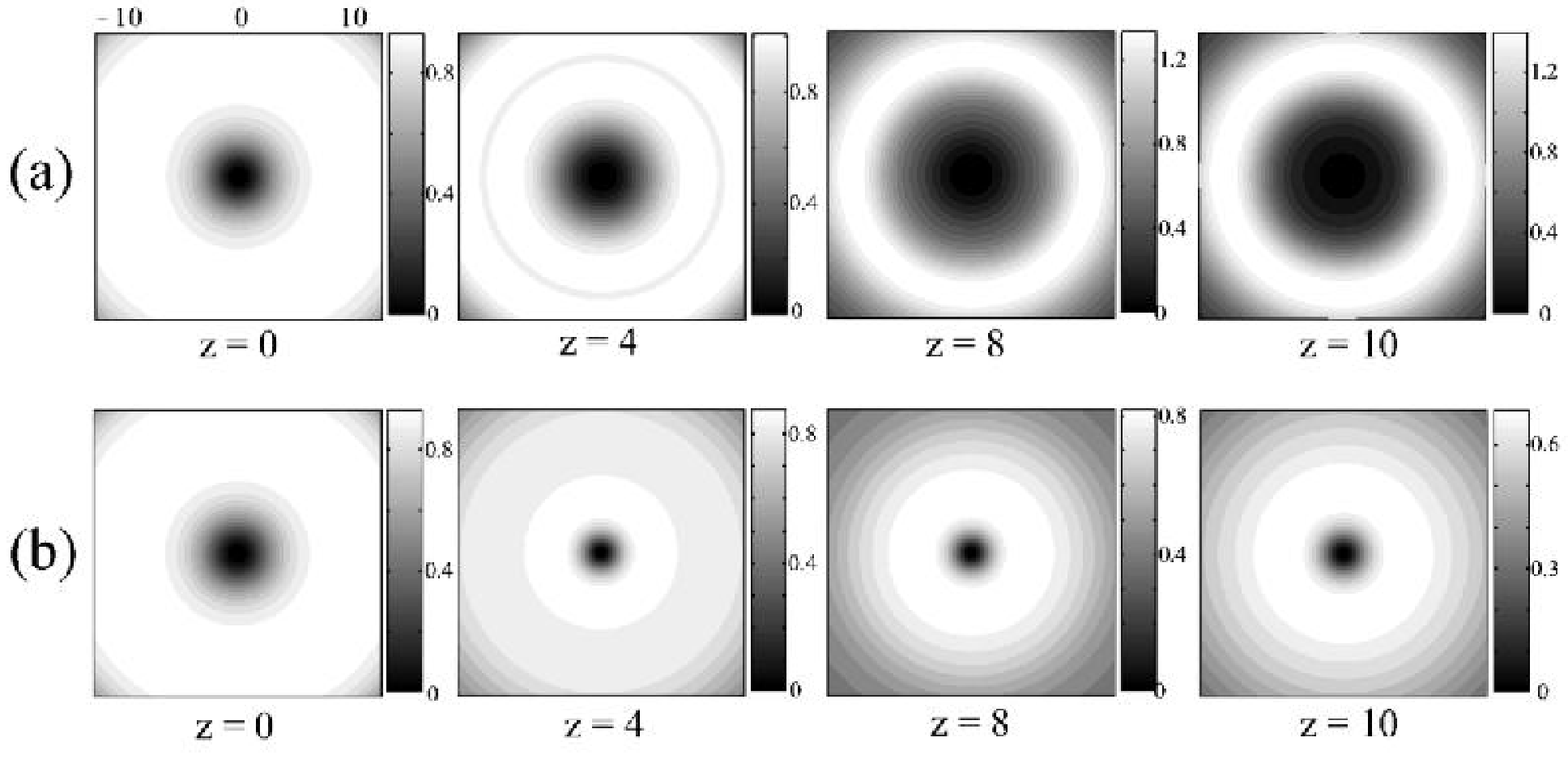}{fig2}{Propagation of the paraxial
super-Gaussian beam with a phase dislocation generated by the
input beam $E(r,\varphi,z=0)=\tanh(r/3)\exp\left
(-(r/18)^{10}/2+i\varphi\right )$, in (a) linear medium,
$\chi^{(3)}=0$, and (b) self-defocusing Kerr medium,
$\chi^{(3)}<0$. Shown is the field intensity.}

A stable bright spatial soliton is radially symmetric, and it has
no nodes in its intensity profile. If, however, a beam with
elaborate geometry carries a topological charge and propagates in
a self-focusing nonlinear medium, it has a doughnut like
structure. However, such a doughnut beam is unstable, and it
decays into a number of more fundamental bright spatial solitons,
such an example is shown in Fig.~\rpict{fig1}. The resulting field
distribution does not preserve the radial symmetry, and the vortex
beam decays into several solitons that repel and twist around one
another as they propagate. This rotation is due to the angular
momentum of the vortex beam transferred to the splinters.

Remarkably, the behavior of a laser beam in {\em a self-defocusing
nonlinear medium} (i.e. when $\chi^{(3)} < 0$) is distinctly
different, see an example in Fig.~\rpict{fig2}. Such a medium
cannot produce a lensing effect and therefore cannot support
bright solitons. Nevertheless, a negative change of the refractive
index can compensate for spreading in light intensity of the dip,
thus creating a dark soliton (\cite{Kivshar:1998-81:PRP}), a
self-trapped, localized low-intensity state (a dark hole) in a
uniformly illuminated background. Vortices of single and multiple
topological charges can be created in both linear and nonlinear
media by use of, e.g., computer-generated holograms or spatial
light modulators. Propagating through a nonlinear self-defocusing
medium, such as a vortex-carrying beam, creates a self-trapped
state, {\em a dark vortex soliton}. Dark vortex solitons have been
observed experimentally in different materials with
self-defocusing nonlinearity, such as slightly absorbent liquids,
vapors of alkali metals, and photorefractive crystals
(\cite{Swartzlander:1992-2503:PRL, Mamaev:1996-4544:PRL,
Chen:1997-2948:PRL}).

In this review paper, we describe the basic concepts of the
nonlinear physics of optical  vortices in the context of the
propagation of singular beams in nonlinear media. In particular,
we overview the recent advances in the study of optical vortices
propagating in different types of self-defocusing and
self-focusing nonlinear media, but concentrate mostly on the case
of self-focusing nonlinearity which leads to the azimuthal
instability of a vortex-carrying beams
(\cite{Kivshar:2003:OpticalSolitons}). We summarize different
physical settings where such a nonlinearity can support novel
types of stable or quasi-stable self-trapped beams carrying
nonzero angular momentum, such as vortex solitons, necklace beams,
soliton clusters, etc. We also describe the properties of the
vortex beams created by partially incoherent light and the
discrete vortices in periodic photonic lattices. In addition, we
present some of the experimental results demonstrating the
propagation of singular optical beams in nonlinear media.

\section[Vortices in $\chi^{(3)} media$]{Self-trapped vortices in Kerr-type media} \lsect{X3}

In this section, we describe the conventional {\em scalar} optical
vortices in Kerr-like nonlinear media. In a defocusing media, a
diffracting core of an optical vortex may get self-trapped and the
resulting beam with a singular core should be classified as a
vortex soliton. In contrast, as discussed above, a vortex-carrying
beam itself become self-trapped in a focusing nonlinear medium,
and it is know to suffer the azimuthal modulational instability.

\subsection[Dark vortex solitons]{Vortices in defocusing nonlinear media}
\lsect{dark}

In a self-defocusing nonlinear medium, a screw dislocation in the
wave phase can create a stationary beam structure with a phase
singularity resulting in a self-trapped vortex beam or {\em a
vortex soliton}. To describe the major properties of vortex
solitons, we consider the propagation of a continuous wave (CW)
beam in a bulk self-defocusing medium governed by a
($2+1$)-dimensional NLS equation. In the specific case of the Kerr
nonlinearity, this equation can be written in the normalized form
\begin{equation}
\leqt{vs.1}
    i\frac{\partial u}{\partial z}+\frac{1}{2}\nabla^2 u - |u|^2 u = 0.
\end{equation}
We can eliminate the background of constant amplitude $u_0$
through the transformation, $z' = u_0^2 z$, $x'=u_0 x$, $y'=u_0
y$, $u =u_0 \psi \exp(-iu_0^2 z)$, and obtain the following
equation for the normalized field $\psi$:
\begin{equation}
\leqt{nls_h}
    i\frac{\partial \psi}{\partial z}+ \frac{1}{2}\nabla^2 \psi + (1-|\psi|^2)\psi=0,
\end{equation}
where we have dropped the primes for simplicity of notation. The
dimensionless field should satisfy the boundary conditions
$|\psi| \to 1$ as $x$ and $y \to \pm \infty$.

The existence of vortex solutions for the ($2+1$)-dimensional
cubic NLS equation \reqt{nls_h} can be established using an
analogy between optics and fluid dynamics. Using the so-called
Madelung transformation (\cite{Spiegel:1980-236:PD,
Donnelly:1991:QuantizedVortices, Nore:1993-154:PD}),
\begin{equation}
    \psi({\bf r}, z) = \chi ({\bf r}, z) \exp[\varphi ({\bf r}, z)],
\end{equation}
where ${\bf r}$ is a two-dimensional vector with coordinates $x$
and $y$, we can transform the NLS equation \reqt{nls_h} into the
following set of two coupled equations:
\begin{eqnarray}
    \frac{\partial \chi^2}{\partial z} &+& \nabla \cdot (\chi^2 \nabla \varphi) = 0,
\leqt{h1} \\
    \frac{\partial \varphi}{\partial z} &+& \frac{1}{2} (\nabla\varphi)^2
        = 1- \chi^2 + \frac{ \nabla^2 \chi}{2\chi}.
\leqt{h2}
\end{eqnarray}
These equations can be viewed as the equations governing the
conservation of mass and momentum for a compressible inviscid
fluid of density $\rho =\chi^2$ and velocity ${\bf v} = \nabla
\varphi$, with the effective pressure defined as $p=\rho^2/2$.
More importantly, this kind of analogy between optics and fluid
mechanics remains valid even for the generalized NLS equation with
an arbitrary form of $g(|u|^2)$, provided the effective pressure
is defined as
\begin{equation}
    p(\rho) = \int \rho \frac{dg(\rho)}{d\rho} \,d\rho.
\end{equation}
The analogy, however, is not exact because, in addition to the
standard pressure,  Eq.~\reqt{h2} includes a second term that has
no analog in fluid mechanics. This term results from the so-called
{\em quantum-mechanical pressure} in the context of superfluids.

\pict[0.708]{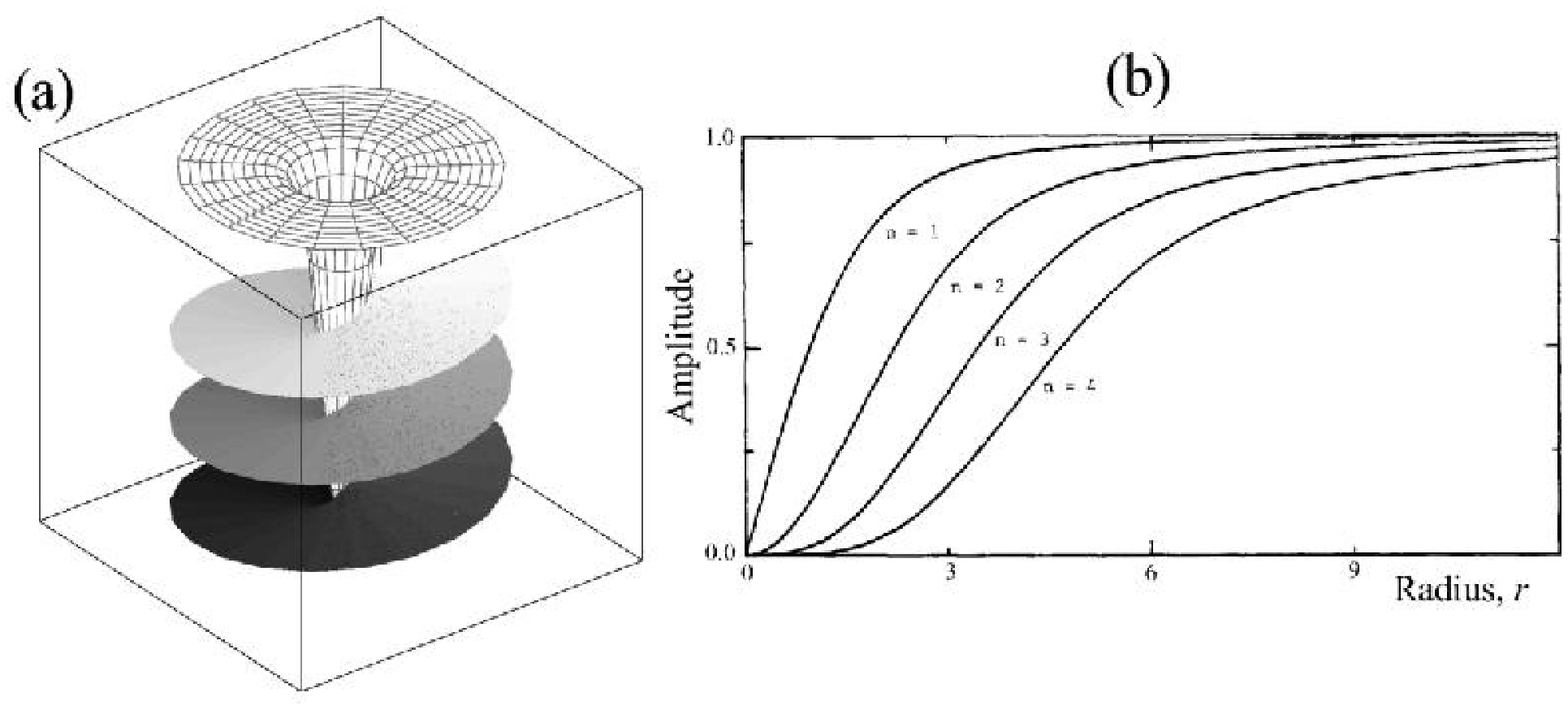}{fig3}{(a) Schematic of the intensity
distribution in an optical beam carrying a vortex (mesh) and its
helical wave front (gray surface). After
\cite{Kivshar:2001-26:OPN}. (b) Vortex profiles in a
self-defocusing Kerr medium for four first values of the integer
vortex charge $m$ (\cite{Neu:1990-385:PD}).}

The Madelung transformation is singular at the points where
$\chi=0$. Around such points on the plane $(x,y)$, the circulation
of ${\bf v}$ is not zero but equals $2\pi$. These points present
{\em topological defects} of the scalar field, and they are called
{\em vortices}. To find the stationary solution corresponding to a
vortex soliton, also called a dark soliton with circular symmetry,
see Fig.~\rpict{fig3}(a), we look for solutions of the cubic NLS
equation  \reqt{nls_h} in the polar coordinates $r$ and $\theta$,
\begin{equation}
    \psi(r, \theta; z) = U(r) e^{im \theta},
\end{equation}
where the integer $m$ is the so-called vortex {\em winding
number}, sometimes also called the {\em vortex charge}, and the
real function $U(r)$ satisfies the amplitude equation,
\begin{equation}
\leqt{U-r}
    \frac{d^2U}{dr^2}+\frac{1}{r}\frac{dU}{dr}-\frac{m^2}{r^2}U+(1-U^2)U=0,
\end{equation}
with the boundary conditions
\begin{equation}
    U(0) = 0, \qquad U(\infty) = 1.
\end{equation}
The continuity of $U$ at $r=0$ forces the first condition, while
$U(\infty)=1$ is consistent with a uniform background of intensity
$U_0^2$ as $r\rightarrow \infty$.

Equation~\reqt{U-r} can be solved numerically to find the shape $U(r)$ of the vortex soliton for different values of $m$, shown in Fig.~\rpict{fig3}(b) (\cite{Neu:1990-385:PD}). Alternatively, the approximate envelopes converge to the stationary states in the numerical simulation of the full Eq.~\reqt{nls_h} (\cite{Velchev:1997-77:OC}). The region in the vicinity of $r=0$, where $U(r)$ is significantly less than 1, is called  the {\em vortex core}. The functional form of $U(r)$ near $r=0$ and $r=\infty$ can be established directly
from Eq.~\reqt{U-r} by taking the appropriate limit and is found to be
\begin{equation}
    U(r) \sim \left\{ \begin{array}{l@{\quad{\rm as}~}l}
        a r^{|m|} + {\rm O}(r^{|m|+2}) & r \rightarrow 0, \\
        1 - \frac{m^2}{2r^2} + {\rm O}(1/r^4) & r \rightarrow \infty.
    \end{array} \right.
\end{equation}

The structure of the vortex soliton for an arbitrary form of the
nonlinearity can be found using the same method and solving
numerically for the amplitude function $U(r)$. No qualitatively
new features are found when the nonlinearity is allowed to
saturate (\cite{Chen:1992-2252:JOSB}). However, the effective
diameter of the vortex core increases almost linearly with the
saturation parameter $s=I_0/I_s$, where $I_s$ and $I_0$ are the
saturation and background intensities, respectively
(\cite{Tikhonenko:1998-79:JOSB}).

Stability of vortex solitons associated with the generalized NLS
equation has not yet been fully addressed. However, it is believed
that vortices with the winding numbers $m= \pm 1$ are {\em
topologically stable} but that those with larger values of $|m|$
are unstable and should decay into $|m|$ single-charge vortices.
In the context of superfluids, the multi-charged vortex solitons
are found to survive for a relatively long time
(\cite{Aranson:1996-75:PRB}). A similar behavior is found to occur
in optics (\cite{Dreischuh:1999-7518:PRE}). For this reason,
multi-charged vortices are usually classified as metastable. As an
example, an intentional perturbation of a triply charged vortex
leads to its incomplete decay to a long-lived doubly charged
vortex and a singly charged vortex
(\cite{Dreischuh:1999-6111:PRE}), confirming that saturation of
nonlinearity can effectively suppress the instability. We should,
however, stress that multi-charged vortices are strongly unstable
in anisotropic nonlinear media (\cite{Mamaev:1997-2108:PRL}).

\pict[0.708]{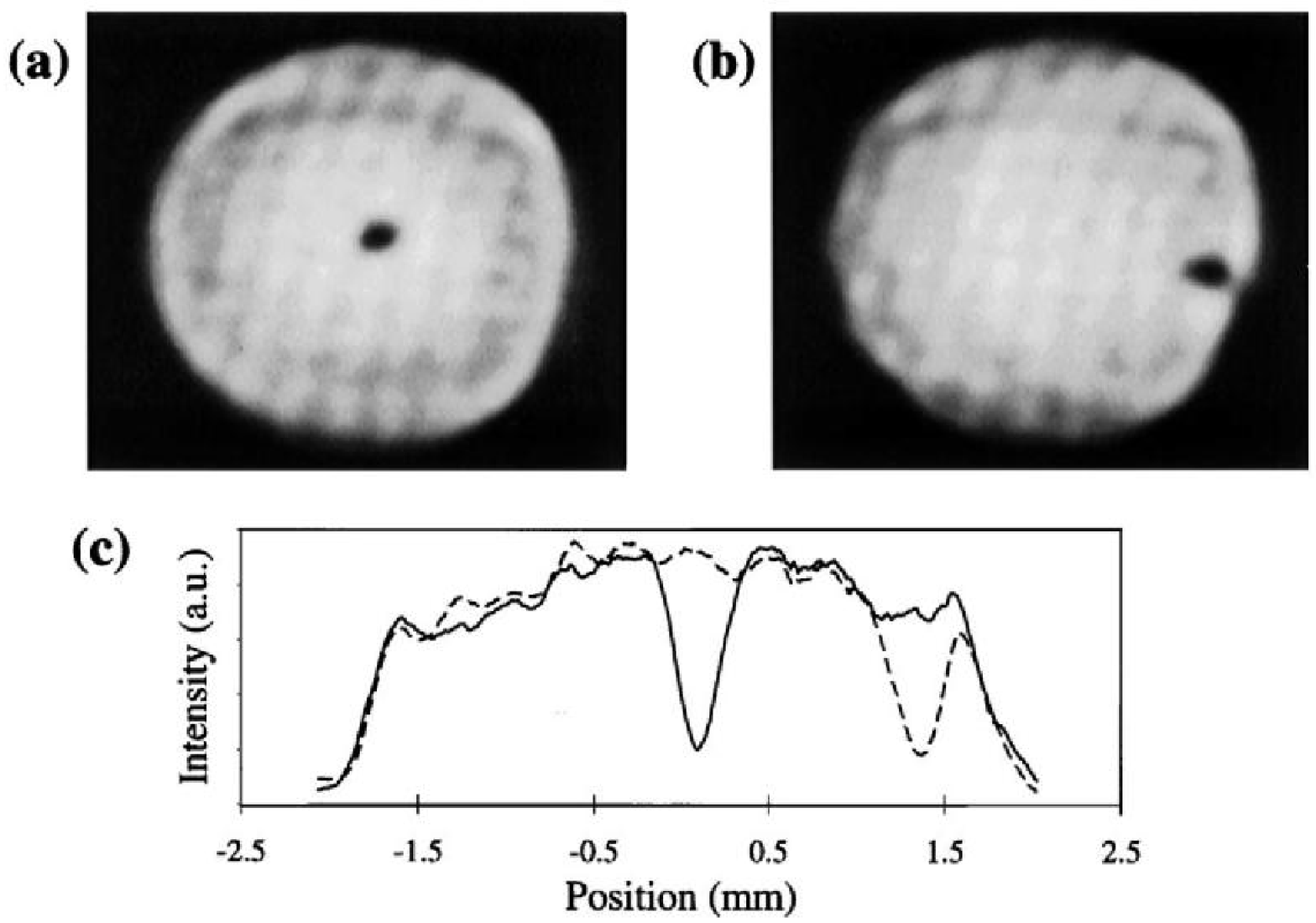}{fig4}{Intensity distributions at the output
of a nonlinear medium (a Rubidium vapor). The position of the
vortex in (a) and (b) was controlled by translation of the vortex
mask. (c) The intensity profile at the cross-section made through
the vortex core in the cases (a) (solid) and (b) (dashed),
respectively (\cite{Christou:1996-1649:OL}).}

Experimentally, a vortex soliton appears as a dark region that
maintains its shape on a diffracting background beam displaying a
nontrivial dynamical behavior (\cite{Swartzlander:1992-2503:PRL,
Tikhonenko:1996-108:OC}). To generate the vortex input beam, one
images the waist of the Gaussian beam onto the surface of a singly
charged phase mask with a telescope. The first diffracted order of
this mask is then imaged onto the plane of the nonlinear cells
input window, providing an initial condition consisting of a
singly charged vortex nested centrally at the waist of a Gaussian
beam. The position of the vortex in the initial field is
controlled by translation of the phase mask across the beam.

Figure~\rpict{fig4}(a) shows a typical intensity distribution at
the output with the vortex nested at the approximate center of the
beam. Figure~\rpict{fig4}(b) shows the output under the same
conditions, apart from a translation of the phase mask. There is
little change in the beam away from the core of the vortex, as is
seen in Fig.~\rpict{fig4}(c), upon which cross sections on a line
through the vortex cores of Figs.~\rpict{fig4}(a)
and~\rpict{fig4}(b) are overlaid. To measure the size of the
output background, we should remove the vortex from the profile
and calculated the average $1/e^2$ radius of undisturbed
background. Phenomena such as the rotation and radial drift of the
vortex relative to the background CW beam are often observed
experimentally, even though they cannot always be predicted from a
casual analysis of the stationary solution of the NLS equation. A
proper theoretical description of these effects requires
analytical techniques capable of analyzing the vortex motion
(\cite{Christou:1996-1649:OL, Kivshar:1998-198:OC}). Physically,
specific dynamical features such as vortex rotation and drift
result from a nonuniform intensity profile of the background field
which is typically a Gaussian beam.

The rotation rate of optical vortices can be controlled by
introducing a modulated phase gradient of the background beam,
when the slope of the helical wave front is not uniform in the
azimuthal direction (\cite{Kim:2003-351:JOSB}). Phase profile
determines not only the dynamics of a single vortex but also
interaction between two vortices, such as attraction and repulsion
of counter- and co-rotating vortices, respectively
(\cite{Luther-Davies:1994-1816:OL, Velchev:1996-385:OC}). A pure
phase modulation, obtained by using computer-synthesized
holograms, was used to create {\em dark ring} solitary waves
(\cite{Kivshar:1994-40:PRE, Baluschev:1995-5517:PRE,
Dreischuh:1996-139:APB, Dreischuh:1996-145:APB,
Kamenov:1997-68:PS, Neshev:1997-429:APB,
Dreischuh:2002-66611:PRE}). Similar examples of phase patterns
with singularities include vortex arrays and lattices in
self-defocusing Kerr type media studied by
\cite{Neshev:1998-413:OC, Kim:1998-131:OC, Kim:1998-308:JKPS,
Dreischuh:2002-550:JOSB}.

\subsection{Ring-like beams in focusing nonlinear media}
\lsect{soliton}

In an isotropic optical medium with a local nonlinear response the
propagation of a paraxial light beam is described by the
well-known generalized nonlinear Schr\"odinger (NLS) equation
(\cite{Kivshar:2003:OpticalSolitons}). In the dimensionless form,
the generalized NLS equation has the form (cf. Eq.~\reqt{vs.1}),
\begin{equation}
\leqt{eq} i\frac{\partial E}{\partial z}+\Delta _{\perp }E + F(I)E=0,
\end{equation}
where $E$ is the complex envelope of the electric field,
$\Delta_{\perp} = \partial^2/\partial x^2+\partial^2/\partial y^2$
is the transverse Laplacian, and $z$ is the propagation distance
measured in the units of the diffraction length $L_D$. Function
$F(I)$ describes the nonlinear properties of an optical medium,
and it is assumed to depend on the total beam intensity,
$I\equiv\left|E\right|^2$. Examples of these Kerr-like materials
include two-level model of resonant gases, $F=I$, or pure Kerr
nonlinearity; the so-called saturable nonlinearity, $F=I(1+\alpha
I)^{-1} $, or its low-intensity expansion, the cubic-quintic
model, $F=I-\alpha I^2$, etc. Here the parameter $\alpha$ defines
the nonlinearity saturation.

In a self-focusing medium, i.e. $F(I)\geq 0$, the diffraction of a
light beam can be compensated by the nonlinearity and the balance
between these two counter-acting ``forces'' corresponds to the
stationary state. {\em Spatial optical solitons} are stationary
spatially localized solutions of the NLS equation \reqt{eq} which
do not change their intensity profile during propagation
(\cite{Segev:1998-503:OQE, Stegeman:1999-1518:SCI,
Kivshar:2002-59:OPN}). Such a definition covers many different
types of stationary beams with a finite power and, in general, the
spatial solitons can be found in a generic form,
\begin{equation}
\leqt{statsol}
 E(x,y,z) = U(x,y) \exp \left[ ikz + i\phi (x,y)\right],
\end{equation}
where the real functions $U$ and $\phi$ are the soliton amplitude
and phase, respectively, and $k$ is the soliton propagation
constant. Substituting Eq.~\reqt{statsol} into Eq.~\reqt{eq}, we
arrive at the system of coupled equations for the soliton
amplitude and phase,
\begin{eqnarray}
\leqt{Seq} \Delta _{\perp}U - kU - \left ( \nabla\phi \right )^2 U +F(U^2)U=0, \\
\leqt{Seq2} \Delta _{\perp} \phi + 2 \nabla \phi \nabla \ln U =0.
\end{eqnarray}

We start with the solutions of the system \reqt{Seq}, \reqt{Seq2}
with a constant phase, taking $\phi = 0$ without restriction of
generality. In this case, it can be shown that the only type of a
structure localized in both transverse dimensions should possess a
radial symmetry, i.e. $U(x,y) = U(r)$, where $r = \sqrt{x^2+y^2}$.
Solutions of this type include {\em the fundamental (bell-shaped)
soliton} [see Fig.~\rpict{fig5}(b) for $m=0$] and higher-order
modes with several rings surrounding the central peak
(\cite{Haus:1966-128:APL, Yanauskas:1966-261:SR,
Soto-Crespo:1991-636:PRA, Edmundson:1997-7636:PRE}). The number of
radial nodes, defined by the index $n$, distinguishes the
higher-order radially-symmetric spatial solitons. The main
parameter characterizing the spatial soliton is its power
\begin{equation}
\leqt{power} P=\int\left|E\right|^{2}d{\bf r} = \int\ U^2 d{\bf r},
\end{equation}
being the integral of motion associated with phase invariance of a
solution to Eq.~\reqt{eq}. Radial modes with $n$-rings in the
intensity profile, $U_n$, belong to the different branches of the
dependance $P(k)$, i.e. they form a discrete set of soliton
families, $P_n(k)$. The generalization of each soliton family
includes transversely moving solitons, obtained by applying the
Galilean transformation, ${\bf r} \rightarrow {\bf r}-2{\bf q}z$
and $\phi \rightarrow \phi + {\bf q} ({\bf r} - {\bf q} z)$, where
${\bf v}=2{\bf q}$ is the soliton transverse velocity. Such moving
solitons can be characterized by the soliton {\em linear momentum}
\begin{equation}
\leqt{linmom} {\bf L}={\rm Im} \int E^{\ast}\nabla E d{\bf r} =\int \nabla \phi \ U^2 d{\bf r},
\end{equation}
which is defined for the fundamental solitons as ${\bf L} = {\bf q}P$.

\pict[0.708]{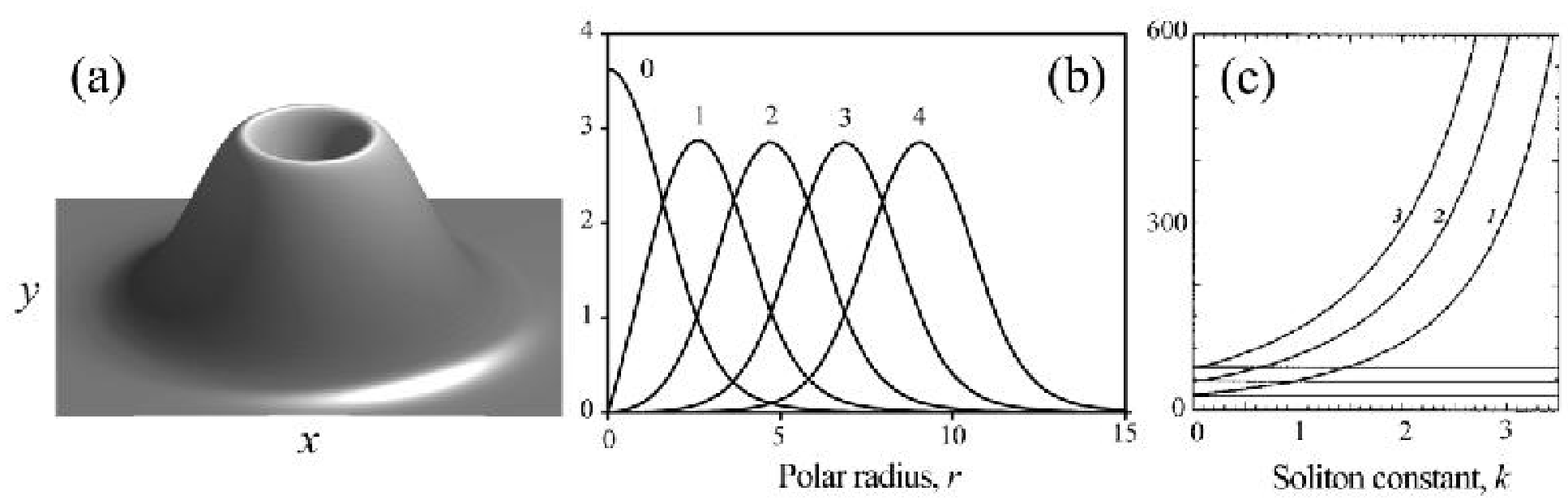}{fig5}{Examples of the stationary radially
symmetric soliton solutions of Eq.~\reqt{eq} with saturable
nonlinearity $F =I/(1+\alpha I)$, characterized by the topological
charge $m$ indicated next to the curves, the case $m=0$
corresponds to the fundamental soliton; (a) intensity distribution
of a single-charged vortex, $m=1$, (b) radial profiles for
$\alpha=0.5$ and $k=1$, (c) soliton power vs. soliton parameter
$k$ for $\alpha=0.2$, after \cite{Skryabin:1998-3916:PRE}.}

A novel class of spatially localized beams in self-focusing
nonlinear media, associated with the rotation of the field phase,
was introduced by \cite{Kruglov:1985-401:PLA}. The beam phase has
a spiral structure with a singularity at the origin, as the one
shown in Fig.~\rpict{fig3}(a), representing a phase dislocation of
the wave front in the form of an optical vortex
(\cite{Kivshar:2001-26:OPN}). The intensity of such a beam
vanishes at the beam center, and, at the same time, the beam
remains localized (i.e. its intensity decays at infinity)
propagating in the form of a ring-like beam, see
Fig.~\rpict{fig5}(a,b). The existence of the ring-profile solitary
waves can be explained intuitively. Indeed, quasi-one-dimensional
solitary waves, i.e. the (1+1)-dimensional solitary waves embedded
into a (2+1)-dimensional bulk medium, undergo {\em the transverse
modulational instability} (\cite{Kivshar:2000-118:PRP}). One of
the possible ways to suppress this instability is to consider a
ring-profile structure created from a (1+1)-dimensional soliton
stripe wrapped around its tail (\cite{Lomdahl:1980-125:PLA}). It
can be shown that such a soliton ring with the phase depending on
the radial coordinate can display a stabilization effect, and even
initially expanding beam can shrink and eventually collapse
(\cite{Lomdahl:1980-125:PLA, Afanasjev:1995-3153:PRE, Anastassiou:2001-911:OL}). The beams
studied by \cite{Kruglov:1992-2277:JMO} provide another example of
nonstationary ring-profile solitary waves.

Introduced by \cite{Kruglov:1985-401:PLA} ring-profile {\em vortex
solitons} represent the first example of a spatial soliton with
the field dependence on the azimuthal coordinate $\varphi =
\tan^{-1} (y/x)$. They can be found as the solutions to
Eqs.~\reqt{Seq},~\reqt{Seq2} with a rotating spiral phase $\phi$
in the form of {\em a linear function of the polar angle}
$\varphi$, i.e. $\phi = m\varphi $. Substituting this expression
into Eqs.~\reqt{Seq},~\reqt{Seq2} we obtain
\begin{equation}
\leqt{R}
\frac{d^2U}{dr^2}+\frac{1}{r}\frac{dU}{dr}-\frac{m^2}{r^2} U -kU+F(U^2)U=0,
\end{equation}
where the radially symmetric amplitude $U(r)$ vanishes at the center, $U\sim r^{|m|}$ at $r\to 0$. The rotation velocity should be quantized by the condition of the field univocacy Eq.~\reqt{S}, and therefore $m$ is integer [see Fig.~\rpict{fig5}(b,c)]. The index $m$ stands for a phase twist around the intensity ring, and it is usually called winding number, topological index or topological charge of the solitary wave. Such winding number distinguishes {\em azimuthal} higher-order stationary states, in addition to the radial modes, so that the full set of radially-symmetric spatial soliton families can be denoted by $P_{n,m}(k)$, with radial and azimuthal quantum numbers $n$ and $m$. Figure~\rpict{fig5}(b) illustrates the envelopes $U(r)$ for single ring ($n=0$) vortex solitons with different topological charges $m$. Corresponding families are divided by the minimal threshold power, necessary for the formation of corresponding higher-order states, the dependencies $P_{0,m}(k)$ for a fixed nonzero saturation $\alpha=0.2$ are shown in Fig.~\rpict{fig5}(c). The threshold power is defined in the limit of zero saturation $\alpha=0$, i.e. in pure Kerr medium $F(I)=I$, where soliton constant $k$ plays a role of a scaling parameter, see horizontal lines in Fig.~\rpict{fig5}(c). Subsequently, vortex solitons were re-discovered in other studies by \cite{Kruglov:1992-2277:JMO, Atai:1994-R3170:PRA, Afanasjev:1995-3153:PRE} and for other types of nonlinear optical media, including quadratic nonlinear media (\cite{Torner:1997-608:ELL, Torner:1997-2017:JOSB, Firth:1997-2450:PRL, Skryabin:1998-3916:PRE}), the latter nonlinear model is discussed in Sect.~\rsect{X2}.

The important integral of motion associated with this type of the
solitary waves is the beam {\em angular momentum},
\[ M{\bf e}_z={\rm Im}\int E^{\ast} ({\bf r} \times \nabla E) d{\bf r}, \]
which can be expressed through the soliton amplitude and phase,
\begin{equation}
\leqt{mom}
M = \int \frac{\partial \phi}{\partial \varphi}\ U^2 d{\bf r}.
\end{equation}
It is important to point out that the nonvanishing angular
momentum is an overall property of the light beam, not necessarily
directly connected to the quantum properties at the single photon
level. Similarly, the angular momentum is not necessarily
associated with optical singularities although in practice the two
phenomena may occur together (\cite{Allen:2002-S1:JOB,Berry:2004-S155:JOA,Padgett:2004-35:PT}).
The angular momentum of a paraxial light beam can be separated
into spin and orbital parts (see, e.g.,
\cite{Cohen:1989:Photons,Barnett:2002-S7:JOB}), where {\em the
spin momentum} is associated with the polarization structure of
the light, and {\em the orbital momentum} is associated with the
spatial structure of the beam, in particular, the beam carrying
optical vortices. Therefore, the angular momentum of a scalar
vortex soliton defined by Eq.~\reqt{mom} should be identified as
an orbital angular momentum. However, as we describe below, the
ring-profile vortex optical beams  experience {\em the azimuthal
instability} in nonlinear media, and they decay into a number of
{\em moving} fundamental solitons. Because the input beams carry
the overall angular momentum given by Eq.~\reqt{mom}, the
splinters fly off the ring along the tangential trajectories.
Therefore, in the soliton community it has become customary to
refer to the {\em soliton spin angular momentum}, and thus to {\em
spinning solitons}, and to use {\em orbital angular momentum} in
the case of several interacting solitons. This notion leads to the
description of the break-up of a vortex due to modulational
instability in terms of the transformation of the initial soliton
spin angular momentum to the net orbital angular momentum of
moving splinters (\cite{Firth:1997-2450:PRL,
Skryabin:1998-3916:PRE}). The ratio of the soliton angular
momentum to its power can be identified with {\em the soliton
spin}, $S = M/P$, and for the vortex solitons the spin is equal to
its nonzero topological charge $S=m$ [cf. Eq.~\reqt{S}]. We
notice, however, that such a notation might be confusing when it
is used in other areas where the concepts of spin angular momentum
and orbital angular momentum are employed in the rigorous, proper
sense.

\subsection{Azimuthal modulational instability}
\lsect{MI}

As was shown in many numerical and analytical studies, the
ring-like vortex beams in self-focusing nonlinear media are
subject to the azimuthal symmetry-breaking modulational
instability, a specific type of transverse modulational
instability similar to one which is responsible for filamentation
of the beams and generation of trains of optical solitons
(\cite{Kivshar:2000-118:PRP}). This effect should not be mixed
with the well-known collapse instability
(\cite{Berge:1998-260:PRP}) which is eliminated in the nonlinear
media with saturable and quadratic nonlinearity. As a result,
stable fundamental solitons have been found in two and three
spatial dimensions and the instability criterion was established
by \cite{Vakhitov:1973-1020:IVR}. It states that the principal
mode of the non-linear wave equation~\reqt{eq} is stable if its
integrated intensity Eq.~\reqt{power} has a positive slope,
$\partial P/\partial k>0$, and the latter is possible if the
nonlinearity growth monotonically with intensity, i.e. for
$dF/dI\ge 0$. Both these conditions are satisfied for the vortex
solitons in, e.g., self-focusing media with saturation, see
Fig.~\rpict{fig5}(c). However, this property does not guarantee
the vortex stability against azimuthal perturbations.

So far, no universal criterion, similar to the Vakhitov-Kolokolov
stability criterion for the fundamental optical solitons, has been
suggested for the stability of vortex solitons, and the vortex
stability should be addressed separately for different models.
Below, we present the stability analysis developed by
\cite{Soto-Crespo:1992-3168:PRA,Firth:1997-2450:PRL,Torres:1998-625:JOSB}.

We assume that there exists a stationary solution $E=E_0$ solving
Eq.~\reqt{eq}. The {\em linear stability} of this solution can be
established by the behavior of an additional small perturbation
$|p|\ll|E_0|$. Substituting the perturbed solution $E=E_0+p$ into
Eq.~\reqt{eq} and linearizing it with respect to a small
perturbation $p$, we obtain
\begin{equation}
\leqt{perteq}
i\frac{\partial p}{\partial z}+\Delta_{\perp}p+(F_0+|E_0|^2F'_0)\,p + E_0^2F'_0\,p^*=0,
\end{equation}
where $F_0=F(|E_0|^2)$ and $F'_0=(dF/dI)|_{I=|E_0|^2}$.
Equation~\reqt{perteq} describes the evolution of initially small
perturbation $p$ corresponding to the solution $E_0$: if $p$ does
not grow with the beam propagation, the stationary solution $E_0$
is linearly stable.

The linear equation~\reqt{perteq} can be solved by the separation
of variables, depending on the geometry of underlying stationary
point $E_0$. For the case of radially symmetric vortices
determined by \reqt{R}, i.e. $E_0=U(r)\exp(im\varphi+ikz)$, the
perturbation $p$ should posses an azimuthal periodicity and,
therefore, it can be represented as a Fourier series
\begin{equation}
\leqt{pert0}
p(r,\varphi,z)=\sum_{n=-\infty}^{\infty}p_n(r,z)\exp(in\varphi).
\end{equation}
Substituting this decomposition into Eq.~\reqt{perteq} and
matching terms with equal angular dependance, we obtain the
infinite set of equations for complex modal functions $p_n$.
However, for any integer $s$, only two modes $p_{m+s}$ and
$p_{m-s}$ are actually coupled and build a closed system:
\begin{equation}
\leqt{pert1}
\left\{i\frac{\partial}{\partial z}+\hat{L}^{\pm}\right\}p_{m\pm s}+\exp(i2kz)Ap_{m\mp s}^*=0,
\end{equation}
where $A\equiv U^2F'_0$ and $\hat{L}^\pm\equiv
d^2/dr^2+r^{-1}d/dr-(m\pm s)^2r^{-2}+F_0 +A$. Solution to these
equations is given by $p_{m+s}(r,z)=u_s(r)\exp(ikz+\gamma_s z)$
and $p_{m-s}(r,z)=v_s^*(r)\exp(-ikz+\gamma^*_s z)$, with complex
perturbation wave number $\gamma_s$ and the modes $u_s$ and $v_s$
that solve an eigenvalue problem
\begin{equation}
i\gamma_s
\left(\begin{array}{c}u_s\\v_s\end{array}\right)=
\left[\begin{array}{lr}
k-\hat{L}^+&-A\\
A &-k+\hat{L}^-
\end{array}\right]
\left(\begin{array}{c}u_s\\v_s\end{array}\right).
\leqt{pertmod}
\end{equation}
In these notations, if the eigenvalue $\gamma_s$ has a positive
real part for some $s$, the perturbation modes $p_{m\pm s}$ grow
exponentially with the {\em growth rate} $\rm{Re}\;\gamma_s>0$;
and such modes are called {\em instability modes}. The equivalent
representation of perturbation
$p(r,\varphi,z)=\exp(im\varphi+ikz)\left\{u_s(r)\exp(is\varphi+\gamma_s
z)+v^*_s(r)\exp(-is\varphi+\gamma^*_s z)\right\}$ guarantees, due
to the completeness of the basis of azimuthal harmonics
Eq.~\reqt{pert0}, that all possible perturbations are taken into
account.

\pict[0.708]{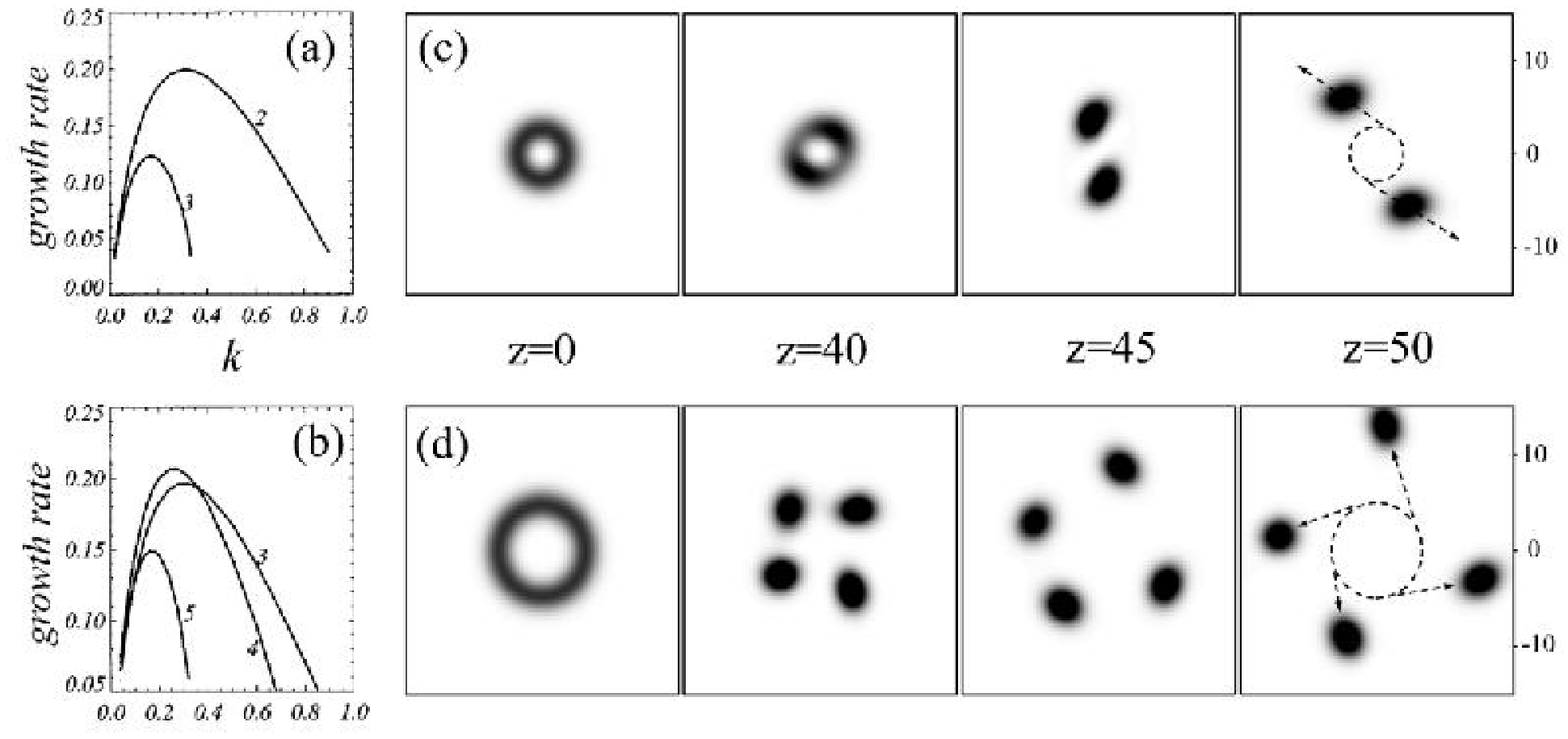}{fig6}{Growth rate, ${\rm Re}\;\gamma_s(k)$,
of the instability modes for (a) single- and (b) double-charged
vortex solitons in saturable medium, $F(I)=I/(1+I)$. Corresponding
values of perturbation index $s$ are shown next to the curves.
After \cite{Skryabin:1998-3916:PRE}. Break-up of the vortex
solitons with the maximum growth rate and the charges (c) $m=1$
and (d) $m=2$. Dashed curves at $z=50$ show the peak intensity of
the initial rings and trajectories of the solitons flying away
after the decay.}

The instability growth rate depends on the vortex power, as is
shown in Figs.~\rpict{fig6}(a,b) for a particular case of spatial
vortex solitons with the topological charges $m=1$ and $m=2$ in a
saturable medium with $\alpha=1$. For any $s$, the growth rate
vanishes in the linear limit $k\to 0$, and also in the opposite
limit of infinite power, $k\to 1/\alpha$, but at least one mode
has a nonzero positive value in the whole domain of the vortex
soliton existence, $k\in[0,1/\alpha]$. Thus, all vortex
solitons are {\em linearly unstable} in saturable media. Similar
results hold for the parametric interaction in a quadratic medium,
as we discuss in detail in Sect.~\rsect{X2}.

The index of the modes with the highest growth rate depends on the
model and the mode topological charge. For example, a
single-charged vortex {\em always} has the $s=2$ mode growing
faster than other modes in a saturable medium, while this can be
the $s=3$ mode in a quadratic medium, and, as shown in
Fig.~\rpict{fig1}(b), also in a pure Kerr medium with $\alpha=0$.
In general, the higher-charge vortices allow for competition
between different modes with close values of the growth rate, such
as the modes with $s=3$ and $s=4$ for $m=2$, see
Fig.~\rpict{fig6}(b).

The symmetry-breaking instability of the ring-like vortex beams
has been observed experimentally in saturable vapors
(\cite{Tikhonenko:1995-2046:JOSB, Tikhonenko:1996-2698:PRL,
Bigelow:2004-83902:PRL}), photorefractive
(\cite{Chen:1997-1751:OL}) and quadratic
(\cite{Petrov:1998-1444:OL}) nonlinear media. In all such cases,
the generation of different numbers of fundamental solitons due to
the ring instability was observed. Figure~\rpict{fig6} shows a
numerical example of the breakup scenario of the ring-like vortex
solitons, when the initial stationary ring-profile structure
decays, under the action of a numerical noise, into two [the case
$m=1$, see Fig.~\rpict{fig6}(a)] or four [the case $m=2$, see
Fig.~\rpict{fig6}(b)] fundamental solitons. The number of
splinters coincides exactly with the topological index of the
instability mode with highest growth rate, see
Fig.~\rpict{fig6}(a,b), thus the predictions of a linear stability
analysis are in an excellent agreement with the numerical solution
of the full system. The detailed stability analysis of solitary
waves with central phase dislocation were reported by
\cite{Skryabin:1998-3916:PRE} for both  saturable self-focusing
and quadratic nonlinear media, and for the latter case also by
\cite{Torres:1998-625:JOSB}. In the experiments, the excitation of
the ring-profile vortex beams is conducted by pumping the
nonlinear media by suitable approximations of Laguerre-Gaussian
modes, and it was shown earlier that not only the spontaneous
symmetry-breaking instability of the vortex solitons could be
observed in that geometry but also that such excitation conditions
offer additional possibilities by inducing suitable instabilities (\cite{Torner:1997-608:ELL, Torner:1997-2017:JOSB}), as we discuss
in detail in Sect.~\rsect{X2}.

An analytical approach to the study of the filament dynamics after
the breakup, based on the conservation of the beam angular
momentum and Hamiltonian, was developed by
\cite{Skryabin:1998-3916:PRE}. Given initial values of the
conserved quantities, it is possible to predict the features of
the filament trajectories and estimate their number. Two different
analytical expressions for the velocity of the filaments in the
transverse plane were derived, both formulas giving a reasonably
good quantitative predictions for the velocities. The formula
based on angular momentum being particularly simple and effective:
the escape velocity can be estimated as $v\approx |m|/R$, where
$R$ is the initial radius of the vortex ring. The important
conclusion, giving an insight into the underlying physics of the
beam breakup, is that when filaments move out along tangents to
the initial ring, they carry away its angular momentum. In our
notations we can describe this effect as the transformation of
initial spin angular momentum of the vortex soliton to the angular
momentum of the splinters spiraling out.

Stabilization of coherent vortex solitons against the azimuthal
instabilities remains a major challenge in the physics of spatial
vortex solitons. Several theoretical models were suggested which
support stable vortex solitons, including the formation of vortex
solitons in the presence of competing nonlinearities (see
Sect.~\rsect{comp}), nonlocal nonlinear media
(\cite{Yakimenko:nlin.PS/0411024:ARXIV, Breidis:2005:OE}), or Bessel photonic
lattices (\cite{Kartashov:2005-00000:PRL}), without experimental
observations so far.

\section[Vector vortices]{Composite spatial solitons with phase dislocations}
\lsect{X3vect}

In this section we describe optical vortices (and other closely
related higher-order spatial solitons) composed of several beams
(components) interacting via cross-phase modulation (XPM). Because
all of those components contribute to the total beam intensity,
the refractive index change is usually refereed as commonly
induced waveguide, and such composite solitons can be thought as
corresponding guided modes. In the simplest case of two
interacting beams, e.g. two orthogonally polarized components of a
vector soliton, this system posses radially-symmetric vector
vortices as well as azimuthally modulated multipole and
necklace-ring vector solitons. For large number of mutually
incoherent components the composite solitons are closely related
to partially coherent self-trapped beams and vortices.

\subsection{Soliton-induced waveguides}
\lsect{WG}

Spatial solitons may be understood as the modes of the effective waveguides they induce in a nonlinear medium (\cite{Kivshar:2003:OpticalSolitons}). A natural extension of this concept is to assume that the waveguide induced by relatively
powerful soliton beam may guide and control another weak beam. This concept may be applicable to bright (\cite{DeLaFuente:1991-793:OL}) as well as dark (\cite{LutherDavies:1992-496:OL, LutherDavies:1992-1755:OL}) spatial solitons.

The possibility of effective waveguiding of a weak probe (or signal) beam via the XPM type interaction can be analyzed within the coupled system of NLS equations for two wave envelopes $E_{1,2}$,
\begin{eqnarray}
\leqt{eq1}
i \frac{\partial E_1}{\partial z}+\Delta_{\perp} E_1 + \sigma\left(c_{11}|E_1|^2+c_{12}|E_2|^2\right)E_1=0,\\
\leqt{eq2} i \frac{\partial E_2}{\partial z}+\Delta_{\perp} E_2 +
\sigma\left(c_{21}|E_1|^2+c_{22}|E_2|^2\right)E_2=0,
\end{eqnarray}
where the coefficient $\sigma= \pm 1$ defines a focusing and
defocusing nonlinear medium, respectively. In a general case with
different carrier frequencies and polarization states of two
interacting beams, the contributions $c_{nn}$ from self-phase
modulation (SPM) and the XPM coefficients $c_{nj}$ are all
different, $c_{nj}\ne c_{jn}\ne c_{nn}$ ($n,j=1,2$ and $j\ne n$).
If the two waves have the same carrier frequency but different
polarization states, the interaction is symmetric, $c_{nj}=c_{jn}$
and $c_{jj}=c_{nn}$, and one of them can be scaled away, e.g. SPM
coefficients $c_{nn}=1$. In the latter case, the XPM coefficients
$c_{nj}=2/3$, for linearly polarized components, $c_{nj}=2$, for
circular polarized components, and in the case of elliptically
polarized components this parameter satisfied the relation:
$2/3<c_{nj}<2$. Transition to the Manakov system $c_{nj}=1$,
completely integrable in 1D geometry
(\cite{Manakov:1974-248:JETP}), occurs at ellipticity angle $35^o$
(\cite{Menyuk:1989-2674:IQE}).

When the signal beam, e.g. $E_2$, is much weaker than the soliton
beam $|E_2|\ll|E_1|$, the system~\reqt{eq1},\reqt{eq2} can be
linearized with respect to $E_2$. Then the equation~\reqt{eq1}
transforms to the NLS equation~\reqt{eq} with the nonlinear
potential $F(I)=\sigma c_{11} |E_1|^2$ which supports stationary
soliton solutions, while the signal wave $E_2$ propagates in
effectively \textit{linear} regime in the waveguide with the
refractive index $\sigma c_{21} |E_1|^2$. Depending of the actual
shape of the soliton beam $E_1$ and its peak intensity, the
induced waveguide can be single-moded, i.e. supporting only the
fundamental bell-shaped mode $E_2$, as well as multi-moded, when
the higher-order guided modes, including vortices, can be guided.

\pict[0.708]{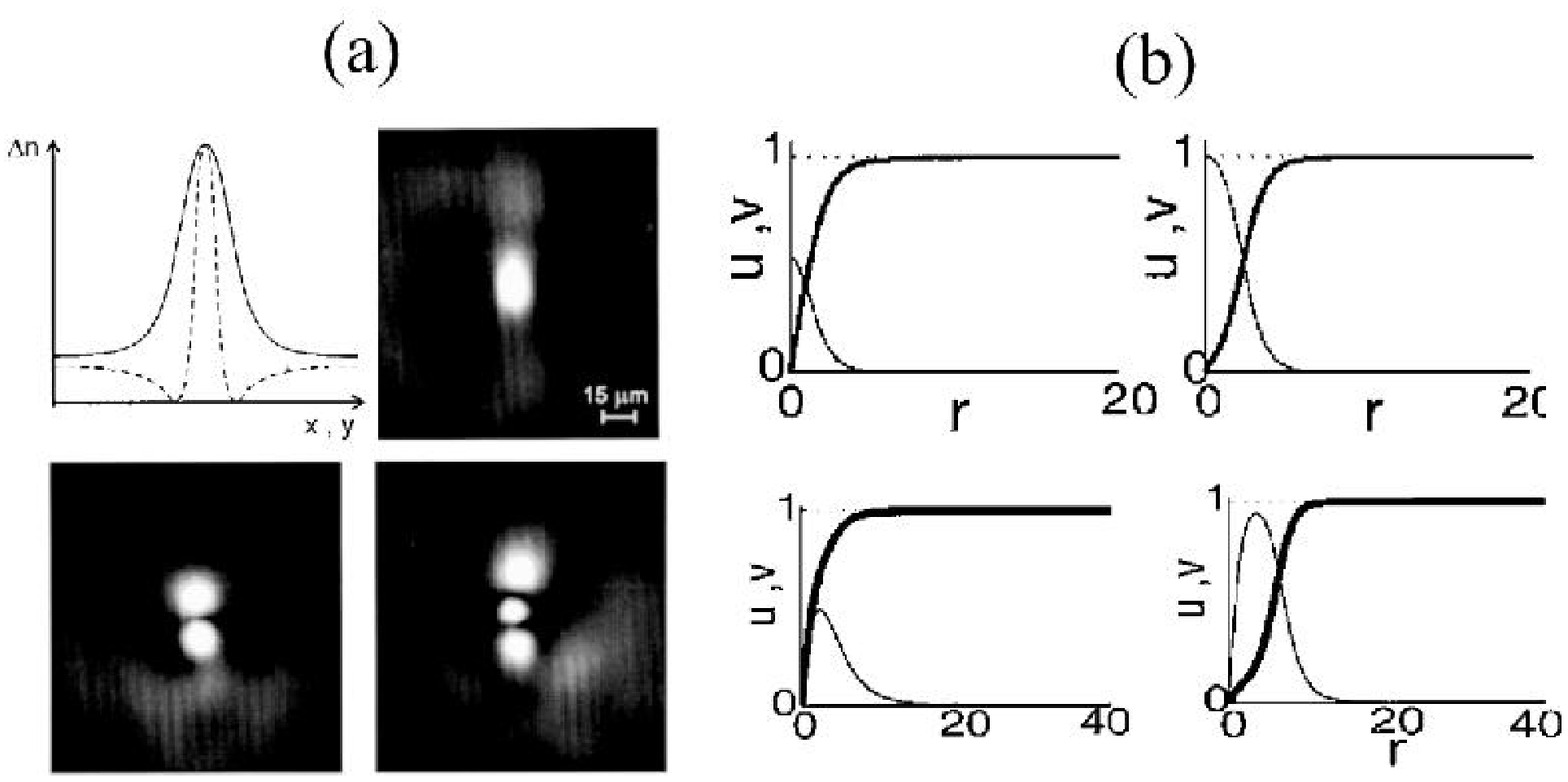}{fig7}{(a) Higher-order modes guided by a fundamental soliton in the self-focusing photorefractive medium: plot shows the calculated distribution of the refractive index induced by the fundamental soliton and experimental photos demonstrate the intensity of three first modes of a guided red probe beam (\cite{Petter:2002-1145:JOSB}). (b) Radially symmetric fundamental (top) and vortex (bottom) modes localized in the waveguide induced by the vortex soliton (thick lines) in self-defocusing medium (\cite{Carlsson:2000-660:OL}).}

In focusing nonlinear media, such as anisotropic photorefractive
crystals, the fundamental soliton has a bell-shaped form
(\cite{Segev:1992-923:PRL}). Similar to the 1D case
(\cite{Morin:1995-2066:OL, Shih:1997-3091:JOSB}), the (grin) light
beam from a frequency-doubled Nd:YAG laser creates a
two-dimensional  spatial soliton and induces an effective
waveguide for the probe beam at less photosensitive wavelength; in
SBN crystal it can be taken from the HeNe laser (red beam). The
effective contribution to the refractive index from the red beam
is negligible, i.e. both the SPM $c_{22}\sim 0$ and XPM
coefficient $c_{12}\sim 0$ in Eqs.~\reqt{eq1},\reqt{eq2}. At the
same time the guidance of a probe beam by a single two-dimensional
soliton as well as by the pair of interacting solitons is indeed
possible, it was demonstrated by \cite{Petter:2001-55:OC}.
Furthermore, because of the anisotropy of photorefractive
screening nonlinearity (\cite{Zozulya:1998-522:PRA}), the
effective waveguide is highly anisotropic and may support
higher-order modes. Figure~\rpict{fig7}(a) shows the induced
refractive index profile and corresponding higher-order TEM guided
modes, generated experimentally by \cite{Petter:2002-1145:JOSB}.

In defocusing nonlinear media, spatial solitons are associated
with a nontrivial phase structure, and a two-dimensional soliton
is a \textit{dark} vortex, as discussed in Sect.~\rsect{dark}. In
this case, the spatial profile of the induced refractive index
(``induced fiber'') allows to support radially symmetric spatially
localized modes (\cite{Snyder:1992-789:OL}). The waveguiding
properties of dark optical vortex solitons in self-defocusing Kerr
media have been analyzed by~\cite{Sheppard:1994-859:OL,
Law:2000-55:OL, Carlsson:2000-660:OL}. It was shown that these
properties depend crucially on the relative strength of the cross-
and self-phase modulation effects. Families of composite solitons
formed by a vortex and its guided mode with or without a
topological charge have been identified. Examples of these modes
are presented in Fig.~\rpict{fig7}(b).

A soliton-waveguiding experiment has been conducted by
\cite{Truscott:1999-1438:PRL} in an atomic vapor. A weak probe
beam tuned near one atomic resonance is guided through a waveguide
written by an intense pump \textit{vortex} beam at a different
atomic resonance. As the pump beam is tuned close to resonance, it
creates a nonlinear refractive index profile in the atomic vapor
with which the weak probe beam interacts. The efficiency of the
guiding is found to depend strongly on the power and frequency of
the guiding beam. Moreover, since the guiding takes place in an
atomic vapor, it is possible to tune to both sides of the atomic
resonance. This has the distinct advantage that it allows the
guiding of light into either bright or dark regions of the guiding
beam. The theory of waveguides electromagnetically induced in Rb
vapors was developed by \cite{Kapoor:2000-53818:PRA}. Their
density matrix approach was based on the three-level V-system, and
it was generalized latter to the five-level model by
\cite{Andersen:2001-23820:PRA}, who took into account the
hyperfine structure of the D-line of rubidium as well as the
presence of the two major isotopes. The results allow one to
deduce which frequency combinations are likely to give successful
guiding.

Systematic analysis of the waveguiding properties of the vortex
solitons and vortex-mode vector solitons in saturable nonlinear
media, for both self-defocusing and self-focusing nonlinearities,
was performed recently by~\cite{Salgueiro:2004-327:UJP}. Following
the earlier analysis of~\cite{Carlsson:2000-660:OL}, the authors
examine two major regimes of the vortex waveguiding. The most
interesting {\em nonlinear regime} corresponds to large
intensities of the guided beam, and it gives rise to composite (or
vector) solitons with a vortex component, that have been
identified and analyzed numerically.

In quadratic media, the simultaneous guidance of both fundamental
and second-harmonic waves by an optical vortex soliton has been
analyzed by \cite{Salgueiro:2004-593:OL}. These authors describe
novel types of three-component vector soliton created by a vortex
beam together with both fundamental and second-harmonic
parametrically coupled localized modes and determine conditions
for a potential enhancement of the conversion efficiency.

\subsection{Higher-order vector solitons}
\lsect{multi}

When the guided beam is weak, i.e. in the linear waveguiding
regime, the analysis of the soliton waveguiding properties can be
carried out by using approximate analytical methods, and reducing
the problem to the well-known analysis of the linear guided-wave
theory (for example, see \cite{Law:2000-55:OL} and also the theory
of optical vortices in optical fibers by~\cite{Volyar:1998-272:OPS,
Volyar:1999-242:OPS}). For a finite-amplitude probe beam, the
linearization in Eqs.~\reqt{eq1},\reqt{eq2} is no longer valid,
and the nonlinear theory of soliton-induced waveguides should be
developed (\cite{Ostrovskaya:1998-1268:OL}); this theory takes
into account \textit{mutual} effect of the interacting waves on
each other.

Several light beams generated by a coherent source can be combined
together to produce a {\em vector} soliton with a complex internal
structure. Properties of two-component vector solitons have been
extensively studied in both self-focusing
(\cite{Manakov:1974-248:JETP, Christodoulides:1988-53:OL,
Snyder:1994-1012:PRL, Christodoulides:1996-1763:APL,
Krolikowski:1996-782:OL}) and self-defocusing
(\cite{Kivshar:1993-337:OL, Haelterman:1994-3376:PRE}) nonlinear
media. The structure of multi-component vector soliton may become
rather complicated and, for example, the soliton intensity profile
may display several peaks. These so-called ``multi-hump'' solitons
propagate as the corresponding higher-order modes of the
soliton-induced waveguides. The first experimental demonstration
of such multi-component solitons has been reported by the
Princeton group (\cite{Mitchell:1998-4657:PRL}), who observed both
single and multi-peak spatial solitons. Linear stability of the
two-component vector solitons has been studied by
\cite{Ostrovskaya:1999-296:PRL}, and it has been shown that the
two-peak structures can be stable in propagation while three-peak
soliton structures are unstable.

\pict[0.708]{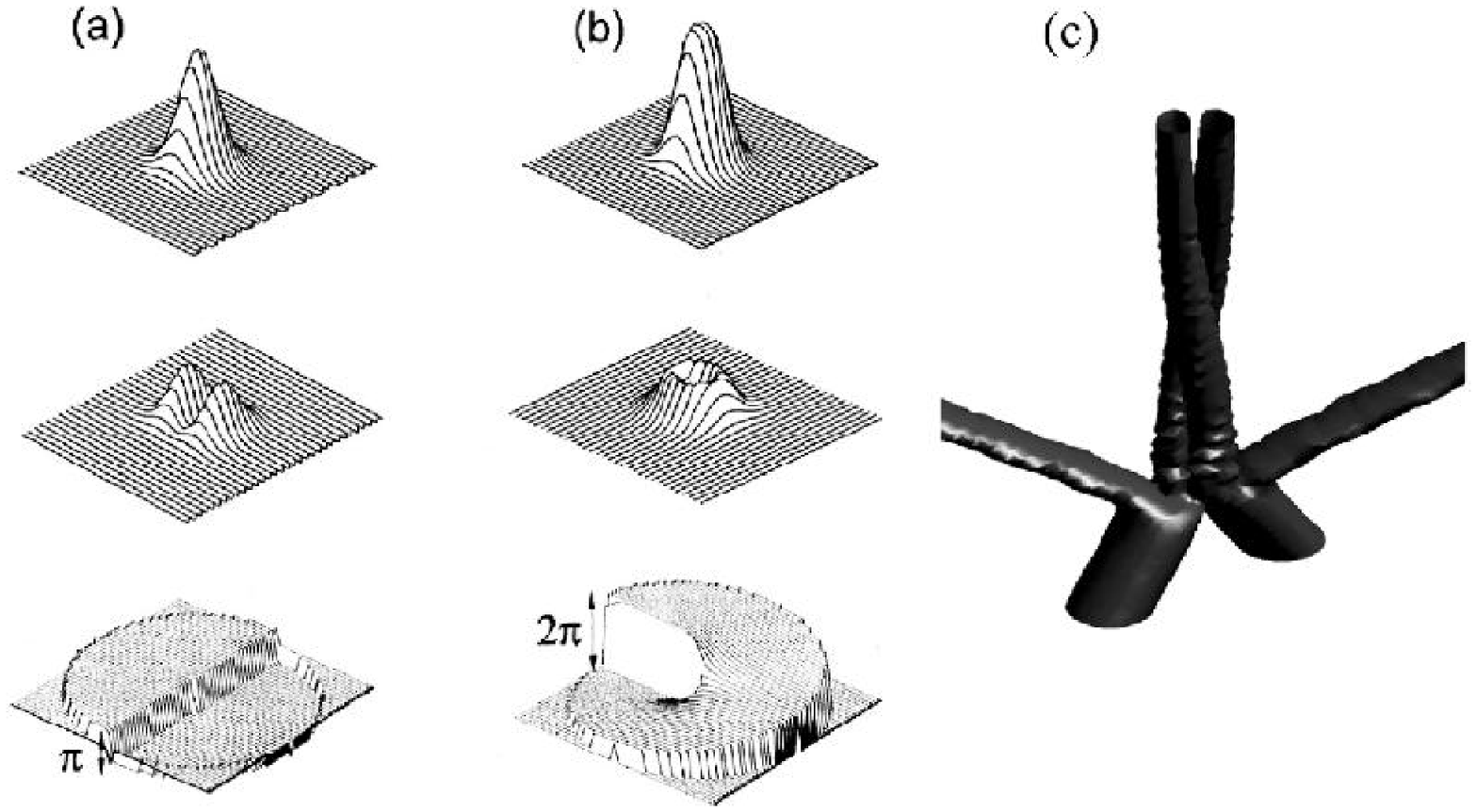}{fig8}{Examples of constituents of (a)
dipole-mode ($s=0.3$) and (b) vortex-mode ($s=0.65$) two-component
vector solitons (top row: $|E_1|^2$, middle row: $|E_2|^2$) shown
with the phase distributions needed to generate the $E_2$-modes
experimentally (bottom row), after
\cite{Krolikowski:2000-1424:PRL}. Delayed-action interaction
between the vortex-mode vector solitons and formation of spiraling
dipole (\cite{Musslimani:2001-799:PRL}).}

Recently, the concept of multi-hump spatial solitons has been
extended to two transverse dimensions. First,
\cite{Musslimani:2000-1164:PRL, Musslimani:2000-61:OL} studied
stationary propagation of the {\em vortex-mode vector soliton}
which has a nodeless shape in one component and a vortex in the
other component, see Fig.~\rpict{fig8}(b). However, it appears
that such a {\em radially symmetric}, ring-like vector soliton
(which is analogous to the Laguerre-Gaussian modes of cylindrical
waveguide) may undergo a {\em symmetry-breaking instability}
(\cite{Garcia-Ripoll:2000-82:PRL, Malmberg:2000-643:OL}) which
transforms it into a {\em radially asymmetric} dipole-mode vector
soliton, even in a perfectly isotropic nonlinear medium, see
Fig.~\rpict{fig8}(a). The dipole-mode vector soliton is a novel
type of an optical vector soliton that originates from trapping of
a dipole ${\rm HG}_{01}$-type mode by a
fundamental-soliton-induced waveguide created by the other,
incoherently coupled, co-propagating beam. Moreover, it has been
shown that, while many other topologically complex structures may
be created, it is only the dipole mode that is expected to
generate a family of {\em dynamically robust vector solitons}.
Very recently, the rigorous stability analysis performed by
\cite{Yang:2003-16608:PRE}, has shown that very close to the
bifurcation line, where the vortex and dipole components are
small, both solutions are linearly stable. Far from the
bifurcation line, the family of vortex solitons becomes
azimuthally unstable, while the dipole-mode solitons remain stable
in the whole domain of their existence.

Dipole-mode vector solitons have been observed experimentally  in
photorefractive media by \cite{Krolikowski:2000-1424:PRL}. Two
different methods have been used, one is based on a phase
imprinting technique, and another uses the symmetry-breaking
instability of vortex-mode soliton. In the latter case, the
initial angular momentum of vortex component is transformed to the
dipole and leads to its rotation during propagation after the
break-up. \cite{Skryabin:2002-55602:PRE} showed that rotational
velocity provides an additional parametrization for the
dipole-soliton family, and, with the help of a generalized
Vakhitov-Kolokolov stability criterion, they predicted stability
thresholds for spiraling solutions. Rotating dipole-mode soliton
can be viewed as an optical ``propeller'' because of the mutually
tilted phase fronts of the dipole lobes
(\cite{Carmon:2001-143901:PRL}). In anisotropic photorefractive
media, however, the rotation of the dipole is limited because of
preferable direction for its orientation
(\cite{Neshev:2001-1185:OL, Motzek:2001-161:OC}).

It appears that the interaction of composite solitons, determined
by the effective interaction potential
(\cite{Malomed:1998-7928:PRE}), depends significantly on their
angular momenta. \cite{Musslimani:2001-799:PRL} demonstrated
numerically that collisions of vortex-mode solitons with opposite
topological charges ("spins") can lead to mutual trapping of two
composite beams and the formation of bound state with a prolonged
lifetime of about $35$ diffraction lengths. The metastable bound
state eventually disintegrates giving rise to new vector solitons,
the process is characterized as the ``delayed action
interaction''. If both colliding vortex-mode solitons have the
same topological charge, the resulting object is a stable rotating
dipole-mode soliton, as is shown in Fig.~\rpict{fig8}(c). These
authors draw an analogy with spin-orbit coupling in interaction of
soliton as effective ``particles''. The comprehensive study of
collisions between vortex-mode solitons was reported by
\cite{Musslimani:2001-66608:PRE}, and out-of-plane scattering of
the dipole-mode solitons was studied by \cite{Pigier:2001-1577:OL,
Krolikowski:2003-16612:PRE}.

The dipole-mode vector soliton is the first example of
\textit{azimuthally-modulated} spatial solitons. As we noted
above, close to the bifurcation line this composite structure can
be described as a linear mode guided by a scalar fundamental
soliton. Natural extension of this approach is to search for
higher-order modes such as \textit{multipole-mode} vector
solitons. Indeed, the vortex-mode soliton can be described as a
coherent superposition of two dipole components twisted in space
and shifted in phase by $\pi/2$ with respect to each other.
Similarly, the higher-charge optical vortex might be constructed
from multi-poles, e.g. double charged vortex consists of two
quadrupoles. This analogy with linear waveguide theory suggests
the general ansatz for the azimuthally modulated component of a
composite spatial soliton,
\begin{equation}
\leqt{multipol}
E_2(x,y,z)=U(r)\left\{\cos m\varphi+i p \sin m \varphi\right\}\exp(ikz),
\end{equation}
where the parameter $p$ is real, $m$ is integer, and $k$ is the
propagation constant. For $p=1$, Eq.~\reqt{multipol} describes a
vortex-mode soliton with the charge $m$ and the radially symmetric
intensity $U^2(r)$. This ansatz has been used by
\cite{Desyatnikov:2001-435:OL} as a variational approximation to
the modes [$E_2$-component in equations similar to
Eqs.~\reqt{eq1},\reqt{eq2}] of the waveguide, induced by
fundamental soliton in the $E_1$ component. Experimental results
of the generation of \textit{quadrupole} and \textit{hexapole}
vector solitons are shown in Fig.~\rpict{fig9}. The corresponding
parameters in the ansatz Eq.~\reqt{multipol} are $m=2$ and $m=3$
correspondingly, with $p=0$ in both cases. For $p\ne 0$, the
angular momentum of the azimuthally modulated solitons is nonzero
and this leads to the soliton rotation.

\pict[0.708]{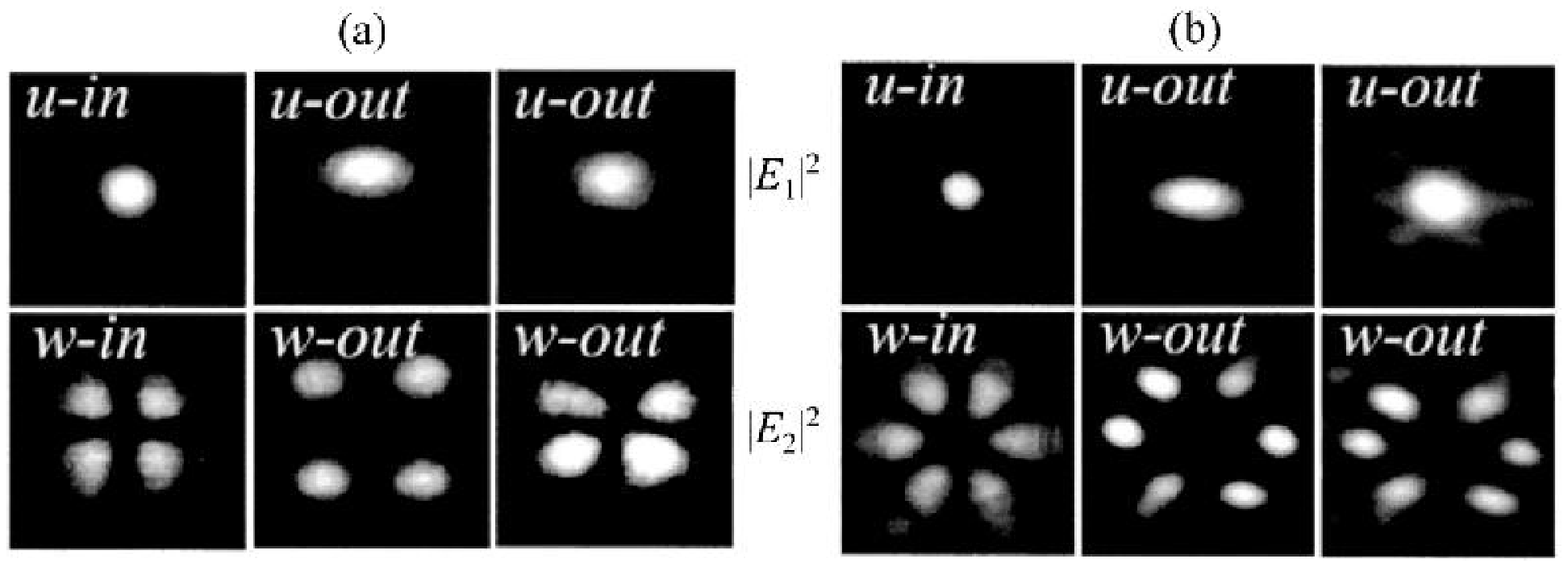}{fig9}{Experimental demonstration of
multipole-mode solitons: (a) quadrupole vector soliton, and (b)
hexapole vector soliton. Intensities of two components shown at
the output after independent (middle columns) and simultaneous
(right columns) nonlinear propagation. In all cases, the
propagation distance is $z=10$mm.
After~\cite{Desyatnikov:2002-586:JOSB}.}

Incoherent interaction between the components of a composite (or
vector) ring-like beam allows to compensate for repulsion of
beamlets, creating a new type of quasi-stationary self-trapped
structure exhibiting the properties of the necklace-ring beams and
ring vortex solitons. The physical mechanism for creating such
composite vector ringlike solitons is somewhat similar to the
mechanism responsible for the formation of the so-called {\em
soliton gluons} (\cite{Ostrovskaya:1999-327:OL}) and multi-hump
vector solitary waves (\cite{Ostrovskaya:1999-296:PRL}), and it is
explained by a balance of the interaction forces acting between
the coherent and incoherent components of a composite soliton. In
that case, the mutual repulsion of out-of-phase beamlets in the
$E_2$-component is balanced by the incoherent attraction of the
mutually coupled $E_1$ component.

\subsection{Multi-component vortex solitons}

The effective optical waveguide induced by the fundamental soliton
or dark vortex soliton in a self-focusing or self-defocusing
nonlinear media, respectively, has a relatively simple bell-like
shape whose modes are well known from the linear guided-wave
theory. In contrast, the doughnut shape of a bright vortex soliton
does not allow simple predictions about its guided modes, and,
consequently, the possible structure of multi-component vortex
solitons. Moreover, similar to the one-dimensional multi-hump
vector solitons, their two-dimensional counterparts have a complex
non-monotonous radial envelope (\cite{Musslimani:2000-1164:PRL}).
For example, in addition to the well-understood ``first''
bifurcation from the scalar fundamental soliton, giving rise to
the vortex-mode solitons, the corresponding stationary solutions
have the second bifurcation point where the ``guided'' component
transforms to the scalar vortex
(\cite{Desyatnikov:2002-586:JOSB}). This situation can also be
regarded as guiding of a simple bell-shaped beam by a ring-like
waveguide. Very recently, the comparison between the so-called
single- and double-vortex solitons has been carried out
by~\cite{Salgueiro:2004-000:PRE}. Here the notation
"single-vortex" is used to distinguish a vortex-mode soliton,
which consists of a strong fundamental component and a weak guided
vortex (see Fig.~\rpict{fig8}(b)), from the other case (``second''
bifurcation) with a strong vortex component and a small
fundamental beam. At the same time, the ring vortex waveguide also
supports double-vortex (or "vortex-vortex") vector solitons.
Several types of the double-vortex solitons are shown in the
Fig.~\rpict{fig10}(a).

Extension of the concept of two-dimensional multi-hump vector
solitons to the case of larger number of mutually incoherent
components was proposed by~\cite{Musslimani:2000-61:OL}, who
studied a radially symmetric potential commonly created by a
strong fundamental soliton and several vortex beams. Similar idea
applied to the ring vortex beams result in the so-called
``necklace-ring'' vector solitons
(\cite{Desyatnikov:2001-33901:PRL}). In general, the interaction
of $N$ paraxial beams via XPM can be described by the system of
coupled NLS equations for the envelopes $E_{n}(x,y,z)$ ($n=1,2,
\dots, N$),
\begin{equation}
\leqt{eqn} i \frac{\partial E_n}{\partial z} +\nabla^2_{\perp} E_n
+F(I)E_n=0,
\end{equation}
where, similar to Eq.~\reqt{eq}, the function $F(I)$ describes the
nonlinear refractive index, and the total beam intensity is
defined as $I=\sum^N_{n=1}|E_n|^2$. In contrast to the
system~\reqt{eq1},\reqt{eq2}, all the SPM and XPM contributions
are taken to be equal here for simplicity. In addition to a
variety of radially symmetric solutions, such as those displayed
in Fig.~\rpict{fig10}(a), there exists a special class of
\textit{azimuthally modulated} stationary states. The simplest
route to find this class of spatial solitons is to search for the
modes of a radially symmetric potential with $I=U^2(r)$. This is
possible if all components have the same radial envelope $U(r)$,
i.e. we are looking for the stationary solution in the form
\begin{equation}
\leqt{En}
E_n(x,y,z)=U(r)\Phi_n(\varphi)\exp(ikz),
\end{equation}
with the complex azimuthal envelopes $\Phi_n(\varphi)=a_n\cos
m\varphi+b_n\sin m\varphi$ selected in such a way that
$\sum^N_{n=1}|\Phi_n|^2=1$. The latter condition requires $\sum
\rm{Re}(a_nb^*_n)=0$ and $\sum |a_n|^2=\sum |b_n|^2=1$. These
equations define exact solutions to the system~\reqt{En} for any
$N$ and, in the particular case $N=1$, they describe a scalar
vortex of the charge $m$ with $a=1$ and $b=i$. Radially symmetric
multi-component vortices correspond to the symmetric case,
$b_n=\pm ia_n$; in this case the total soliton spin is integer $m$
or zero, for the counter-rotating vortices. In a general case, the
components resemble azimuthally modulated optical necklaces,
introduced by~\cite{Soljacic:1998-4851:PRL} (see
Sect.~\rsect{neck}), and the total spin may take a fractional
value.

\pict[0.708]{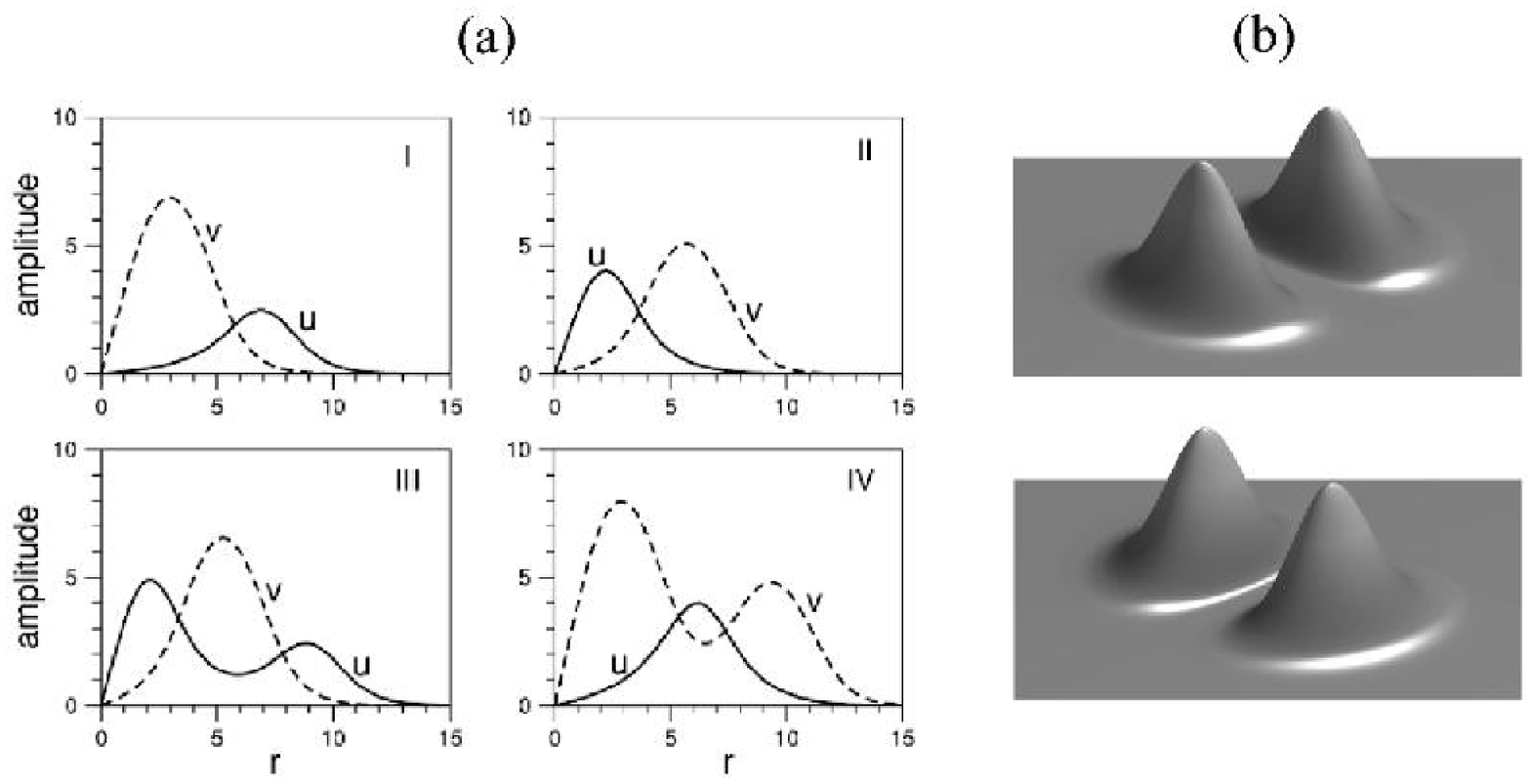}{fig10}{(a) Envelopes of different types of
double-vortex vector solitons (\cite{Salgueiro:2004-000:PRE}). (b)
Intensities of two components of a dipole-dipole vector soliton.
The total intensity of this soliton is a perfect ring as in
Fig.~\rpict{fig5}(a) (\cite{Desyatnikov:2001-33901:PRL}).}

The simplest two-component solution of the necklace-ring type with
$m=1$ is given by $a_1=b_2=1$, $a_2=b_1=0$, and it represents two
crossed dipoles, shown in Fig.~\rpict{fig10}(b). Although such a
structure has no vorticity and carries zero angular momentum, the
total intensity has a profile of a single-charge scalar vortex
soliton. This perfect symmetry helps to find such solutions, but
it is not crucial for their existence. Indeed, similar solutions
have been found numerically and generated experimentally
by~\cite{Ahles:2002-557:JOSB} in anisotropic photorefractive
medium, which posses no radially symmetry. The main outcome of
these studies, presented by~\cite{Desyatnikov:2001-33901:PRL}, is
that the vectorial interaction allows for additional stabilization
of otherwise nonstationary beams, such as expanding necklace
beams. Nevertheless, no linearly stable necklace-ring solitons
have been found so far.

Similar stabilization of \textit{counter-rotating} vector vortices
was reported for the case of a self-focusing saturable medium
by~\cite{Bigelow:2002-46631:PRE}, and was demonstrated
experimentally in a self-defocusing photorefractive medium
by~\cite{Mamaev:2004-S318:JOB}. The total angular momentum for
interacting vortices with opposite topological charges is less
than that of co-rotating vortices, and this was found to be the
main reason for long lifetimes. Theoretically, the maximal growth
rate of the azimuthal instability is significantly smaller for
zero-spin vector solitons (\cite{Ye:2004-219:OC}). The azimuthal
instability can be eliminated completely in the so-called
cubic-quintic (CQ) model, as discussed in Sect.~\rsect{CQ}, and
there exists also the stability domain for the co-rotating
vortices even in presence of four-wave mixing
(\cite{Mihalache:2002-615:JOA}). The counter-rotating vortices
with zero total angular momentum have smaller stability domain and
exhibit an interesting ``internal'' instability dynamics in CQ
medium (\cite{Desyatnikov:2004-000:PRE}). Due to the exchange of
the angular momentum between interacting components, the soliton
slowly reverse the topological charges, keeping the zero total
angular momentum and perfectly stable total intensity. This
phenomenon is related to ``charge-flipping'' effect predicted
by~\cite{Alexander:2004-63901:PRL} to occur for the discrete
vortices in two-dimensional optical lattices (see
Sect.~\rsect{lattice}).

Very recently, \cite{Park:2004-21602:PRA} showed that the
necklace-type solutions exist for the model of two-component
Bose-Einstein condensates where the symmetry between the SPM and
XPM contributions to the nonlinear interaction is broken. These
nontopological vortices exhibit ``spin'' dynamical behavior and
may accumulate the Berry phase under an adiabatic change of
external fields that control the trapping potential (see also
Sect.~\rsect{BEC}).

The generalization of the concept of necklace-ring vector solitons
include the temporal effects in dispersive media, where the
three-dimensional spatiotemporal vortex solitons, or spinning
light bullets, have been predicted to exist
(\cite{Desyatnikov:2000-3107:PRE}). Corresponding solutions were
found by \cite{Andersen:2002-376:JOSB, Kovachev:2004-949:IJMS} for
nonlinear Maxwell equations and by \cite{Kovachev:2004-78:PD} for
Maxwell-Dirac equations. The internal structure of the
three-dimensional vector vortices with radially symmetric total
intensity is described in this case by spherical harmonics.

Finally, combining the counter-rotating vortices with strong
guiding from the fundamental soliton leads to a possibility to
stabilize completely the necklace-ring type solutions, as was
shown by \cite{Desyatnikov:2002-634:OL, Motzek:2002-501:OC} in the
case of three-component composite solitons. With further
increasing the number of interacting components, the composite
beams posses a complex internal structure, and they can serve as
the modal approximation for spatially incoherent and partially
coherent beams. The latter however posses the distinctive features
which we describe below.

\subsection{Partially coherent vortices}

As was demonstrated in all examples presented above, optical
vortices occur in coherent systems having a vanishing intensity at
the vortex position and well-defined phase front topology being
associated with the circulation of momentum around the helix axis.
If a vortex-carrying beam is {\em partially incoherent}, the phase
front topology is not well defined, and statistics are required to
quantify the phase. In the incoherent limit neither the helical
phase nor the characteristic zero intensity at the vortex center
is observable. However, several recent studies have shed light on
the question how phase singularities can develop in incoherent
light fields and how these phase singularities can be unveiled
(\cite{Gbur:2003-117:OC, Schouten:2003-968:OL,
Palacios:2004-143905:PRL}). In particular,
\cite{Palacios:2004-143905:PRL} used experimental and numerical
techniques to explore how a beam transmitted through a vortex
phase mask changes as the transverse coherence length at the input
of the mask is changed. Assuming a quasi-monochromatic,
statistically stationary light source and ignoring temporal
coherence effects, they demonstrated that robust attributes of the
vortex remain in the beam, most prominently in the form of a ring
dislocation in the cross-correlation function.

Propagating in nonlinear coherent systems, optical vortices become
highly unstable when the nonlinear medium is self-focusing, see
Sect.~\rsect{MI}. However, when spatial incoherence of light
exceeds a certain threshold, the stable propagation of optical
vortices in self-focusing nonlinear media is possible and has been
recently demonstrated in experiments conducted with a biased
photorefractive SBN crystal. \cite{Jeng:2004-43904:PRL} generated
partially incoherent vortices and vortex solitons, and then
inspected their stability. First, a cw laser light beam (at 488
nm) of the extraordinary polarization was made partially
incoherent by passing it through a lens and then through {\em a
rotating diffuser}. The rotating diffuser introduced {\em
random-varying phase and amplitude} on the light beam every
$1\mu$s, which is much shorter than the response time (about $1$
s) of photorefractive crystal. By adjusting the position of the
diffuser to near (away from) the focal point of the lens in front
the diffuser, it is possible to increase (decrease) the degree of
the light coherence, and collect the light after the rotating
diffuser by a second lens and then pass is through a
computer-generated hologram to imprint a vortex phase. Since the
partially incoherent light beam can be considered as {\em a
superposition of many mutually-incoherent light beams}, the
first-order diffracted light beam after the hologram becomes a
superposition of many mutually-incoherent vortex beams. Then, the
partially coherent vortex beam was launched into the SBN crystal
along its a-axis. The total power of the vortex beam is of
0.17$\mu$W, which results in the nonlinearity of the
photorefractive crystal falling into the Kerr region for the peak
intensity of the vortex beam to the background intensity is much
less than unity. A lens was used to project the images at the
input and output faces onto a CCD camera.

\pict[0.708]{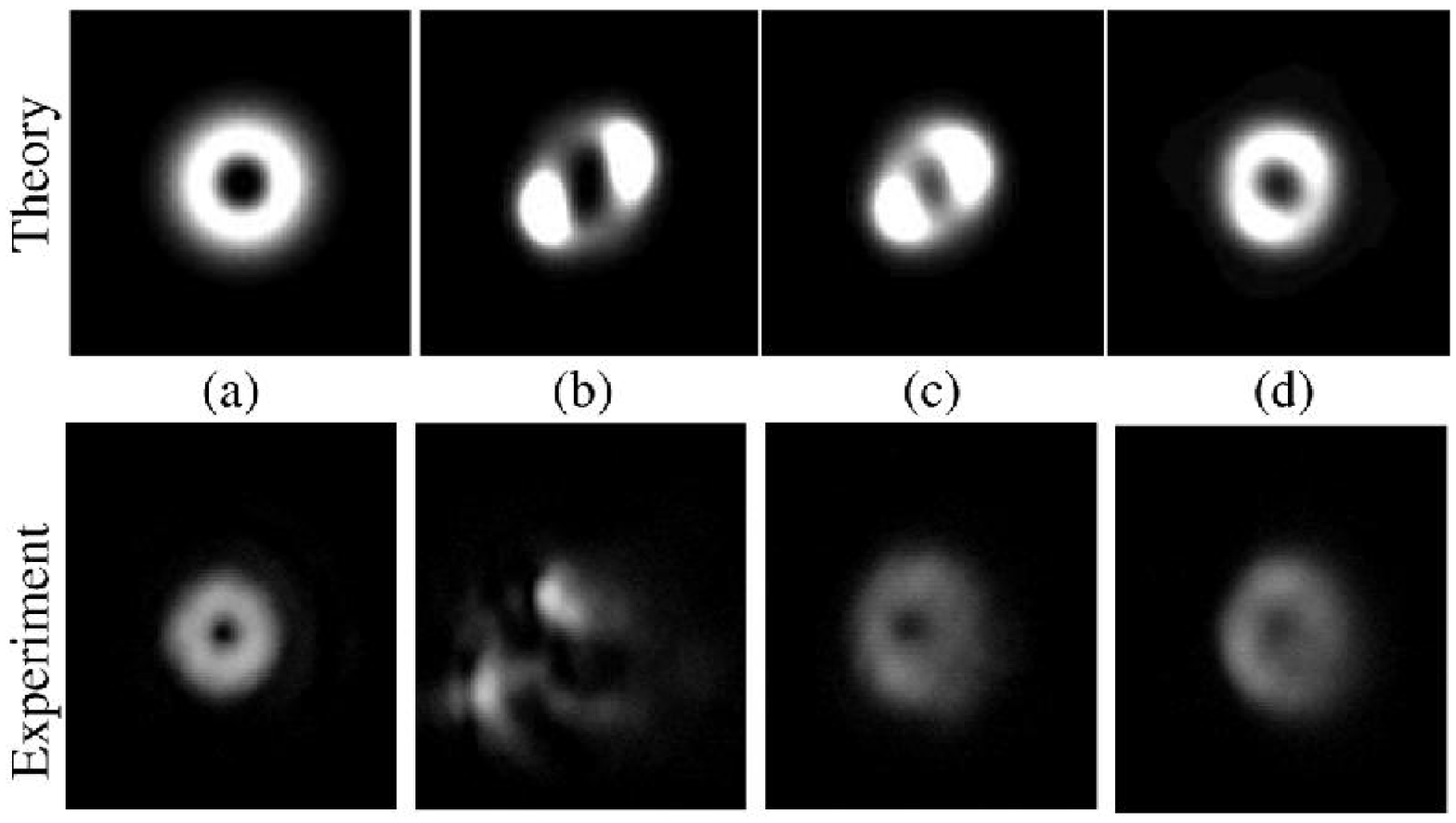}{fig11}{Numerical and experimental results
showing the stabilization of the vortex with growing incoherence:
(a) input intensity, (b) vortex after 9mm of propagation for the
coherent case, (c) vortex after 9mm for the partially incoherent
case, $\theta_0=0.14$ (less coherent), and (d) vortex after 9mm
for the partially incoherent case, $\theta_0=0.29$ (least
coherent). The incoherence stabilize the vortex soliton when
voltage of 2.5 kV is applied.}

When the diffuser is removed from the experimental setup, the
vortex beam at the input face of the crystal is shown as
Fig.~\rpict{fig11}(a). While a 2.5 kV biasing voltage is applied
on the photorefractive crystal creating a Kerr-type self-focusing
nonlinear medium, the vortex beam breaks up into two pieces
[Fig.~\rpict{fig11}(b)]. This vortex break-up is due to the
azimuthal instability. As the rotating diffuser is used,
Fig.~\rpict{fig11}(c) clearly shows that the vortex light beam is
stabilized by the reduction of the degree of coherence though two
very unclear bright spots still can be seen on the opposite sides
(top and bottom) of the ring-like intensity distribution. The
rotating diffuser has been than moved further away from the focal
point of the lens to make the light more incoherent,
Fig.~\rpict{fig11}(d) shows the generated stable {\em partially
incoherent vortex soliton} at the output face of the crystal.

The propagation of partially incoherent optical vortices in a
photorefractive medium has been studied numerically
by~\cite{Jeng:2004-43904:PRL} using the coherent density approach
developed by~\cite{Christodoulides:1997-646:PRL,
Anastassiou:2000-4888:PRL}. The coherent density approach is based
on the fact that partially incoherent light can be described by a
superposition of mutually incoherent light beams that are tilted
with respect to the $z$-axis at different angles. One thus makes
the ansatz that the partially incoherent light consists of many
coherent, but mutually incoherent light beams $E_j$: $I=\sum_j
|E_j|^2$. By setting $|E_j|^2=G(j\vartheta) I$, where
$G(\theta)=(1/\sqrt{\pi}\theta_0) \exp (-\theta^2/\theta_0^2)$ is
the angular power spectrum, one obtains a partially incoherent
light beam whose coherence is determined by the parameter
$\theta_0$, i.e. less coherence means larger $\theta_0$. Here,
$j\vartheta$ is the angle at which the $j$-th beam is tilted with
respect to the $z$-axis. A set of 1681 mutually incoherent
vortices was used in simulations, all initially tilted at
different angles.

The top row in Figs.~\rpict{fig11}(a-d) shows numerical results
for the propagation of an input Gaussian beam carrying a phase
dislocation [(a)] after the total propagation (9 mm) in a
nonlinear medium for the coherent light [(b)] and two different
partially incoherent beams [(c,d)], corresponding to the values
$\theta_0=0.14$ and $\theta_0=0.29$, respectively. The most
obvious difference to the scenario of the propagation the a
coherent vortex is that the vortex decay undergoes a visible delay
when the degree of incoherence grows. Furthermore, in the
incoherent case the vortex changes its profile only very slowly as
it propagates and thus can be considered as being in a transition
stage between the decay and stabilization.

Spatial coherence properties of optical vortices created in
partially coherent light were studied
by~\cite{Motzek:nlin.PS/0407043:ARXIV,
Motzek:nlin.PS/0410051:ARXIV}, who revealed the existence of phase
singularities in the spatial coherence function of a vortex field
that can characterize the stable propagation of vortices through
nonlinear media. Thus, the phase singularities of the spatial
coherence function predicted to exist in incoherent vortices
propagating in linear media (\cite{Palacios:2004-143905:PRL}) also
survive the propagation through nonlinear media. The intensity
distribution in the far field shows a local minimum in the center
of the beam, contrary to what one would obtain if the vortex was
propagating through a linear medium, and also in contrast to the
result we would obtain if we were propagating a light beam without
topological charge. This emphasizes the importance of the
interaction between coherence and nonlinearity. Not only the phase
structure, but also the intensity distribution strongly depends on
the initial form of the coherence function of the light beam.

The interaction of a coherent vortex beam with partially coherent
fundamental soliton, similar to the vortex-mode soliton discussed
in Sect.~\rsect{multi}, was considered recently
by~\cite{Motzek:2004-2285:OL}. Strong destabilization and
enhancement of azimuthal instability of vortex component is
observed for a low-amplitude incoherent beam. In the opposite
limit, vortex can be stabilized by a large-amplitude fundamental
beam with the value of its incoherence above a certain threshold.
These results are consistent with the stabilization dynamics of a
coherent vortex- and dipole-mode solitons
(\cite{Yang:2003-16608:PRE}).

\section[Vortices in $\chi^{(2)}$ media]{Multi-color vortex solitons}
\lsect{X2}

Similar to Kerr-type (or $\chi^{(3)}$) nonlinear media,
self-induced trapping of light occurs in quadratic ($\chi^{(2)}$)
nonlinear media (\cite{Karamzin:1974-734:PZETF,
Karamzin:1975-834:ZETF, Kanashov:1981-122:PD}). In this case, both
spatial and temporal multi-color solitons form through the mutual
focusing and trapping of the waves parametrically interacting in
the nonlinear medium. Occurrence of self-focusing effects in
quadratic nonlinear processes were sporadically suggested under
specific conditions, namely when the parametric interaction is
weak resulting in an effective third-order effect for the pump
wave (\cite{Ostrovskii:1967-331:PZETF, Flytzanis:1976-271:PQE}).
However, it took two decades before such effective
nonlinearity-induced phase shift was identified experimentally
(\cite{Belashenkov:1989-1383:OPSR, Desalvo:1992-28:OL}), and until
the importance of the associated, so-called \textit{cascaded}
nonlinearities was properly appreciated by
\cite{Stegeman:1993-13:OL} (for a review, see
\cite{Stegeman:1996-1691:OQE}). Since then, formation of spatial
and temporal multi-color solitons has been observed experimentally
in a variety of physical settings (\cite{Torner:2001-36:OPN,
Buryak:2002-63:PRP}) after the pioneering observations in the case
of second-harmonic generation (SHG) by
\cite{Torruellas:1995-5036:PRL} in a crystal of potassium titanium
phosphate (KTP), and by \cite{Schiek:1996-1138:PRE} in a planar
waveguide made of lithium niobate (LiNbO$_3$).

Cascaded nonlinearities can be modelled by the standard theory of
the $\chi^{(2)}$-mediated three-wave mixing described in detail in
several books on nonlinear optics
(\cite{Shen:1984:PrinciplesNonlinear,
Butcher:1992:ElementsNonlinear, Boyd:1992:NonlinearOptics}),
including more complex multistep parametric processes in quadratic
media (\cite{Saltiel:nlin.PS/0311013:ARXIV}). A comprehensive
review on optical  \textit{quadratic solitons} was published by
\cite{Buryak:2002-63:PRP}, and can be also found in a recent book
by \cite{Kivshar:2003:OpticalSolitons}. A number of overview
papers on the theory and experimental generation of spatial
parametric optical solitons in quadratic nonlinear media were
published during the last years
(\cite{Stegeman:1997-133:AdvancedPhotonics,
Kivshar:1997-451:AdvancedPhotonics, Torner:1998-229:BeamShaping,
Etrich:2000-483:ProgressOptics, Stegeman:2001:SolitonDriven,
Sukhorukov:2001-423:SolitonDriven,
Torruellas:2001-127:SpatialOptical}), and summaries of the
advances in the field were reported by \cite{Kivshar:1998-571:OQE,
Stegeman:1999-19:PSN, Torner:2001-36:OPN, Torner:2002-42:OPN,
Torner:2003-22:IQE}.

In this section, we describe briefly the physics and the salient
properties of the so-called \textit{quadratic vortex solitons},
i.e., self-trapped multi-color optical beams, composed of several
waves carrying nested optical vortices and parametrically
interacting in a phase-matchable quadratic crystal under
conditions close to phase-matching. Most studies and experiments
have been conducted for the simple case of SHG, or frequency
doubling, and parametric down-conversion, thus we concentrate on
the corresponding families of ring-like vortex solitons. We
discuss the spontaneous and induced modulational-instabilities
that affect the ring-shaped beams, the potential stabilization of
quadratic vortices by competing nonlinearities, and we summarize
the result of the available experimental observations. Vortex
solitons are two-dimensional light beams, therefore we concentrate
in quadratic vortex solitons in bulk media.

\subsection{Model}\lsect{X2model}

The SHG process for generating a double-frequency wave is a
special case of a more general three-wave mixing parametric
processes which occur in a dielectric medium with a quadratic
nonlinear response. The typical three-wave mixing and SHG
processes require only one phase-matching condition to be
satisfied and, therefore, they  can be classified as {\em single
phase-matched parametric processes}. Following the detailed
derivations presented by \cite{Menyuk:1994-2434:JOSB,
Bang:1997-51:JOSB}, and in \cite{Buryak:2002-63:PRP}, we consider
parametric interaction between three stationary
quasi-monochromatic waves with the electric fields ${\bf E} = \sum
{\bf e}_jE_j(x-\rho_jz,y,z)\exp(ik_jz- i\omega_jt)+c.c.$ (where
$j$ = 1,2,3), with the three frequencies satisfying the
energy-conservation condition, $\omega_1+\omega_2 = \omega_3$. We
assume that these three beams propagate along the same direction
$z$. However, when appropriate, the {\em energy walk-off} due to
birefringence should be taken into account by introducing the
angles $\rho_j\ll 1$ between the Pointing vector and the wave
vectors. For concreteness, we take the special case of Type II SHG
when two fundamental waves $\omega_1=\omega_2=\omega$ with the
ordinary ($E_1$, $\rho_1=0$) and extraordinary ($E_2$,
$\rho_2\neq0$) polarizations interact with the second-harmonic
(SH) wave, $E_3(\omega_3=2\omega)$, $\rho_3\neq0$. We assume that
waves propagate under conditions close to perfect phase-matching,
with a small mismatch between the three wave vectors given by the
parameter $\Delta k = k_1(\omega_1)+k_2(\omega_2)- k_3(\omega_3)$.
In the slowly varying envelope approximation, one can derive the
following set of three parametrically coupled equations:
\begin{eqnarray}
i\frac{\partial a_1}{\partial z}+ \frac{1}{2}\Delta a_1 + a_3 a_2^*e^{-i\beta z} & = & 0,\leqt{a1}\\
i\frac{\partial a_2}{\partial z}-i\delta_2\frac{\partial a_2}{\partial x}+\frac{\alpha_2}{2} \Delta a_2 +a_3 a_1^*e^{-i\beta z} & = & 0, \\
i\frac{\partial a_3}{\partial z}-i\delta_3\frac{\partial
a_2}{\partial x}+\frac{\alpha_3}{2} \Delta a_3+ a_1 a_2e^{+i\beta
z}&=&0. \leqt{a3}
\end{eqnarray}
Here the Laplace operator $\Delta$ acts on a transverse
coordinates $(x,y)$ normalized to a characteristic beam width $r$,
and the propagation coordinate is measured in units twice the
diffraction length, so that the normalized phase mismatch is
$\beta=k_1r^2\Delta k$. A typical value of the dimensionless
mismatch parameter, $\beta=\pm 3$, is obtained for a focused beam
with $r\simeq 15 \mu$m, and the mismatch $\pi/|\Delta k|\simeq
2.5$mm. Other parameters are: $\delta_j=k_1r\rho_j$,
$\alpha_j=k_1/k_j$ (we use $\alpha_1\equiv 1$); in practice,
$\alpha_2\simeq1$ and $\alpha_3\simeq0.5$. Finally, a normalized
field amplitude of some $|a|^2\sim 10$ corresponds to an actual
power flow in the range of $1-10$GW/cm$^2$ in a typical quadratic
nonlinear crystal, such as KTP. Of course, the above estimates
greatly depend on the particular properties of the material
employed on the pump light conditions, thus for a detailed
derivation of the governing equations and estimates for different
materials, we refer to the several reviews mentioned above.

In the case of {\em Type I} SHG, only a single beam at the pump
frequency $\omega$ interacts with a field at the frequency
$2\omega$. Equations \reqt{a1}--\reqt{a3} can then be reduced by
setting $a_1 = a_2=u$, and $\delta_2=0$, $\delta_3=\delta$, and
$v=a_3\exp(i\beta z)$. One obtains:
\begin{eqnarray}
i\frac{\partial u}{\partial z} + \frac{1}{2}\Delta u+ u^* v &=& 0, \leqt{u}\\
i\frac{\partial v}{\partial z} -i\delta \frac{\partial v}{\partial x} -\beta v +\frac{1}{4} \Delta v+ u^2 &=& 0. \leqt{v}
\end{eqnarray}

Indeed, in what is sometimes refereed as the {\em cascading}
limit, corresponding to large values of the phase mismatch
parameter, $\beta\gg1$, and to weak SH signals,  Eq.~\reqt{v}
approximately reduces to a simple relation, $v\simeq u^2/\beta$,
and the fundamental harmonic satisfies the familiar cubic NLS
equation,
\begin{equation}
i\frac{\partial u}{\partial z} +\frac{1}{2} \Delta u+\frac{1}{\beta}|u|^2u \simeq 0,
\leqt{X2NLS}
\end{equation}
which poses well-known stable soliton solutions in one-dimensional
geometries. However, it is worth stressing that most quadratic
solitons occur under conditions where the above reduction
\textit{does not hold}. One obvious example is the case analyzed
here of ring-profile vortex solitons, which exist in
two-dimensional geometries. Similarly, the above derivation does
not hold near phase-matching, whenever the second-harmonic waves
are intense, and in general with high enough light intensities.
However, these are the conditions where most quadratic solitons
are generated in practice. Therefore, the analogy indicated by
Eq.~\reqt{X2NLS} must be used with the proper understanding of its
important limitations in the interpretation of most experiments.

\subsection{Frequency doubling with vortex beams}
\lsect{X2dbl}

Second-harmonic generation is a particular case of
\textit{frequency conversion} processes associated with the energy
transfer between several waves propagating in a nonlinear medium.
The nonlinear wave-mixing obeys conservation of energy, linear
momentum, and, under proper conditions, angular momentum. In the
general case of frequency mixing in which two fields of optical
frequencies $\omega_1$ and $\omega_2$ combine to produce a third
field of frequency $\omega_3$, conservation of energy requires the
condition $\omega_1+\omega_2=\omega_3$. Conservation of the linear
momentum leads to the phase-matching requirement ${\bf k}_1+{\bf
k}_2={\bf k}_3$.

Under proper conditions and definitions, the angular momentum
carried by the light beam must also be conserved. Conservation of
spin angular momentum imposes constraints in the polarization of
the input and output light beams. In the case of co-linear
interaction of paraxial light beams in a quadratic medium,
conservation of the paraxial beam orbital angular momentum holds
too, and affects the spatial shape of the generated beams. The
simplest situation occurs in the case of Type I frequency doubling
of single Laguerre-Gaussian (LG) pump modes
(\cite{Basistiy:1993-422:OC, Dholakia:1996-3742:PRA,
Soskin:1998-301:PAO, Allen:1999-291:ProgressOptics}). In this
case, the winding number of the LG pump modes must double, because
of the azimuthal symmetry of the governing equations. In the
general case of Type II phase-matching or frequency-mixing of
single LG modes with winding numbers $m_j$, $j=1,2,3$, the
symmetry of the equations leads to $m_1+m_2=m_3$, as it is
observed experimentally by \cite{Berzanskis:1997-273:OC}.

However, conservation of orbital angular momentum in parametric
processes with arbitrary input beams and general phase-matching
geometries does not necessarily translate into simple algebraic
rules between the winding numbers of the vortices present in the
beam. Illustrative cases are multi-mode and complex pump beams
(\cite{Petrov:1998-7903:PRE, Berzanskis:1998-372:OC,
Berzanskis:1999-263:OpticalVortices, Petrov:1999-357:OC,
Molina-Terriza:2000-1197:JOSB, Jarutis:2000-159:OC,
Stabinis:2001-419:OC}), or the generation of vortex-streets by the
presence of Poynting-vector walk-off
(\cite{Molina-Terriza:1999-899:OL, Molina-Terriza:2002-625:OL}).
Actually, the orbital angular momentum of a light beam is not
necessarily directly related to the properties of the vortices
that it contains.

Notice also that, in general, the angular momentum at the
classical level is an overall property of the light beam, and must
be clearly distinguished from the quantum angular momentum at the
single photon level. That in the latter case the conservation of
orbital angular momentum in parametric processes is not
necessarily given by simple algebraic rules is most clearly
illustrated in non-collinear geometries
(\cite{Molina-Terriza:2003-155:OC}), or in the so-called
transverse-emitting processes (\cite{Torres:2004-1939:OL}).

In what follows we concentrate in vortex solitons, which are
intense light beams carrying energies which might exceed tens of
$\mu J$ at visible or near-infrared wavelengths, thus far from any
single-photon effects. In the next subsections, we discuss the
families and stability properties of {\em quadratic vortex
solitons} --- nonlinear optical beams with strong energy exchange
between its constituents.

\subsection{Families of the vortex solitons}
\lsect{X2sol}

Because the Type II phase-matching process involves three beams,
there is a wider variety of different solutions than in the case
of the Type I SHG geometries, which we will regard as a limit of
degeneracy of a former model when both fundamental frequency (FF)
beams are of the same polarization, see Sect.~\rsect{X2model}. The
Type II model described by Eqs.~\reqt{a1}--\reqt{a3} defines an
infinite-dimensional Hamiltonian system with a conserved
Hamiltonian which in the absence of walk-off ($\delta_j=0$) is
given by
\begin{eqnarray}
H=\tfrac{1}{2}\int\left(\tfrac{1}{2}\sum\alpha_j|\nabla_\perp A_j|^2+\beta|A_3|^2-
A_1^*A_2^*A_3-A_1A_2A_3^*\right)dr_\perp,
\leqt{X2H}
\end{eqnarray}
where $A_{1,2}=a_{1,2}$, and $A_3=a_3\exp(-i \beta z)$. We use two
additional conserved quantities of the beam evolution, namely the
total beam power or energy flow $I$,
\begin{eqnarray}
I=\sum I_j=\int\left(\tfrac{1}{2}|A_1|^2+\tfrac{1}{2}|A_2|^2+|A_3|^2\right)dr_\perp,
\leqt{X2I}
\end{eqnarray}
and the energy imbalancing $I_i$,
\begin{eqnarray}
I_i=\tfrac{1}{2}\int\left(|A_1|^2-|A_2|^2\right)dr_\perp.
\leqt{X2Iu}
\end{eqnarray}
The conservation of $I_i$ means that the energy transfer between
the fundamental waves and the second-harmonic wave cannot favor
any of the two orthogonal polarizations that compose the
fundamental beam. Notice that strictly speaking  the above
expressions hold only for continuous-wave light propagation, and
that the temporal effects on pulsed light might introduce
important new features in the imbalancing of the quadratic
solitons (\cite{Minardi:2003-123901:PRL}). In the absence of the
Poynting vector walk-off, the total beam orbital angular momentum,
defined as
\begin{eqnarray}
M=\frac{1}{4i}\int \left\{ \left[ {\bf r}_\perp \times \sum
\left(A_j^*\nabla_\perp A_j -A_j\nabla_\perp A_j^*\right)\right]{\bf e}_z\right\}dr_\perp,
\leqt{X2M}
\end{eqnarray}
is also conserved during the beam evolution. For our purposes, it
is convenient to investigate configurations without walk-off, and
we hereafter set $\delta_j=0$.

\pict[0.708]{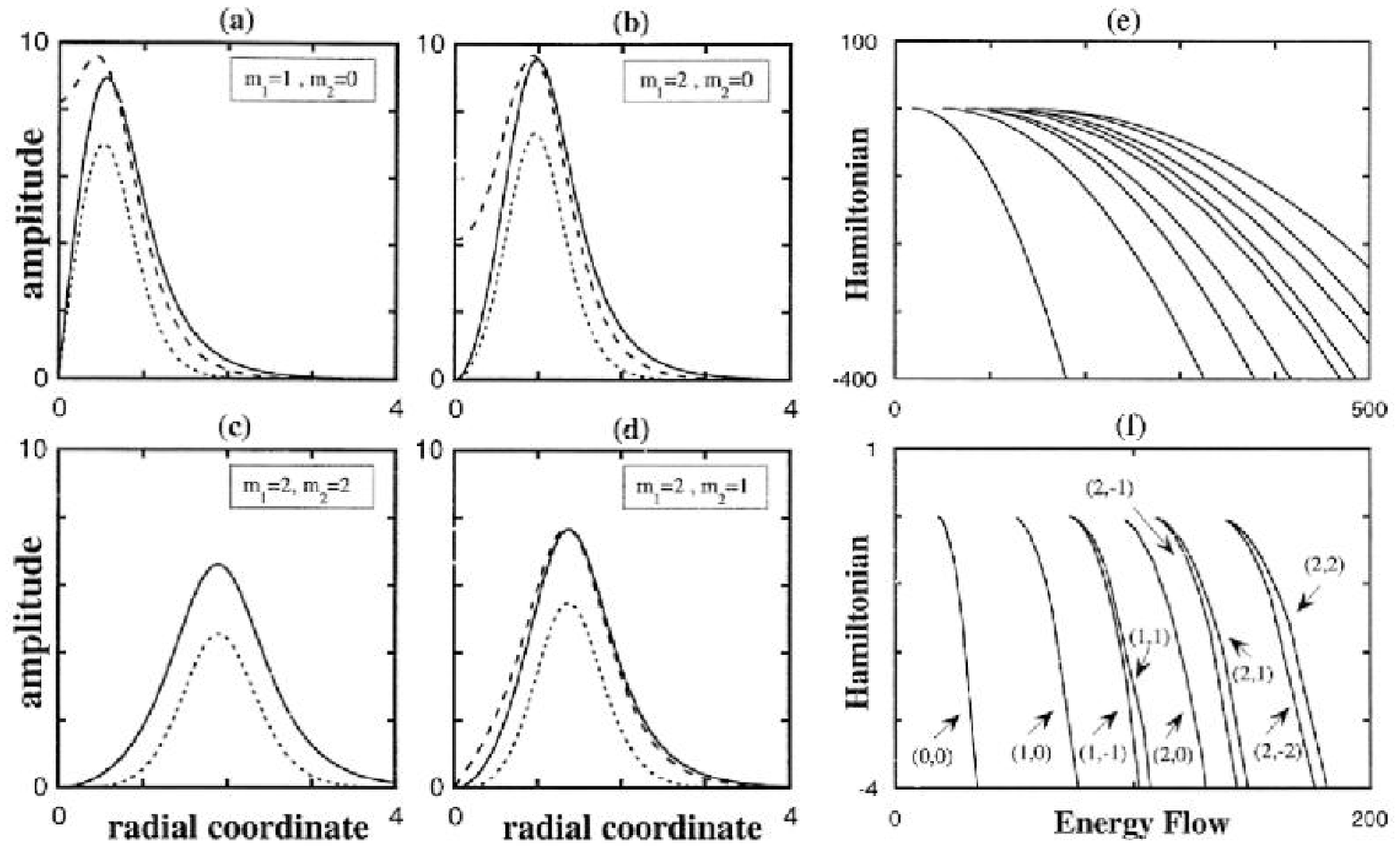}{fig12}{(a)-(d) Typical solutions for the
vortex solitary waves with different combinations of topological
charges, $m_3=m_1+m_2$. Solid line:  ordinary-polarized
fundamental beam; dashed line: extraordinary-polarized fundamental
beam; dotted line: second-harmonic field. Parameters are:
$\beta=3$, $\kappa_1=2$, and $I_u=0$. Notice that in (c) the
curves corresponding to two fundamental beams are identical. (e)
Families of the lowest-order vortex solitons presented through the
energy flow-Hamiltonian diagram. Plot (f) is a zoom of the
corresponding region of plot (e). Numbers in parentheses stand for
the topological charges of the two fundamental beams, i.e.,
$(m_1,m_2)$. The curve labelled (0,0) corresponds to the family of
lowest-order, vorticity-less bright
solitons~\cite{Torres:1998-77:OC}.}

Spatial optical solitons are optical beams with constant
transverse profile along the propagation direction, defined as
stationary solutions of the corresponding propagation equations
Eqs.~\reqt{a1}--\reqt{a3}. From this definition, it follows that
we may  look for the soliton solutions in a form of the generic
ansatz, $a_j=V_j(x,y)\exp(i\kappa_jz)$, where $\kappa_j$ are the
nonlinear corrections to the corresponding propagation constants.

Stationary propagation of the multi-color solitons requires
vanishing power exchange between the fundamental and
second-harmonic waves; to avoid this exchange the condition
$\kappa_3=\kappa_1+\kappa_2+\beta$ applies. Radially symmetrical
solutions are found by separation of variables in the form
\begin{eqnarray}
V_j(x,y)=U_j(\rho)\exp(im_j\varphi),
\leqt{X2V}
\end{eqnarray}
with real amplitudes $U_j$ depending on the polar radius
$\rho=\sqrt{x^2+y^2}$ and phases being linear functions of the
azimuthal coordinate $\varphi=\arctan(y/x)$. Using this ansatz and
the algebraic constraint to topological charges $m_1+m_2=m_3$, we
obtain the $z$-independent (stationary) version of
Eqs.~\reqt{a1}--\reqt{a3}:
\begin{eqnarray}
\frac{\alpha_j}{2}\left(\frac{d^2}{d\rho^2}+\frac{1}{\rho}\frac{d}{d\rho}-\frac{m_j^2}{\rho^2}\right)U_j+U_pU_q=\kappa_jU_j,
\leqt{X2U}
\end{eqnarray}
with $j,p,q=1,2,3$, and $j\ne p\ne q$. For a fixed set of the
topological charges, solutions depend on two parameters (e.g.
$\kappa_1$ and $\kappa_2$), which  correspond to different total
and relative (imbalancing) energy flows between the three
interacting waves~(\cite{Buryak:1996-5210:PRL,
Buryak:1997-3286:PRL, Peschel:1997-7704:PRE}).

Properties of the fundamental (bell-shaped) quadratic spatial
solitons have been described in a number of studies (see, e.g.,
the review paper by~\cite{Buryak:2002-63:PRP}), and instabilities
of higher-order \textit{vorticityless} modes with central peak and
one or more surrounding rings are also known
(\cite{Skryabin:1998-1252:PRE}). We do not consider these issues
here and focus on the ``doughnut''--shaped vortex solitons.

Families of vortex solitons as solutions of Eqs.~\reqt{X2U}, for
different combinations of the topological charges, wave vector
mismatches, and both zero and nonzero imbalancing $I_i$, have been
found by \cite{Firth:1997-2450:PRL, Torres:1998-77:OC,
Skryabin:1998-3916:PRE, Molina-Terriza:1998-170:OC}.
Figure~\rpict{fig12} shows typical shapes of vortex solitary waves
with different combinations of topological charges. We choose a
particular case of zero imbalancing $I_i=0$ because it also covers
the solutions with equal amplitudes for both FF beams, such as
that shown in Fig.~\rpict{fig12}(c). This particular branch
corresponds to stationary solutions for the Type I geometry
Eqs.~\reqt{u}--\reqt{v}, consisting of only two components. The
latter ones have been studied in detail by
\cite{Firth:1997-2450:PRL, Torres:1998-625:JOSB,
Skryabin:1998-3916:PRE}.

Useful information about the soliton families is given by
integrals of motion defined above; for the stationary solutions
under consideration the angular momentum and Hamiltonian can be
expressed in terms of the beam powers, \begin{eqnarray}
M&=&\,\tfrac{1}{2}\left \{(m_1+m_2)I+(m_1-m_2)I_i\right \},\\
H&=&-\tfrac{1}{4}\left
\{(\kappa_1+\kappa_2)I+(\kappa_1-\kappa_2)I_i -\beta I_3\right \}.
\leqt{X2MH}
\end{eqnarray}
Note that the vectorial nature of three-wave multi-color vortex
solitons allows combinations including vorticity-less beam, e.g.
$m_1=1$ and $m_2=0$, or having zero total angular momentum $M=0$
(e.g. $m_1=-m_2$ and $I_i=0$), similar to their $\chi^{3}$
counterparts described in Sect.~\rsect{X3vect}. However, the
two-component solutions in the Type I model are always limited by
the constraints $\kappa_{1,2}\equiv\kappa$, $m_3=2m$ with
$m_{1,2}\equiv m$, and $I_u\equiv0$, therefore $M=mI$, similar to
the scalar $\chi^{3}$ spatial soliton. Figures~\rpict{fig12}
(e)-(f) show some examples of the Hamiltonian dependencies
Eq.~\reqt{X2MH} in the case $I_i=0$, the similar plots for the
nonzero imbalancing are available
in~\cite{Molina-Terriza:1998-170:OC}.

\subsection{Spontaneous break-up: azimuthal instability}

Generation of different vortex patterns due to the frequency
conversion, described in Sect.~\rsect{X2dbl}, occurs for the input
powers of the FF pump beam below some threshold. For higher input
powers, which are sufficient for the soliton formation, the
generation of sets of simple fundamental solitons was predicted
numerically for SHG by \cite{Petrov:1997-1037:OQE,
Torner:1997-608:ELL, Torner:1997-2017:JOSB}. Such phenomenon is
related to the azimuthal modulational instability of the
corresponding stationary states -- ring-shaped vortex solitons.
The linear stability analysis of the Type I vortex solitons was
performed by \cite{Firth:1997-2450:PRL, Skryabin:1998-3916:PRE}
and \cite{Torres:1998-625:JOSB}. The experimental confirmation of
the spontaneous break-up of optical vortex solitons in quadratic
crystal was reported by \cite{Petrov:1998-1444:OL}.

\pict[0.708]{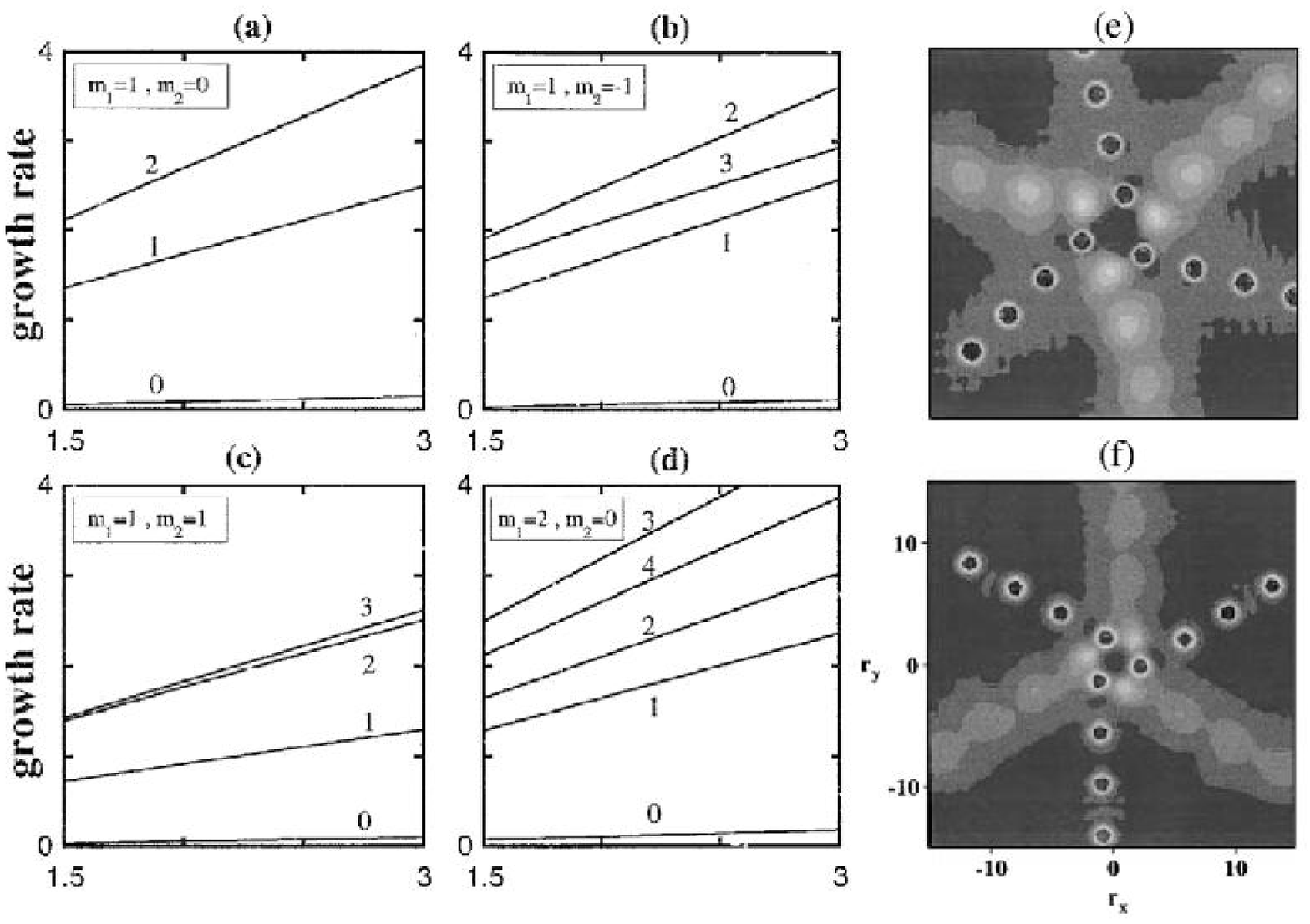}{fig13}{(a)-(d) Instability growth rate for
perturbations with different azimuthal indices as a function of
the nonlinear wave number shift $\kappa_1$, for the various
families of vortex solitary waves shown in Fig.~\rpict{fig12}
(\cite{Torres:1998-77:OC}). (e)-(f) Stroboscopic view of the decay
of an exact vortex solitary wave solution in the presence of the
corresponding exact azimuthal perturbation of a given index. The
plots show the light patterns of the ordinary polarized
fundamental beams calculated at $z=4,8,12,16$ propagation units,
when the input is the field of Fig.~\rpict{fig12} plus the
corresponding symmetry-breaking perturbation with azimuthal index
$s=3$. Amplitude of the added perturbation: in (e) $\epsilon=
10^{-2}$ and in (f) $\epsilon=10^{-3}$. The extraordinary
polarized fundamental and the second-harmonic beams exhibit
similar features and thus are not shown
(\cite{Torner:1998-809:OQE}).}

To outline briefly the main steps of these calculations, we
examine the stability of the vortex solitary waves against
azimuthal perturbations and seek the perturbed solutions of the
form
\begin{eqnarray}
a_j=\left\{U_j(r)+\epsilon\left[f_{j,s}(r,z)\exp(is\varphi)+g_{j,s}(r,z)\exp(-is\varphi)\right]\right\}\exp(i\kappa_j
z+im_j\varphi),
\leqt{X2pert}
\end{eqnarray}
where $s$, $f_{j,s}$, and $g_{j,s}$ stand for the azimuthal index
and the envelopes of the perturbation eigenfunctions,
respectively. Inserting Eq.~\reqt{X2pert} into
Eqs.~\reqt{a1}--\reqt{a3} and linearizing the equations in respect
to small perturbations, we obtain a set of six coupled linear
partial differential equations for $f$ and $g$ at a given value of
$s$. Such equations have many different solutions; some of them,
the so-called instability modes, display exponential growth along
the propagation direction. To obtain such solutions, one can use
the method of averaging the growth rate of perturbation over the
propagation direction described by \cite{Soto-Crespo:1991-636:PRA,
Soto-Crespo:1992-3168:PRA}, or further reduce the problem by
setting $\{f;g\}=\exp(\Gamma z)\{\tilde{f}(r);\tilde{g}(r)\}$ and
solving the corresponding boundary value problem for $\tilde{f}$
and $\tilde{g}$ (see also the discussion in Sect.~\rsect{MI}).
Both methods give identical results obtained for the Type I
vortices by \cite{Firth:1997-2450:PRL, Skryabin:1998-3916:PRE} and
\cite{Torres:1998-625:JOSB}.

In Fig.~\rpict{fig13}, we show typical values of the instability
growth rate $\rm{Re}(\Gamma)$ for the solutions with different
sets of topological indices. Instability induced break-up of the
ring-like  solitons that follows qualitatively the similar
scenario in all the cases, namely the splitting of the initial
ring to a number of the fundamental solitons flying off the ring
with the further propagation. Varying an initial perturbation
results in different light patterns, where positions of the
soliton splitters differ, as is seen in Figs.~\rpict{fig13}
(e)-(f), however, their number corresponds exactly to the
azimuthal index of the instability mode with the largest growth
rate. Number of solitons created through the decay is generally
robust to small imbalancing of energy $I_i\neq 0$, while for large
imbalancing, several perturbations with similar growth rate can
come into play (\cite{Molina-Terriza:1998-170:OC}).

\subsection{Induced break-up: soliton algebra}

The process of vortex beam break-up by azimuthal modulational
instabilities is spontaneous, and thus governed by the
perturbations that happen to have the highest growth rate. This
process produces beautiful patterns of solitons flying off the
input vortex ring, but leaves little control over the number of
spots present in such soliton pattern. Vortex ring-shaped solitons
might be broken in a controllable manner by inducing their
splitting in a way that favors predetermined azimuthal symmetries.

One way to favor a given azimuthal symmetry is to impose a
specific phase-pattern to the input beam. In the case of
parametric interactions in quadratic media, such goal can also be
accomplished by seeding the SHG process with a vortex in the SH
frequency: The azimuthal phase-varying relation between the
interacting beams generates a prescribed azimuthal symmetry of
energy exchange between the beams that produces the desired
soliton pattern (\cite{Torner:1998-809:OQE}).

\pict[0.708]{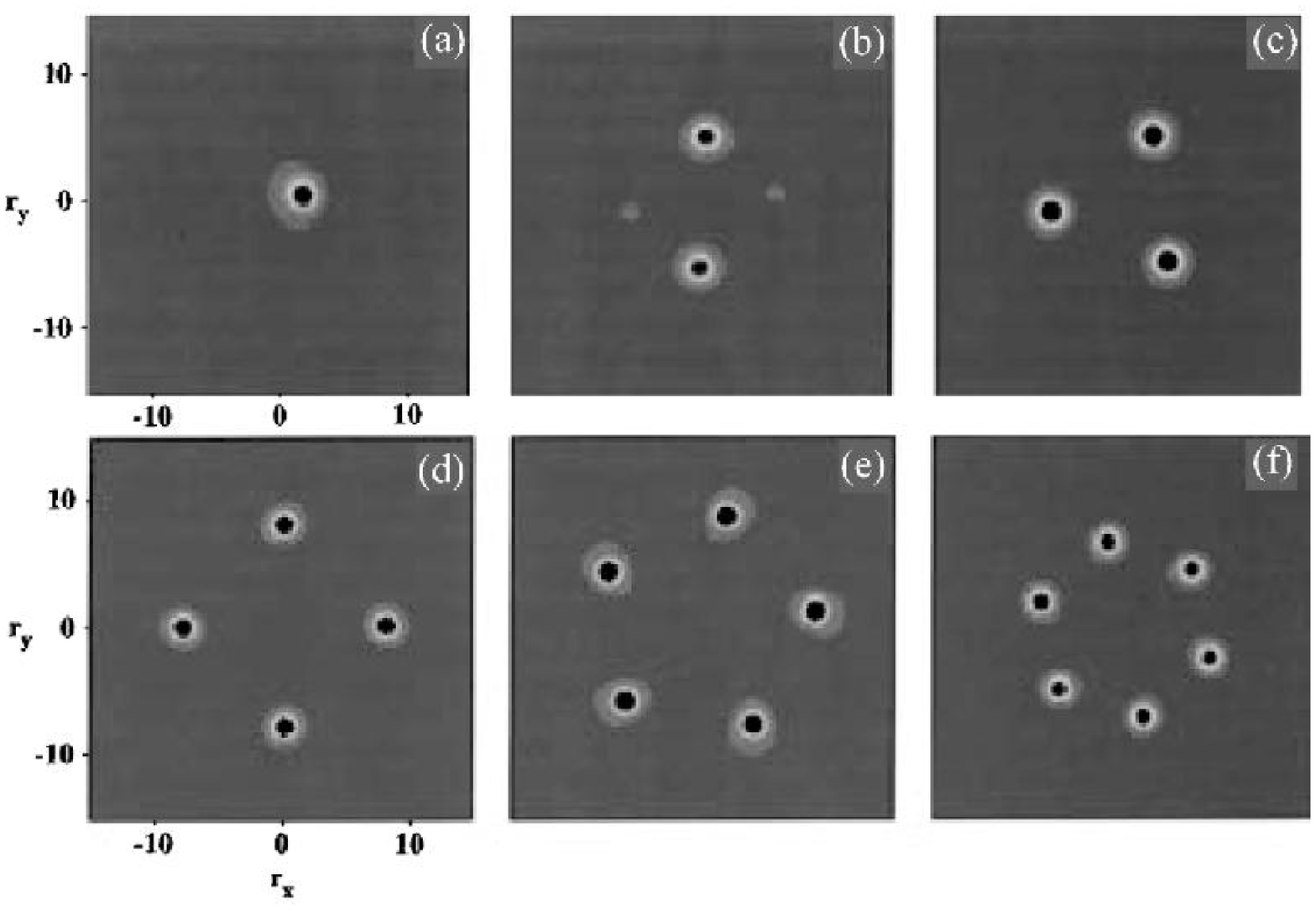}{fig14}{Results of the induced break-up of
the input beams containing topological charges: (a) [0,0,1], (b)
[1,1,0], (c) [1,1,-1], (d) [2,2,0], (e) [2,2,-1], and (f)
[2,2,-2]. Input energy flows: at the fundamental frequency
$I_1=I_2=36\pi$ for (a)-(c) and $I_1=I_2=128\pi$ for (d)-(f); at
the second harmonic $I_3=2\pi$. The plots show the
ordinary-polarized beams at $z=10$ (\cite{Torner:1998-809:OQE}).}

Such a process was termed \textit{soliton algebra}, and is
implemented by changing the topological charge of a weak,
co-linear input SH seed beam. If this topological charge $m_3$
satisfies the condition $m_3=m_1+m_2$, where $m_{1,2}$ are the
charges of two orthogonally polarized FF pump beams, the
generation and subsequent \textit{spontaneous} azimuthal
instability of three-wave vortex solitons occurs. However, if the
condition above is violated, $m_3\neq m_1+m_2$, the beam break-up
into solitons is induced by the local, azimuthally-varying phase
difference that exists between the pump and the seed signals and
hence by the initial local direction of the energy flow between
the FF and SH waves. The number of solitons formed in each portion
depends on the input light conditions, such as the total energy
flow and SH seed intensity. As a result, the information coded in
the value of the input array $(m_1,m_2,m_3)$ is transformed into a
certain number of output soliton
spots~(\cite{Torner:1998-809:OQE}). Different typical output
patterns are presented in Fig.~\rpict{fig14}.

The experimental demonstration of the concept of soliton algebra
were performed in a Type I SHG geometry by
\cite{Minardi:2001-1004:OL}. Under properly chosen operating
conditions, the number of generated solitons $n$ was shown to
follow the rule $n=|2m_{FF}- m_{SH}|$, with $m_{FF}$ and $m_{SH}$
being the topological charges of the input FF and SH beams,
respectively. When $2m_{FF}=m_{SH}$, the beam break-up is known to
occur through spontaneous azimuthal modulation instability,
therefore the input was designed in such a way that $2m_{FF}\neq
m_{SH}$. Both processes, spontaneous or induced, require features
similar to those needed for the generation of solitons with
Gaussian-like beams and thereby is found to be very robust.

In the context of soliton control by the presence of phase
dislocations in the input beams, we notice the observation of the
deflection of multi-color solitons generated by edge-like
topological amplitude and phase dislocations reported by
\cite{Petrov:2003-1439:OL}. The experiments were conducted near
phase-matching in a bulk potassium titanium phosphate crystal
pumped with picosecond light pulses at 1064 nm, and the angular
deflection of the solitons was found to be controllable through
the position of the edge dislocation.

\subsection{Dark multi-color vortex solitons}

Existence of multi-color dark vortex solitons has been discussed
by {\cite{Alexander:1998-670:OL} who analyzed also some basic
properties of such beams which were found to be highly unstable
against modulational instabilities of their nonvanishing
background (\cite{Buryak:2002-63:PRP}). Such instability is known
to be suppressed by a strong effect of competing nonlinearities
(\cite{Alexander:1998-670:OL, Alexander:2000-2042:PRE}), but such
prediction did not find yet experimental verification.
Nevertheless, significant efforts have been put to overcome
modulational instability due to parametric wave interaction and to
generate dark vortex-carrying beams. The most important advance
was reported by \cite{DiTrapani:2000-3843:PRL}, who used a large
walk-off between the components to quench the modulational
instabilities. In the reported experimental observations, the SH
beam broke-up and formed many spikes, having an energy content
much larger than the rest of the SH beam carrying a phase
dislocation. Because of the large walk-off, the spikes propagated
out of the beam rapidly. The generation of such spikes was claimed
to be essential for self-quenching of the instability process. The
experiment was performed with {\em negative} large phase-mismatch,
where at moderate powers the cascading nonlinearities leads to an
effective {\em defocusing} Kerr nonlinearity (Eq.~\reqt{X2NLS}),
known to support stable dark vortices. In the experiments reported
by \cite{DiTrapani:2000-3843:PRL}, the generated SH vortex
dislocation exhibited a size much smaller than the host
diffracting beam, indicating at least a transient trapping effect.

The corresponding numerical simulations performed by
\cite{DiTrapani:1999:NLGW} confirmed such transient self-trapping
and the propagation of the dislocation in the {\em fundamental
frequency} beam for a distance of the order of several diffraction
lengths without core spreading. In contrast, the linear
diffraction of the beam results in strong spreading of the vortex
core. The vortex beams observed in the experiments exhibited some
features expected from a dark vortex soliton, but a comprehensive
investigation of these important observations is not yet fully
developed. Thus, the existence of true stable dark vortices in a
quadratic nonlinear medium remains an open problem.

\section[Competing nonlinearities]{Stabilization of vortex solitons}
\lsect{comp}

In this section we discuss several theoretical predictions of the stabilization of bright vortex solitons in nonlinear media. Several models supporting stable vortex solitons were suggested, e.g. the Kerr media made of alternating self-focusing and
self-defocusing layers (\cite{Towers:2002-537:JOSB, Montesinos:2004-133901:PRL, Montesinos:nlin.PS/0405059:ARXIV, Adhikari:2004-63613:PRA}) and nonlocal self-focusing medium (\cite{Yakimenko:nlin.PS/0411024:ARXIV, Breidis:2005:OE}). In particular, we
discuss here two distinct models with so-called competing nonlinearities. The first model includes self-focusing cubic and self-defocusing quintic terms in the power-law Kerr-type nonlinearity, whereas the second model includes phase-dependent quadratic and self-defocusing cubic nonlinear interaction. We also
summarize numerical results demonstrating stable spatiotemporal vortex solitons in the $(3+1)$-dimensional geometry, the so-called spinning light bullets.

\subsection{Cubic-quintic nonlinearity}
\lsect{CQ}

Nonlinear models discussed in previous sections correspond to the
lowest order nonlinearities available, namely to the first two
nonlinear terms in expansion of optical medium polarization $P =
\chi^{(1)}E + \chi^{(2)}E^2 + \chi^{(3)}E^3+\dots$. For the
centro-symmetric media all even terms vanish and taking into
account higher-order terms one can represent the refractive index
as a \textit{power-law} Kerr-type nonlinearity,
$n=n_0+n_2I+n_4I^2+\dots$, here the intensity $I\equiv |E|^2$.
Obviously, if nonlinear coefficients $n_2$ and $n_4$ have the same
signs, corresponding models exhibit simple increasing of strength
of nonlinear self-action, self-focusing for $n_{2,4}>0$ or
self-defocusing for $n_{2,4}<0$. More interesting situation occurs
when two nonlinear contributions have opposite signs, $n_2n_4<0$,
this case usually refereed as ``competing'' cubic-quintic (CQ)
nonlinearity: corresponding nonlinear terms in propagation
equation being of the third and fifth orders:

\begin{eqnarray}
i \frac{\partial E}{\partial z} + \Delta E +n_2|E|^2E+n_4|E|^4E=0.
\leqt{CQeq}
\end{eqnarray}

Let $n_4$ be of self-focusing type, $n_4>0$. Then, for any sign of
Kerr contribution $n_2$, there will be a threshold power of
(transversely two-dimensional) light beam $E$, when higher-order
self-focusing will predominate both the linear diffraction and
$n_2$ contribution. In this case the beam will collapse, similar
to the pure Kerr case with $n_2>0$ and $n_4=0$ (see recent review
by \cite{Berge:1998-260:PRP}). However, if $n_4<0$, than the
collapse can be stopped, because the parts of light beam with high
enough intensity $I>I_{th}$ experience effectively self-defocusing
environment, $dn/dI<0$. Here the threshold intensity is given by
$I_{th}=-n_2/(2n_4)$. Furthermore, the CQ nonlinearity can be
regarded as a power-law expansion for any collapse-free
nonlinearity with saturation, for example the phenomenological one
discussed above, $n=n_0+n_2I/(1+sI)$. In this case $n_4=-sn_2$,
and the CQ medium is refereed also as a saturable one.

Saturable nonlinearity offered a great advantage over the
conventional cubic one because it supports \textit{stable} spatial
solitons, free of collapse instability. That is why its simplest
version, the CQ nonlinearity, attracts significant attention of
theoreticians from the early days of nonlinear optics
(\cite{Zakharov:1971-136:ZETF}). Numerical studies have confirmed
the stability of fundamental spatial solitons in this system
(\cite{Wright:1995-2481:OL, Dimitrevski:1998-369:PLA,
Quiroga-Teixeiro:1999-1697:JOSB}). Experimentally a CQ nonlinear
dielectric response with positive cubic and negative quintic
contributions has been observed  in chalcogenide glasses
(\cite{Smektala:2000-232:JNCS, Boudebs:2003-427:OC}), and in
organic materials (\cite{Zhan:2002-369:JOSB}). However, in all
these cases the quintic nonlinearity is accompanied by significant
higher-order multiphoton processes such as two-photon absorption,
therefore the validity of the CQ models to light propagation in
these materials requires additional explorations. Very recently, the criteria for the experimental observation of multidimensional solitons in CQ type saturable media were developed by~\cite{Chen:2004-46610:PRE}.

\pict[0.708]{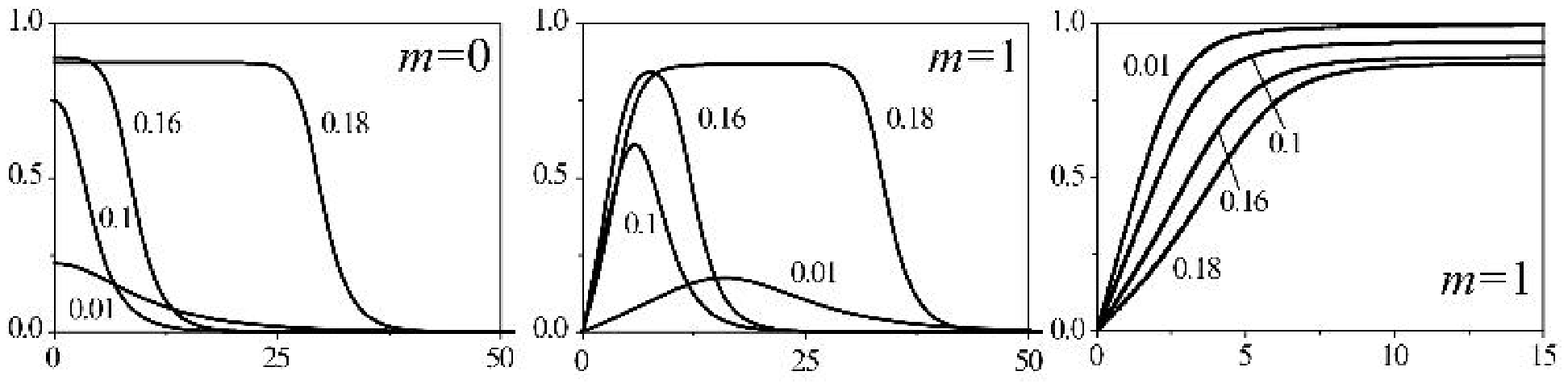}{fig15}{Stationary solutions to the
Eq.~\reqt{CQeq} for different topological charges $m$ and soliton
parameter $\kappa$ (shown next to the corresponding curves). For
$m=1$, we also show the example of two types of optical vortices,
localized (bright) and non-localized (dark), to coexist in media
with competing nonlinearities. Similar solutions exist for any
number $m$. Note that the slopes of envelopes for both types, i.e.
the size of the vortex core, is sufficiently different for the
case of $\kappa=0.1$, and almost exactly the same for
$\kappa=0.18$, indicating the transformation of bright to dark
vortex with increase of power, see text for the details.}

The ``auto-waveguide'' propagation of the ``spiral beams'' with
nonzero topological charge has been predicted by
\cite{Kruglov:1985-401:PLA} for pure Kerr model. As early as 1988,
it was found that saturation in the form of CQ nonlinearity
stabilize optical vortices against collapse, and ``the data from
the computer experiment show that these beams are stable'' (after
\cite{Kruglov:1988-4381:JPA}). Further study, however, reveal that
the azimuthal instability of CQ vortices may take place
(\cite{Kruglov:1992-2277:JMO}). A key insight was put forward by
\cite{QuirogaTeixeiro:1997-2004:JOSB}, who found by numerical
simulations that the vortex solitons with charge $m=1$ could be
stable provided that the power of the beam is over some critical
value. Fig. \rpict{fig15} shows several examples of the bright and
dark vortex soliton solutions to Eqs.~\reqt{CQeq}
\textit{coexisting} in the CQ model
(\cite{Berezhiani:2001-57601:PRE}). Note that with increase of
power the amplitude of the solutions saturates and starting from
some value of the soliton parameter (or, equivalently, some value
of soliton power), the slope of the bright vortex soliton coincide
with the one for dark vortex. That may indicate the transition
from unstable vortices in self-focusing regime to the stable ones
in effectively self-defocusing regime. The critical power for this
transition has been found analytically by
\cite{Michinel:2001-314:JOB}, it was shown to exceed four times
the threshold power of the generation for vortex soliton with
charge $m=1$.

The issue of the stability of vortex solitons in CQ model was put
on more solid mathematical grounds by calculating the linear
stability spectrum by \cite{Towers:2001-292:PLA} and
\cite{Skarka:2001-124:PLA}. The growth rate of small azimuthal
perturbations was found to be nonzero for a limited domain
$0<\kappa<\kappa_{stab}$, with linearly stable solutions above the
critical value $\kappa>\kappa_{stab}$. In addition, very small
instability with respect to shift of the dislocation core (central
dark spot) was found by \cite{Towers:2001-292:PLA,
Malomed:2002-187:PD}. Nevertheless, the supercritical vortex
solitons appear to be strong attractors, the gaussian beam with
nested phase dislocation may initially break to several splitters
but than restores radially symmetric vortex shape
(\cite{Skarka:2001-124:PLA}). A mathematically rigorous stability
analysis was performed by \cite{Pego:2002-347:JNS}, who predicted
the stability of higher-order vortex solitons with $m>1$ as well,
this issue was also addressed recently by
\cite{Davydova:2004-S197:JOA}. A detailed study of the stability
of (2+1) dimensional vortex solitons in both conservative and
dissipative CQ models can be found in
\cite{Crasovan:2001-1041:RAR}, the related issue of the stability
of \textit{spatio-temporal} (3+1)-dimensional spinning light
bullets we discuss in Sect.~\rsect{bullet}.

Formal analogies between the CQ vortex solitons and quantum fluids
have been discussed by \cite{Michinel:2002-66604:PRE} (see also
comments by \cite{Coffey:2002:LFW, Weiss:2003-6:LFW}). In
particular, analogies have been drawn between the collisional
dynamics of vortex solitons and surface tension properties by
\cite{Paz-Alonso:2004-56601:PRE}. Such features can be accurately
explained by the internal oscillations of spatial solitons in the
domain of their stability (\cite{Dong:2004-219:PD}).

To conclude this section we note that competing nonlinearities of
different kinds have been suggested. Examples include thermal
mechanism, studied by \cite{Kruglov:1992-2277:JMO}, or
nonlinearity of the form $n=1+n_2I-n_KI^K$
(\cite{Skarka:2003-317:PLA}). In the latter case, it was suggested
that for intense laser pulses in air the parameter $K$ might be as
high as $K=20$, albeit propagation of intense pulses in air
usually involves a variety of strong multiphoton processes not
captured by the above reduced model.

\subsection{Quadratic-cubic nonlinearity}

As discussed previously, bright doughnut-shaped vortex solitons in
pure $\chi^{(2)}$ media are unstable against azimuthal
symmetry-breaking perturbations, similar to their $\chi^{(3)}$
counterparts in self-focusing media. This property might be a
generic feature of bright vortices in the nonlinear models where
the balance between counteracting self-focusing type nonlinearity
and repulsive diffraction ``forces'' allows stationary
radially-symmetric states, but it is not sufficient to damp the
azimuthal modulational instability along the ring. The dark
vortices in self-defocusing $\chi^{(3)}$ medium, however, are the
stable entities provided their nonzero constant background, the
stationary \textit{plane} wave solutions, is stable against small
modulation.

The \textit{competing} $\chi^{(3)}-\chi^{(5)}$ nonlinearities,
discussed in Sect.~\rsect{CQ}, were shown to posses both, the
ability to localize the bright solitons, and to guarantee
additional stabilization in certain parameters region. Expanding
the analogy to the case of parametric solitons, one may expect
stable vortex solitons to exist in $\chi^{(2)}$-mediated wave
mixing, if there will be possible additional nonlinearity to
compete, for example of the $\chi^{(3)}$ self-defocusing type.

To start with, the dynamical equations that govern the interaction
between a weakly modulated plane wave and its second harmonic for
materials with asymmetric crystal structure, in which the effects
of both the quadratic and the cubic nonlinear susceptibility
tensors must be considered, were derived by
\cite{Bang:1997-51:JOSB}. Following these derivations and taking
into account the diffraction in two transverse dimensions and
paraxial approximation, one can describe soliton-like propagation
of narrow beams:
\begin{eqnarray}
i\frac{\partial u}{\partial z} + \Delta u-\kappa_1 u+ u^* v
-\left(|u|^2/4+2|v|^2\right)u &=& 0, \leqt{X2X3u}\\
2i\frac{\partial v}{\partial z} + \Delta v-\kappa v+ u^2/2 -\left(4|v|^2+2|u|^2\right)u &=& 0,\leqt{X2X3v}
\end{eqnarray}
where $\kappa_1$ is the nonlinear contribution to the propagation
constant for the fundamental wave $u$, and parameter $\kappa$
combines it with the phase-mismatch $\Delta k$, $\kappa=2(\Delta k
+2 \kappa_1)$.

Stable fully localized $(2+1)$-dimensional ring solitons with
intrinsic vorticity in optical media with competing quadratic and
self-defocusing cubic nonlinearities have been found by
\cite{Towers:2001-55601:PRE}. It is noteworthy that properties of
the stationary solutions to Eqs.~\reqt{X2X3u}-\reqt{X2X3v} are
very similar to those shown in Fig.~\rpict{fig15}: with increasing
of power, the amplitude saturates and soliton width diverges to
the dark state. Stability windows for sufficiently broad ring
solitons with the spin $m=1$ and $2$ have been found, both in
direct dynamical simulations and analyzing eigenvalues of the
linearized equations. Similar to their $\chi^{(3)}-\chi^{(5)}$
counterparts, stable two-color vortex solitons survive strong
perturbations such as collisions, as it was shown by
\cite{Malomed:2001-1061:RAR}.

It is necessary to say that conventional nonlinear materials with
strong $\chi^{(2)}$ nonlinearity do not satisfy the requirement of
the model to have a negative $\chi^{(3)}$ coefficient at both the
fundamental and second-harmonic frequencies. Different
possibilities to create a necessary effective $\chi^{(3)}$
nonlinearity have been proposed. For example,
\cite{Malomed:2001-1061:RAR} suggested by creating a layered
medium in which layers providing for the $\chi^{(2)}$ nonlinearity
periodically alternate with others that account for the
self-defocusing Kerr nonlinearity. Engineering $\chi^{(2)}$
quasi-phase-matched gratings (\cite{Bang:1999-1413:OL}) also
produces effective $\chi^{(3)}$ nonlinearities. However, notice
that in this case the higher-order nonlinearities are induced {\em
on average} over all the Fourier components associated to the
quasi-phase-matching modulation, thus the models only hold under
proper conditions when averaging is justified and thus these
models might not be able to stabilize otherwise unstable solitons.

A modulationally stable branch of a plane two-wave solutions to
the system Eqs.~\reqt{X2X3u}-\reqt{X2X3v} and the corresponding
stable dark vortex solitons were found by
\cite{Alexander:1998-670:OL}. Latter on,
\cite{Alexander:2000-2042:PRE} reported existence of novel vortex
states on infinite background, the so-called ``halo-vortex'' and
``ring-vortex''. It is interesting to note that in a similar model
there exist a kind of bright vortex states, localized in
transverse plane by additional harmonic trapping potential; such
system might be perhaps used to describe some features of hybrid
atomic-molecular Bose-Einstein condensates
(\cite{Alexander:2002-S33:JOB}).

\subsection{Spatiotemporal spinning solitons}\lsect{bullet}

Three-dimensional optical spatiotemporal solitons, the so-called
``light bullets'' (LB, this term was introduced by
\cite{Silberberg:1990-1282:OL}), attract a growing interest, as
they represent a new fundamental physical object. They have been
suggested to implement ultra-fast all-optical switching in bulk
media (\cite{McLeod:1995-3254:PRA, Liu:2000-35:rar,
Wise:2002-28:rar}). Physical content of this new object lies in
the spatiotemporal analogy, which allows one to consider in the
same way both, the temporal dispersion of the short light pulse,
and the diffraction (or ``spatial dispersion'') of the narrow beam
(see, e.g., paper by \cite{Kanashov:1981-122:PD}). In the presence
of group-velocity dispersion, the evolution (along propagation
direction $z$) of slowly-varying envelope of the electromagnetic
field $E(x,y,z;t)$ is described by the paraxial equations similar
to the Eq.~\reqt{CQeq} and Eqs.~\reqt{X2X3u}-\reqt{X2X3v}, but
with time-dependant Laplacian $\Delta=\nabla_\perp^2+\kappa
DE_{TT}$. Here $\kappa$ is the propagation constant (wave number),
$D=-d^2\kappa/d\omega^2>0$ is the coefficient of the temporal
dispersion assumed \textit{anomalous}, $T\equiv t-z/v_g$ ($v_g$
being the group velocity of the carrier wave) is the ``reduced
time'', and $\nabla_\perp^2=\partial^2/\partial x^2
+\partial^2/\partial y^2$ represents spatial diffraction.
Normalizing reduced time $\tau=T/\sqrt{\kappa D}$, one obtains
\textit{spatio-temporal} Laplacian $\Delta=\partial^2/\partial x^2
+\partial^2/\partial y^2 +\partial^2/\partial\tau^2$, completely
symmetrical with respect to the spatial coordinates $(x,y)$ and
reduced time $\tau$.

In nonlinear media, the self-focusing may balance out both, the temporal and the spatial broadenings of sufficiently short pulses of light. The combination of these two effects, responsible for the formation of (1+1) temporal and (2+1) spatial solitons, supports stationary states known as (3+1)-dimensional solitons, or light bullets. For the details of physics of LB, the review on theoretical and experimental progress in this field, as well as for the description of fundamental (bell-shaped) three-dimensional solitons, we refer to the recent paper by \cite{Malomed:2004-00:RMP}. Here we consider only the higher-order LB with phase dislocations, or optical vortices in spatiotemporal domain.

Similar to the (2+1)-dimensional case, stationary solutions corresponding to the fundamental (3+1)D soliton can be obtained using radially-symmetrical ansatz of the form $E=U(r)\exp(ikz)$, where $k$ is a soliton parameter and the radius is given by $r^2=x^2+y^2+\tau^2$. The same ansatz describes higher-order \textit{spherically-symmetrical} modes consisting of several concentric shells surrounding inner core (\cite{Edmundson:1997-7636:PRE}). Angle-dependent higher-order states, however, do not allow simple separation of variables, such as $E=U(r)\exp(im\phi+ikz)$, used for two-dimensional vortex solitons. Therefore, the search for higher-order LB with phase dislocations requires solving the full multidimensional stationary equation, quite a nontrivial task. The so-called ``three-dimensional spinning solitons'', introduced by \cite{Desyatnikov:2000-3107:PRE}, is only known example of LB with phase dislocation, and it may represent much broader class of possible (and yet not known) spatiotemporal optical vortices. In other systems, such example incude the ``smoke rings'' of vortex lines in non-degenerate optical parametric oscillator (\cite{Weiss:1999-151:APB}), and the ``parallel vortex rings'' in matter waves trapped by the three-dimensional external potential (\cite{Crasovan:2004-33605:PRA}).

The structure of the spinning LB can be well understood by using the approximate (e.g. variational) solutions. We suppose that stationary optical pulse $E=A(x,y,\tau)\exp(ikz)$ (localized in time $\tau$, so that $A \rightarrow 0$ for $\tau \rightarrow \pm \infty$) has a phase-dislocation located in the transverse plane, $A(x,y,\tau)=V(x,y,\tau)\exp(im\varphi)$ with azimuthal angle $\varphi=\tan^{-1}(y/x)$, similar to the two-dimensional CW vortex beams. Because of this dislocation, field should vanish in the origin and produce a ``doughnut'' shape in transverse plane, $V\rightarrow 0$ for $\rho \rightarrow 0$, here $\rho$ is a polar radius, $\rho^2=x^2+y^2$. Thus spinning LB can be thought as a CW beam modulated in time by an additional multiplier, e.g. of the $\sech(\tau)$ shape,
\begin{equation}
V=U_{cyl}(\rho)\sech(\tau),\leqt{D3c}
\end{equation}
and we note that this ansatz offers the separation of variables in \textit{cylyndrical} coordinates, which does not satisfy the nonlinear equation of course. Physically, the ansatz Eq.~\reqt{D3c} can follow as a result of temporal (longitudinal) modulational instability of initially continuous (CW) optical vortex beam, responsible for generation of soliton trains (this so-called ``neck''- instability in the spatial domain was experimentally observed by \cite{Fuerst:1997-2756:PRL}). Similar shape can be also modelled by a \textit{spherical harmonic},
\begin{equation}
V=U_{sph}(r)\cos(\theta),\leqt{D3s}
\end{equation}
with spherical coordinates, $(x,y,\tau)\rightarrow(r,\varphi,\theta)$. Two model envelopes Eqs.~\reqt{D3c} and \reqt{D3s} can be used as the trial functions for variational method. The advantage of this method is that it actually allows one to separate the variables in corresponding nonlinear equation and greatly simplifies the problem (variational methods in optics were reviewed recently by \cite{Malomed:2002-71:ProgressOptics}). Partial-differential equation for stationary envelope $V(x,y,\tau)$ is than reduced to ordinary differential equations for the envelope $U_{cyl}(\rho)$ or $U_{sph}(r)$. Solutions to these equations and their analysis for the model with cubic-quintic nonlinearity show that radial envelopes $U_{cyl}(\rho)$ and $U_{sph}(r)$ are qualitatively similar to the ones shown in Figs.~\rpict{fig15} for two-dimensional case, and they define very close values for the parameters of stationary spinning LB solution, for example the minimal threshold energy of soliton formation (\cite{Desyatnikov:2000-3107:PRE}). However, variational solutions do not provide a full information about stationary states family and their stability, and direct numerical modelling is necessary.

\pict[0.708]{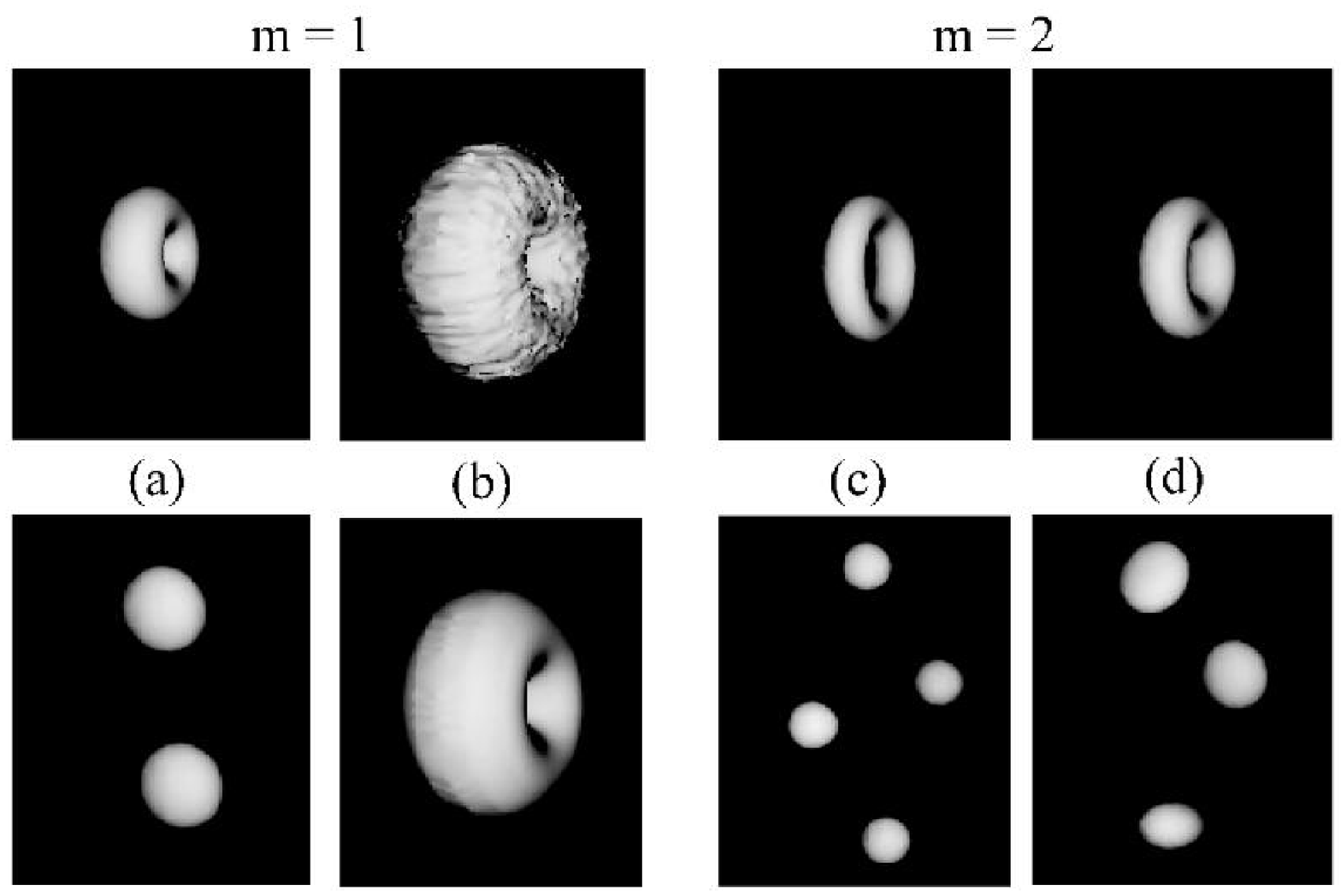}{fig16}{Examples of stable and unstable spinning light bullets in CQ model for topological charges $m=1$ in (a), (b), and $m=2$ in (c), (d). Top row -- soliton solutions (perturbed in (b)) at $z=0$. Bottom row -- after propagation of $z=60$ in (a); $z=100$ in (b); $z=50$ in (c), and $z=90$ in (d). After \cite{Mihalache:2003-56608:PRE}.}

Exact numerical solutions for spinning LD were obtained by
\cite{Mihalache:2000-1505:PRE} in the Type I SHG model, and by
\cite{Mihalache:2000-7142:PRE} for the CQ medium. Authors tested
the stability of solutions in numerical propagation and observed
azimuthal modulational instability which leads to the breakup of
doughnut solitons into several fragments, each being a stable
moving zero-spin soliton (see Fig.~\rpict{fig16}). The general
conclusion based on direct numerical simulations was that the
spinning LBs are \textit{always} unstable against azimuthal
perturbations (\cite{Crasovan:2001-1041:RAR}). Later on, however,
more accurate study of the associated linear stability problem in
the CQ model revealed first ever found completely stable
spatiotemporal vortex soliton (\cite{Mihalache:2002-73902:PRL}).
Example of stable propagation of initially strongly perturbed
spinning LB is shown in Fig.~\rpict{fig16} (b). The reason for
stabilization of spinning LBs was found in the competition between
nonlinearities. Following these predictions, stable spinning LBs
were found in the model with competing quadratic and cubic
nonlinearities (\cite{Mihalache:2002-16613:PRE}) and in
two-component vectorial CQ system
(\cite{Mihalache:2003-56608:PRE}).

\section[Multi-soliton spiraling]{Other optical beams carrying angular momentum}

In this section we summarize some theoretical and experimental
results on the study of self-trapped optical beams carrying
angular momentum, which differ from the optical vortex beams
discussed above. In general, such beams do not necessarily
correspond to the stationary states, and their angular momentum
manifests itself in the complex interaction of simple spatial
solitons and leads to their spiraling. Multi-soliton complexes
with an imposed angular momentum, such as necklaces and soliton
clusters, can also be regarded as multi-soliton spiraling beams.

\subsection{Soliton spiraling}

Spiraling of two spatial solitons was suggested theoretically by
\cite{Poladian:1991-59:OC}. This should occur when two fundamental
solitons collide with trajectories that are not lying in a single
plane, so that they form a two-body system with nonzero orbital
angular momentum. Then, if the mutual interaction is attractive,
the centrifugal repulsive force can be balanced out, and two
solitons orbit about each other in a double-helix structure, as
illustrated in Fig.~\rpict{fig17}. It is interesting to note, that
similar predictions concerning three-dimensional solitons, or
light bullets, were made by \cite{Edmundson:1993-1609:OL}, based
on the particle-like nature of soliton mutual interaction
(\cite{Edmundson:1995-2491:PRA}). In parallel, the experimental
and numerical study of azimuthal instability of vortex solitons,
described in Sect.~\rsect{MI}, revealed the spiraling behavior of
splitters, first demonstrated experimentally in rubidium vapors by
\cite{Tikhonenko:1995-2046:JOSB, Tikhonenko:1996-2698:PRL}.

In quadratic media, soliton spiraling was predicted and studied in
detail theoretically by \cite{Steblina:1998-156:OL,
Buryak:1999-245:JOSB}. The mechanical model, based on potential of
soliton interaction (\cite{Malomed:1998-7928:PRE}) has been
derived, with extremal points of effective potential corresponding
to spiraling bound states. Similar to any phase-sensitive soliton
interaction the effective mass corresponding to the phase degree
of freedom was shown to be always negative. As a result,
corresponding stationary points are of saddle type, i.e. spiraling
bound states are unstable. Depending on initial soliton states,
such as soliton velocities, relative phase, and the impact
parameter, soliton can reflect, spiral, and fuse. More recently,
experimental observation of related phenomena has been reported by
\cite{Simos:2002-211:JP4}.

The spiraling of \textit{mutually incoherent} spatial solitons was
observed experimentally (\cite{Shih:1997-2551:PRL}) and studied
theoretically (\cite{Desyatnikov:1998-1101:JETP,
Schjodt-Eriksen:1998-423:PLA}) as a possible scenario for a
\textit{dynamically stable} two-soliton bound state formed when
two solitons are launched with initially twisted trajectories.
Mutually incoherent solitons always attract each-other in
isotropic Kerr-type medium, independently of their relative phase,
and effective potential minimum corresponds to stable bound state.
Note the very similar results obtained by~\cite{Ren:2000-2124:PRL}
for ``braided light'' in plasmas. As a matter of fact, the soliton
spiraling due to an effectively vectorial beam interaction is
associated with large-amplitude oscillations of a dipole-mode
vector state generated by the interaction of two initially
mutually incoherent optical beams
(\cite{Skryabin:2002-55602:PRE}).

\pict[0.708]{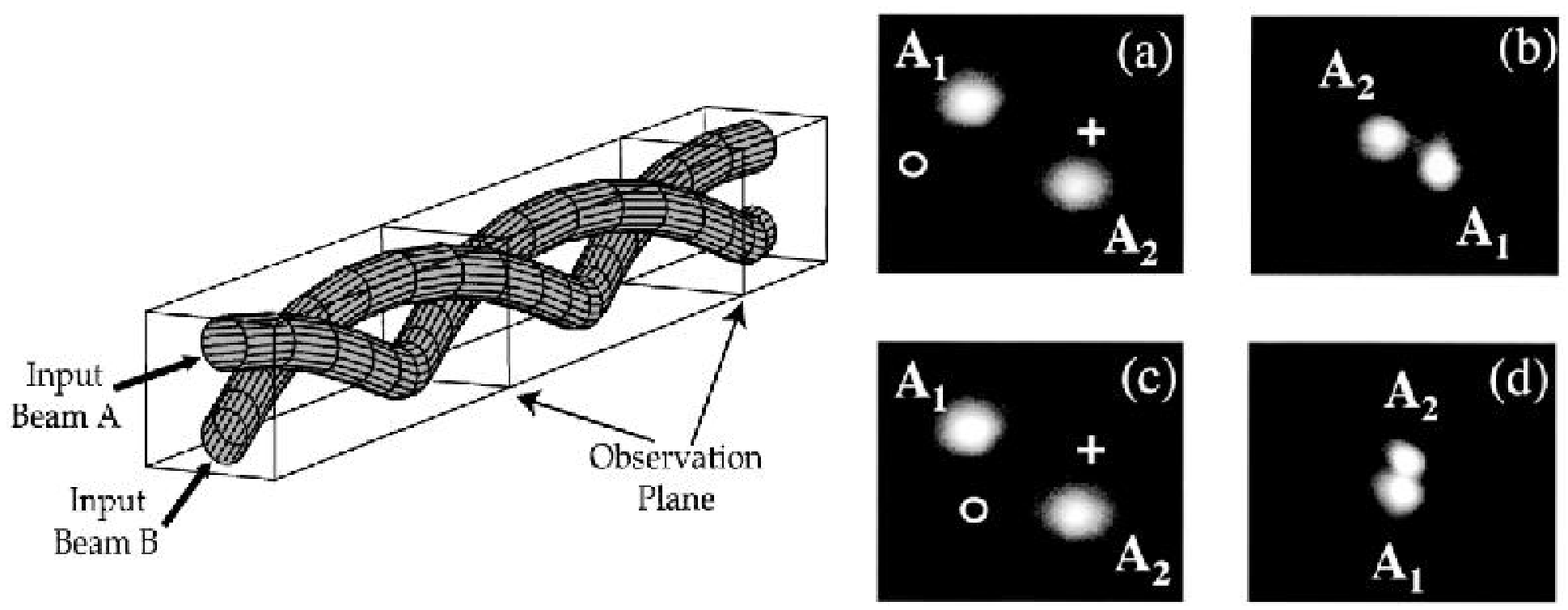}{fig17}{(Left) An illustration of the
soliton spiraling process. The arrows indicate the initial
direction of the two soliton beams. After
\cite{Shih:1997-2551:PRL}. (Right) Spiraling of solitons with
initially skewed trajectories in photorefractive crystal. (a),(c)
Initial position of the beams; cross and circle denote output
positions of solitons A1 and A2, respectively, during individual
propagation. (b),(d) Output positions of the solitons during
simultaneous propagation. After \cite{Stepken:1999-540:PRL}.}

In anisotropic photorefractive medium, however, the mutually
incoherent solitons demonstrate much complicated anomalous
interaction (\cite{Krolikowski:1998-3240:PRL,
Stepken:1998-4112:PRE, Krolikowski:1998-823:QSO}). Anomalous
interaction results in complex trajectories which typically show
partial mutual spiraling, followed by damped oscillations and the
fusion of solitons. The rotation can be propelled to prolonged
spiraling by the skewed launching of beams. This nontrivial
behavior is caused by the anisotropy of the nonlinear refractive
index change in the crystal, as was shown by
\cite{Stepken:1999-540:PRL, Belic:1999-544:PRL} and summarized by
\cite{Krolikowski:1999-975:APB, Denz:1999-6222:PRE}. Nevertheless,
the fascinating analogy between spiraling solitons and mechanical
two-body system is applicable if one takes into account the
anisotropic nature of spatial screening solitons interacting in
photorefractive medium (\cite{Belic:2000-83:PRL}), the comparison
of the above model to the isotropic one was published recently by
\cite{Belic:2002-66610:PRE}.

\subsection{Optical necklace beams}
\lsect{neck}

Since the decay of the ring-profile vortex solitons is associated
with the growing azimuthal modulation of their intensity and the
symmetry-breaking instability, one may try to stabilize the ring
structure by imposing the initial intensity and phase modulation.
The azimuthally modulated rings resemble ``optical necklaces'';
they are closely related to the higher-order guided modes, as we
discussed in Sect.~\rsect{WG}, and also to suitable superpositions
of Laguerre-Gaussian beams.

In Kerr media, the first experimental results on the self-trapping
of necklace-type beams were reported by
\cite{Barthelemy:1993-299:UltrafastPhenomena}.
Figure~\rpict{fig18}(a) shows the experimental data for the case
in which an input beam in the form of a higher-order
Laguerre-Gaussian mode was launched at the input of a CS$_2$ cell.
The beam diameter was 260~$\mu$m, with a petal thickness of about
80~$\mu$m. The output pattern after 5~cm of propagation is shown
when the intensity was low enough that the nonlinear effects were
negligible (middle). As expected, the beam diameter increased to
265~$\mu$m, because of diffraction, while the petal thickness
remained close to 80~$\mu$m. As the beam power was gradually
increased, the petal thickness decreased, because of
self-focusing. The petal size reduced to 30~$\mu$m at an intensity
level of $5\times 10^7$~W/cm$^2$
(\cite{Barthelemy:1994-104:SPIE}). Such self-trapped structures
are remarkably stable and allow one to transport optical beams
with powers several times the critical power at which a Gaussian
beam would otherwise collapse because of self-focusing; they
disintegrate for input intensities lower than the self-trapping
intensity.

\pict[0.708]{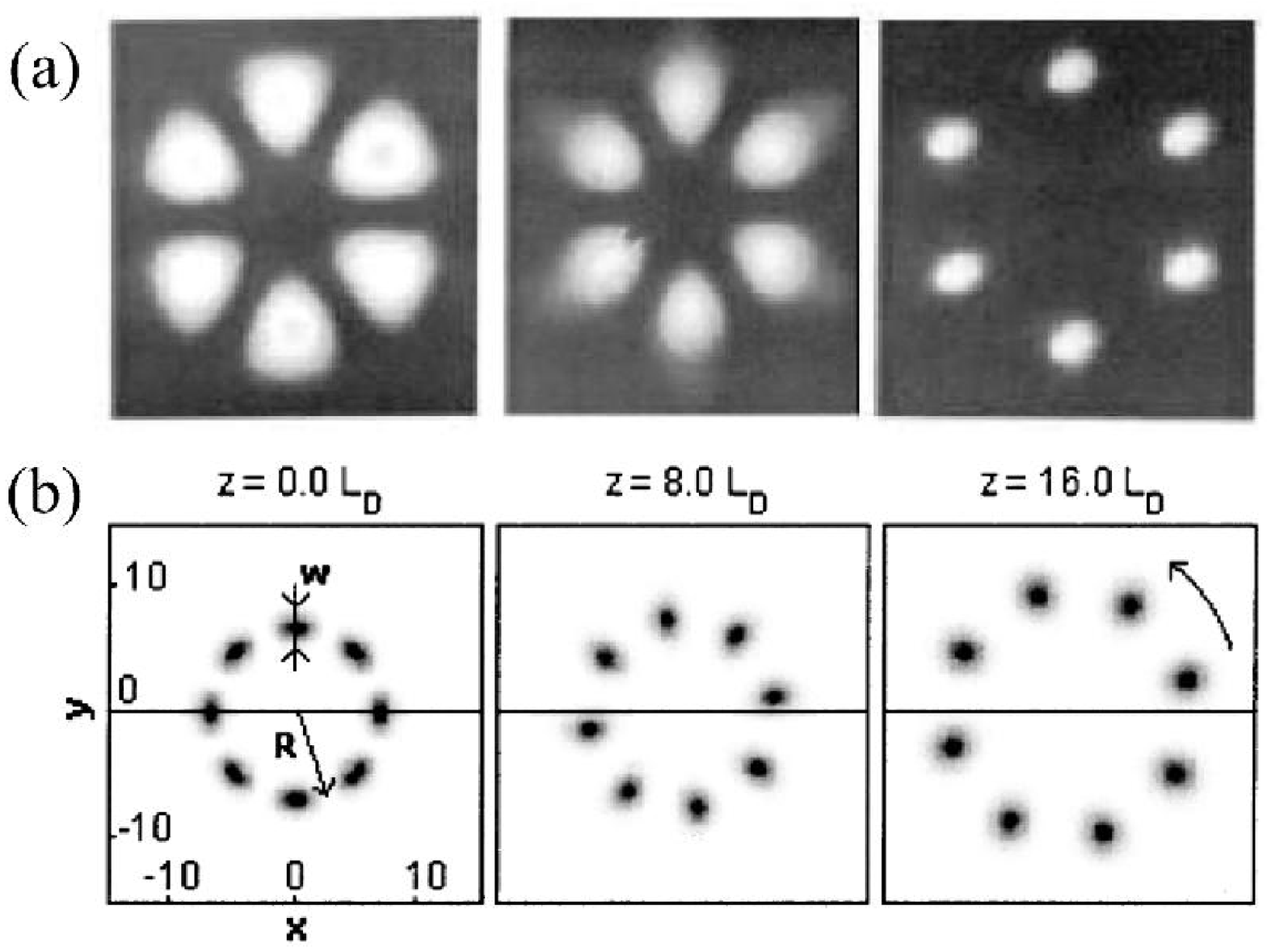}{fig18}{(a) Experimental demonstration of
the necklace beams. Propagation of a Laguerre-Gaussian beam inside
a Kerr medium: (from left to right) input beam, diffracted beam at
low intensity, and the self-trapped necklace beam at high
intensities (\cite{Barthelemy:1993-299:UltrafastPhenomena}). (b)
Rotating and expanding necklace with integer spin.
After~\cite{Soljacic:2001-420:PRL}.}

The concept of necklace beams was developed
by~\cite{Soljacic:1998-4851:PRL}, who studied numerically the
propagation of azimuthally modulated bright rings in Kerr medium.
These authors demonstrated that, in contrast to all previous
studies in this model, the self-trapped beams localized in
transverse plain can preserve their shape during the propagation
and escape the collapse instability. It is possible if such beams
are constructed as a ``necklace'' of the out-of-phase ``pearls'',
each carrying the power less than critical power of catastrophic
self-focusing. Each petal thus slowly diffracts, if it propagates
alone, and this diffraction is greatly suppressed within the ring,
because of collective self-trapping of the ring as a whole. Due to
the repulsion between neighboring petals the ring expands
self-similarly.

The shape of the necklace can be approximated by the ansatz
similar to Eq.~\reqt{multipol} with $p=0$. To slow down the
expansion of the ring, its radius should be taken as large as
possible, so that the radial envelope $U(r)$ in
Eq.~\reqt{multipol} can be approximated by the corresponding 1D
soliton of $\sech$ shape, which is a bright stripe in
two-dimensional  spatial domain. Extensive numerical simulations
and semi-analytical analysis performed
by~\cite{Soljacic:2000-2810:PRE} showed that the dynamics of the
necklace can be controlled and reduced to the quasi-stationary if
the radius of the ring, its width, the amplitude, and the order of
azimuthal modulation (the winding number $m$ in
Eq.~\reqt{multipol}), minimize corresponding action integral.
These ``quasi-solitons'' have a shape of a thin modulated stripe
wrapped to a large ring, and they propagate stably over several
tenths of diffraction lengths.

Necklaces with additional phase modulation, introducing a nonzero
angular momentum, exhibit a series of phenomena typically
associated with rotation of rigid bodies and centrifugal force
effects (\cite{Soljacic:2001-420:PRL}). The simplest way to
explore these novel features is to consider the ansatz
Eq.~\reqt{multipol} with $p\ne 0$, which, from one hand, describes
a vortex with the topological charge $m$ at $p=1$, and, from the
other hand, it includes the varying modulation parameter $p$
($0<p<1$). The {\em spin} of this nonstationary structure is
defined as
\begin{equation}
\leqt{frac} S = \frac {2mp}{1+p^2},
\end{equation}
and it vanishes for $p\rightarrow 0$ (see the definition of spin
after Eq.~\reqt{mom}). When the ring vortex is only slightly
modulated ($p\approx 1$), it decays into a complex structure of
filaments because of a competition between different instability
modes of the corresponding vortex soliton
(\cite{Desyatnikov:2002-S58:JOB}).

When the modulation becomes deeper, e.g. for $p\approx 0.5$, the initial vortex transforms into a necklace-like structure [see Fig.~\rpict{fig18}(b)], and its dynamics is modified dramatically. The modulated ring-profile structure does not decay but, instead, it expands with small rotation, the rotation is much weaker
because the initial angular momentum (spin) is much smaller than that in the case of $p\approx 1$. These ringlike structures were introduced by~\cite{Soljacic:2001-420:PRL} as the first example of optical beams with {\em fractional spin} [see Eq.~\reqt{frac}] and {\em rotating intensity}. Note, that the {\em anticlockwise} direction of the rotation is determined by the gradient of phase, which grows {\em anticlockwise} for $m>0$, see Fig.~\rpict{fig18}(b). The angular velocity vanishes as the ringlike structure expands, in analogy with Newtonian mechanics and ``scatter on ice'' effect.

\subsection{Soliton clusters}
\lsect{cluster}

In saturable media, the fundamental solitons are stable and
demonstrate the particle-like robust interaction, e.g. spiraling
out of the initial vortex ring after it breaks, see
Sect.~\rsect{MI}. The analogy with particles and forces between
them applied to spatial solitons allows one to search for the
bound states of several solitons corresponding to the balance
between all acting forces, as in classical mechanics. In order to
create non-expanding configurations of $N$ solitons in a bulk
medium, first we recall the basic physics of coherent interaction
of two spatial solitons. It is well known
(\cite{Kivshar:2003:OpticalSolitons}) that such an interaction
depends crucially on the relative soliton phase, say $\theta$, so
that two solitons attract each other for $\theta =0$, and repel
each other for $\theta = \pi$. For the intermediate values of the
soliton phase, $0<\theta<\pi$, the solitons undergo an energy
exchange and inelastic interaction.

Here we follow the original paper by
\cite{Desyatnikov:2002-53901:PRL} and analyze possible {\em
stationary configurations} of $N$ coherently interacting solitons
for a ringlike geometry. It is easy to understand that such a
ringlike configuration will be {\em radially unstable} due to an
effective tension induced by bending of the soliton array. Thus, a
ring of $N$ solitons will collapse, if the mutual interaction
between the neighboring solitons is attractive, or expand
otherwise, resembling the expansion of the necklace beams.
Nevertheless, a simple physical mechanism will provide
stabilization of the ringlike configuration of $N$ solitons, if we
introduce {\em an additional phase} on the scalar field that
twists by $2\pi m$ along the soliton ring. This phase introduces
an effective centrifugal force that can balance out the tension
effect and stabilize the ringlike soliton cluster. Due to {\em a
net angular momentum} induced by such a phase distribution, the
soliton clusters will rotate with an angular velocity which
depends on the number of solitons and phase charge $m$.

\pict[0.708]{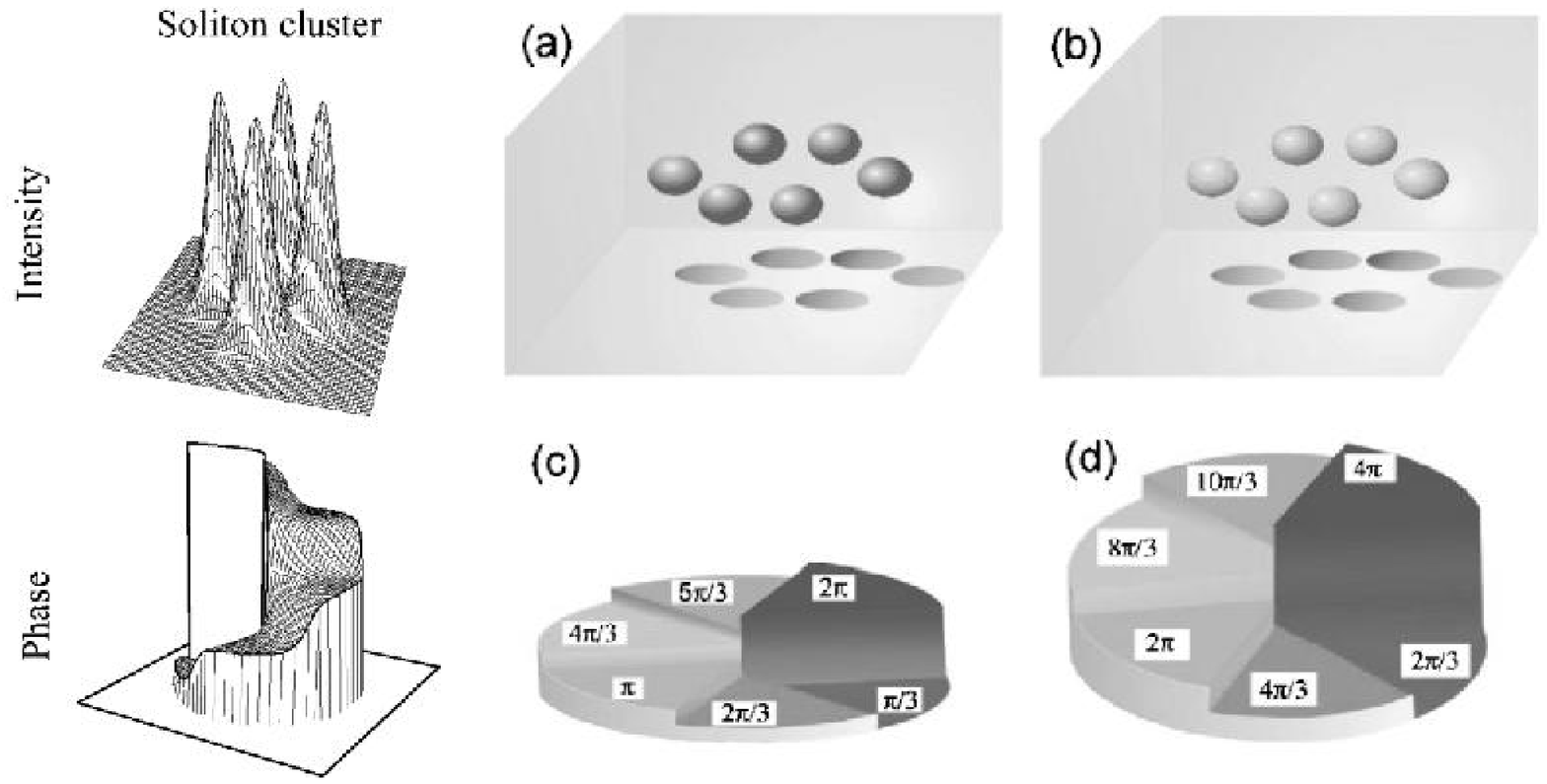}{fig19}{Left: Intensity and phase
distribution for a four-soliton cluster in saturable medium. Note
that in terms of the azimuthal coordinate
$\varphi=\tan^{-1}(y/x)$, the vortex phase is given as a linear
function $m\varphi$ with integer $m$, while the staircase-like
phase of the cluster is a \textit{nonlinear} phase dislocation
(\cite{Desyatnikov:2004-S209:JOA}). Right: Cluster composed of six
spatiotemporal two-color solitons. The topological charge $m$ of
the soliton cluster is equal to one. (a) The fundamental frequency
field and (b) the second harmonic field. (c) The phase
distribution at fundamental frequency and (d) the phase
distribution at the second harmonic
(\cite{Crasovan:2003-46610:PRE}).}

To describe the soliton clusters analytically, we consider a
coherent superposition of $N$ solitons with the envelopes
$G_{n}(x,y,z)$, $n=1,2..N$, propagating in a self-focusing bulk
nonlinear medium. The equation for the slowly varying field
envelope $E = \sum G_{n}$ can be written in the form of the NLS
equation \reqt{eq}. For a ring of identical {\em weakly
overlapping} solitons launched in parallel, we can calculate the
integrals of motion employing a Gaussian ansatz for a single beam
$G_{n}$,
\begin{equation}
\leqt{ans}\ G_{n}=A\exp\left(-\;\frac{|{\bf r} -{\bf r}_{n}|^{2}} {2a^{2}}+ i\alpha_{n}\right),
\end{equation}
where ${\bf r}_n= (x_n; y_n)$ describes the soliton location, and
$\alpha_n$ is the phase of the $n$-th beam. We assume that the
beams $G_n$ are arranged in a ring-shaped array of radius $R$,
i.e. ${\bf r}_n=\lbrace{R}\cos\varphi_n; {R}\sin\varphi_n \rbrace$
with $\varphi_n=2\pi n/N$.

Analyzing many-soliton clusters, we remove the motion of the
center of the mass and put ${\bf L}=0$, here the linear momentum
is given by Eq.~\reqt{linmom}. Applying this constraint, we find
the conditions for the soliton phases,
$\alpha_{i+n}-\alpha_i=\alpha_{k+n}-\alpha_k$, which are satisfied
provided the phase $\alpha_n$ has a linear dependence on $n$, i.e.
$\alpha_n=\theta n$, where $\theta$ is the relative phase between
two neighboring solitons in the ring. Then, we employ the
periodicity condition in the form $\alpha_{n+N}=\alpha_n+2\pi m$,
and find:
\begin{equation}
\leqt{phase}
\theta=\frac{2\pi m}{N}.
\end{equation}

In terms of the field theory, Eq.~\reqt{phase} gives the condition
of the vanishing energy flow ${\bf L} = 0$, because the linear
momentum ${\bf L}=\int {\bf j} d{\bf r}$ can be presented through
the local current ${\bf j} = {\rm Im}(E^{\ast}\nabla E)$.
Therefore, Eq.~\reqt{phase} determines a nontrivial phase
distribution for the effectively elastic soliton interaction in
the ring. In particular, for the well-known case of two solitons
($N=2$), this condition gives only two states with the zero energy
exchange, when $m$ is even ($\theta = 0$, mutual attraction) and
when $m$ is odd ($\theta =\pi$, mutual repulsion).

For a given $N>2$, the condition~\reqt{phase} predicts the
existence of a discrete set of allowed states corresponding to a
set of the values $\theta = \theta ^{(m)}$ with $m=0,\pm 1,
\ldots, \pm (N-1)$. Here, two states $\theta ^{(\pm |m|)}$ differ
by the sign of the angular momentum, similar to the case of vortex
solitons. Moreover, for any positive (negative) $m_+$ within the
domain $\pi <|\theta| < 2\pi$,  one can find the corresponding
negative (positive) value $m_-$ within the domain
$0<|\theta|<\pi$, so that both $m_+$ and $m_-$ describe the same
stationary state. For example, in the case $N=3$, three states
with zero energy exchange are possible: $\theta^{(0)}=0$,
$\theta^{(1)}=2\pi/3$, and $\theta^{(2)}=4\pi/3$, and the
correspondence is $\theta ^{(\pm 1)} \leftrightarrow \theta ^{(\mp
2)}$. The number $m$ determines the full phase twist around the
ring, and it plays a role of the topological charge of the
corresponding phase dislocation, see Fig.~\rpict{fig19}.

Applying the effective-particle approach,
\cite{Desyatnikov:2002-53901:PRL} derived an effective interaction
energy for the soliton cluster and have shown that it can be
classified in a simple way by its extremal points. The existence
of a minimum point suggests that such a configuration describes a
stable or long-lived ring-like cluster of a particular number of
solitons. This prediction was verified by a series of numerical
simulations for different $N$-soliton rings and their propagation
in a saturable medium.

For example, the effective potential is always attractive for
$m=0$, and thus the ring of in-phase solitons exhibit oscillations
and fusion. Another scenario of the mutual soliton interaction
corresponds to the repulsive potential, e.g., for the case
$\theta>\pi/2$. In the numerical simulations corresponding to this
case, the ringlike soliton array expands with the slowing down
rotation, similar to the rotating necklace beams, see
Fig.~\rpict{fig18}(b). Finally, the stationary soliton bound state
that corresponds to a minimum of the effective potential is shown
in Fig.~\rpict{fig19} for particular case $N=4$ and $m=1$. Here
the angular momentum is nonzero, and it produces a repulsive
centrifugal force that balances out an effective attraction of
$\pi/2$-out-of-phase solitons. The general rule, generalizing a
two-soliton phase sensitive interaction, predicts the existence of
a bound state of $N$ solitons if the nonzero phase step $\theta$
is equal or less $\pi/2$, i.e. the interaction is attractive and
can be balanced out by the net centrifugal force for $0<\theta\le
\pi/2$. Therefore, soliton cluster with topological charge $m$ can
be quasi-stationary only for $N\ge 4m$, see Eq.~\reqt{phase}. In
addition, there possible ``excited'' states with the radius of the
cluster oscillating near the minimum of interaction potential
during the propagation.

\pict[0.708]{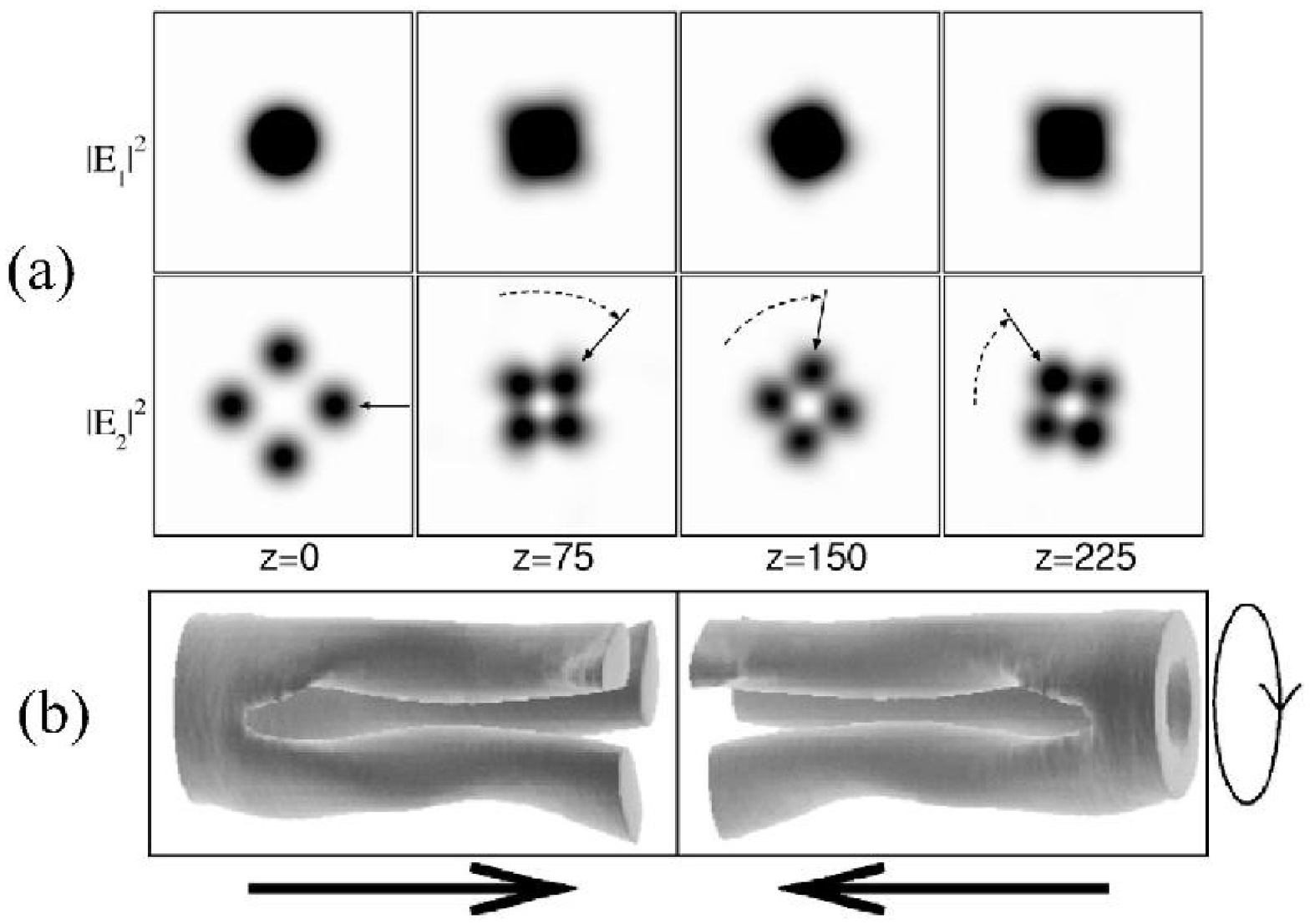}{fig20}{(a) Rotating \textit{vector}
cluster consisting of four fundamental solitons. The angle of
rotation, corresponding to the distance $z=225 L_D$, is $\sim
11.25 \pi $. Note that the direction of rotation is opposite to
the one of necklace in Fig.~\rpict{fig18}(b), despite the fact
that in both cases the angular momentum is positive.
After~\cite{Desyatnikov:2002-S58:JOB}. (b) Iso-surface plot of two
counter-propagating vortices. Breakup into three rotating beamlets
is visible. Because of the topological charge, the beamlets start
to spiral, which is only weakly visible due to the short
propagation distance. The beamlets rotate in the direction
indicated by the arrow on the right
(\cite{Motzek:2003-66611:PRE}).}

The concept of soliton clusters was extended to the case of light
propagation in quadratically nonlinear media
by~\cite{Kartashov:2002-2682:JOSB}, where it is intimately related
to the concept of induced beam break-up or soliton algebra
discussed above. Due to the phase relation between the fundamental
wave and the second harmonic beam (see Sect.~\rsect{X2}), the
topological charge in the second-harmonic field is double the
fundamental wave charge. Therefore, the corresponding phase jump
between neighboring solitons Eq.~\reqt{phase} is doubled, see
Fig.~\rpict{fig20}. Different regimes of cluster propagation were
found to be possible.

Metastable, robust propagation of clusters in media with competing
quadratic and self-defocusing nonlinearities was reported
by~\cite{Kartashov:2002-273902:PRL}. These results were followed
by similar findings in the cubic-quintic model as well
by~\cite{Mihalache:2003-46612:PRE}. Similar to the extension of
the concept of two-dimensional  vortex solitons into
spatiotemporal domain (see paper
by~\cite{Desyatnikov:2000-3107:PRE} and Sect.~\rsect{bullet}), the
bound states of three-dimensional solitons, or light bullets, can
be constructed using the approach described above for
two-dimensional  \textit{cw} beams. Tree-dimensional, robust
``soliton molecules'' were introduced by
\cite{Crasovan:2003-46610:PRE, Mihalache:2004-S333:JOB} using this
concept. In Fig.~\rpict{fig20} such a molecule is shown,
constructed from six two-color ``atoms''.

It turns out that the special initial condition for $N$ solitons
in the ring, namely the ``inverse'' picture of the vortex
splitting Fig.~\rpict{fig6}, allows one to reconstruct the vortex
soliton. \cite{Desyatnikov:2004-S209:JOA} demonstrated how several
initially well separated solitons, being launched toward the
target ring in the absence of perturbations, can excite a
metastable vortex ring. Mutual trapping of several solitons on the
collision can be regarded as a synthesis of soliton molecules, and
it corresponds to a transfer of an initial angular momentum of a
system of solitons to angular momentum stored by the optical
vortex. Similar results were obtained
by~\cite{Mihalache:2004-S333:JOB} for 3D clusters of light
bullets, launched with additional azimuthal tilts which mimics the
vortex phase.

The analogy with particles and their bound states, employed to
develop the concept of \textit{scalar} soliton clusters, can be
applied to the composite solitons, constructed from incoherently
coupled beams, see Sect.~\rsect{X3vect}. In this case, two
mutually incoherent beams can be regarded as atoms of different
sorts, always attracting each other (see, e.g., papers
by~\cite{Segev:1998-503:OQE, Stegeman:1999-1518:SCI}). Now, the
combination of \textit{three} types of forces acting between
solitons allows to construct a great variety of bound states, or
quasi-solitons. These forces are of different origin, namely the
coherent phase-dependant attraction/repulsion, incoherent
attraction, and the repulsive centrifugal force in clusters with
angular momentum. The important issue of the effective interaction
potential was investigated by~\cite{Malomed:1998-7928:PRE,
Maimistov:1999-179:PLA}, and conservation of angular momentum for
interacting two-dimensional and three-dimensional solitons
discussed by~\cite{Desyatnikov:2000-1009:QE}.

Combining the ringlike clusters of $N$ fundamental solitons with
the staircase phase distribution \textit{and} the stabilizing
effect of the vectorial beam interaction,
\cite{Desyatnikov:2002-S58:JOB} introduced a concept of the {\em
vector soliton clusters}. Indeed, if we take a scalar cluster of
solitons and combine it with the other beam that {\em interacts
incoherently} with the primary beam supporting the cluster
structure, the resulting vector soliton will demonstrate the
surprising long-lived rotational dynamics, as clearly seen in
Fig.~\rpict{fig20}(a). Moreover, the vectorial interaction can
allow to trap not only a large number of solitons with $N\geq 4$,
but also three and even two solitons. In the latter case, this
structure is, as a matter of fact, the rotating dipole-mode vector
soliton or optical propeller (\cite{Krolikowski:2000-1424:PRL,
Carmon:2001-143901:PRL}), where two out-of-phase solitons are
trapped by the other beam interacting incoherently. Therefore, the
rotating structure presented in Fig.~\rpict{fig20}(a) is somewhat
similar to the {\em four-soliton propeller}.

Similar ideas are applicable in other field, such as light in plasmas (\cite{Ren:2000-2124:PRL, Berezhiani:2002-46415:PRE}) or atomic mixtures of Bose-Einstein condensates (\cite{Perez-Garcia::2003-61804:PRE}). The latter case of trapped matter-waves is especially interesting because of recent breakthroughs in experimental realization of vortices, we discuss it in Sect.~\rsect{BEC}. Recent results of interaction of counter-propagating vortices presented by~\cite{Motzek:2003-66611:PRE}, also point out to the existence of robust stationary points resembling soliton clusters, see Fig.~\rpict{fig20}(b). This particular system is known to be a reach source of the dynamic instabilities and chaos, thus the clustering of vortices resulted from this instability may indicate
the presence of ``islands'' of order attracting unstable system. Finally, we mention here clusters of dissipative solitons in lasers and externally driven cavities (\cite{Vladimirov:2002-46606:PRE, Skryabin:2002-44101:PRL}), however, these bound states have different physical origin and we briefly explain this difference in Sect.~\rsect{dissipative}.

\section[Lattice vortices]{Discrete vortices in two-dimensional lattices}
\lsect{lattice}

The optical vortices discussed above propagate in homogeneous nonlinear media. When refractive index is periodically modulated, it modifies the wave diffraction properties, and it can affect strongly both nonlinear propagation and localization of light (\cite{Kivshar:2003:OpticalSolitons, Christodoulides:2003-817:NAT}). As a result, periodic photonic structures and photonic crystals recently attracted a lot of interest due to the unique ways they offer for controlling light propagation. In particular, many nonlinear effects, including formation of lattice solitons, have been demonstrated experimentally for one- and two-dimensional optically-induced photonic lattices (\cite{Fleischer:2003-23902:PRL, Neshev:2003-710:OL, Fleischer:2003-147:NAT}). The concept of optically-induced lattices arises from the possibility to modify the refractive index of a nonlinear medium with periodic optical patterns, and use a weaker probe beam to study scattering of light from the resulting periodic photonic structure. Current experiments employ photorefractive crystals with strong electro-optic anisotropy to create a \textit{linear} optically-induced lattice with a polarization orthogonal to that of a probe beam, which also eliminates the nonlinear interaction between the beam and the lattice.

A vortex beam propagating in an optical lattice can be stabilized by the effective lattice discreteness in a self-focusing nonlinear media creating a two-dimensional \textit{discrete vortex soliton}. This has been shown in several theoretical studies of the discrete (\cite{Johansson:1998-115:PD, Malomed:2001-26601:PRE, Kevrekidis:2002-16605:PRE, Kevrekidis:2002-16609:PRE, Kevrekidis:2004-56612:PRE}) and continuous models with an external periodic potential (\cite{Yang:2003-2094:OL, Baizakov:2003-642:EPL, Yang:2004-47:NJP}), and such vortices have been also generated experimentally by~\cite{Fleischer:2004-30:OPN}. Below, we discuss some of the basic properties of the discrete vortices and summarize the major experimental observations.

We consider two-dimensional optically-induced lattices created in photorefractive crystals. In this case, the evolution of a laser beam can be described by the generalized nonlinear Schr\"odinger-type equation,
\begin{equation}
\leqt{nls}
   i \frac{\partial \Psi}{\partial z} + D \left(\frac{\partial^2 \Psi}{\partial x^2}
   +  \frac{\partial^2 \Psi}{\partial y^2}\right) - {\cal G}( x, y, |\Psi|^2) \Psi = 0,
\end{equation}
where $\Psi(x,y,z)$ is the normalized envelope of the electric field, the transverse coordinates $x,y$ and the propagation coordinate $z$ are  normalized to the characteristic values $x_0$ and $z_0$, respectively, $D = z_0 \lambda / (4 \pi n_0 x_0^2)$ is the beam diffraction coefficient, where $n_0$ is the average medium refractive index and $\lambda$ is the vacuum wavelength. The function ${\cal G}(x,y,|\Psi|^2)$ accounts for both lattice potential and nonlinear beam self-action effects,
\begin{equation}
\leqt{photorefr}
 {\cal G}=\gamma \left\{I_b+I_0 \sin^2( \pi x/d ) \sin^2( \pi y/d )+|\Psi|^2 \right\}^{-1},
\end{equation}
where $\gamma$ is proportional to the external biasing field, $I_b$ is the dark irradiance, and $I_0$ is the intensity of interfering beams that induce a square lattice of period $d$ through the photorefractive effect (see details by~\cite{Fleischer:2003-147:NAT, Neshev:2004-123903:PRL, Fleischer:2004-123904:PRL}). Similar mathematical models appear for describing the self-action effects in nonlinear photonic crystals (\cite{Mingaleev:2001-5474:PRL}), and the nonlinear dynamics of atomic BEC in optical lattices (\cite{Ostrovskaya:2003-160407:PRL}).

\pict[0.708]{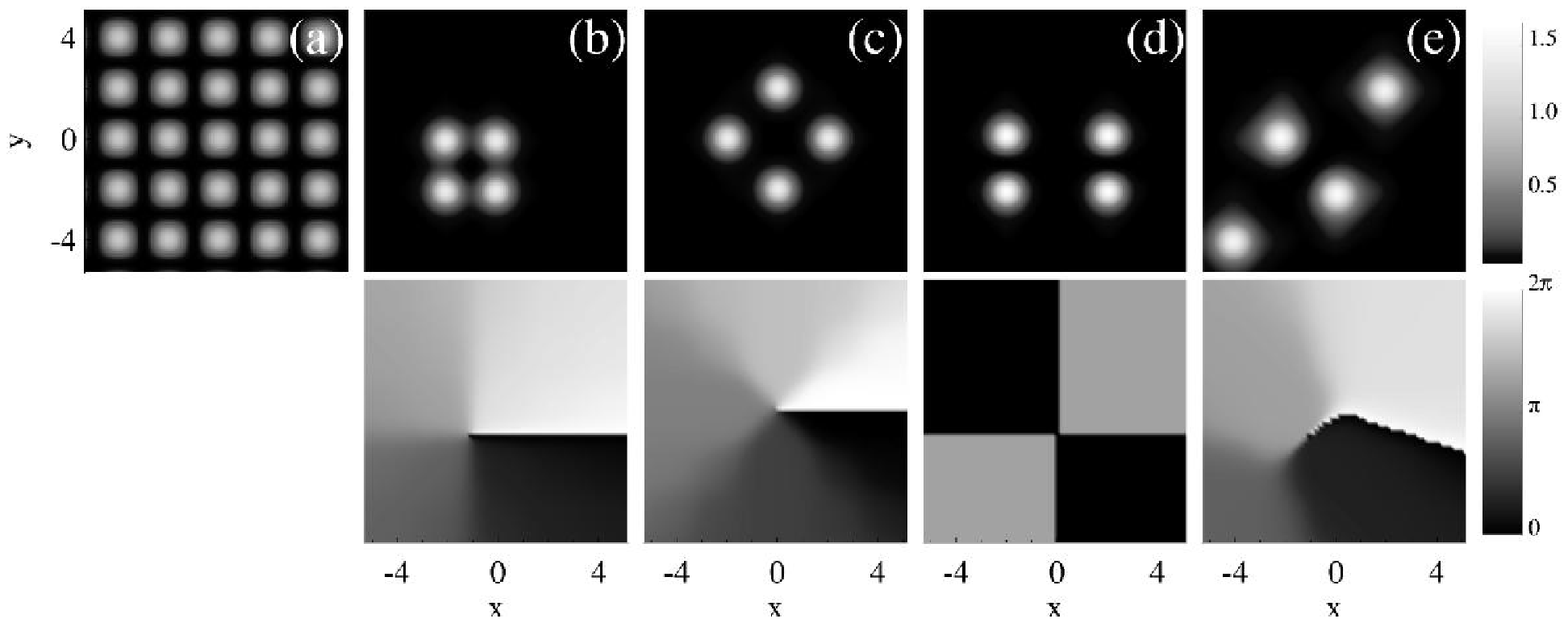}{fig21}{Examples of the vortex-type soliton structures with various symmetries in a square lattice potential (a), where (b,c)~symmetric square vortex solitons; (d) rectangular structure that can only have a trivial phase profile; (e) rhomboid configuration with a topological charge $+1$. Shown are the intensity profiles (top) and phase structure (bottom). After \cite{Alexander:2004-63901:PRL}.}

Self-focusing nonlinearity in Eq.~\reqt{nls} can compensate for the diffraction-induced beam spreading in the transverse directions, leading to the formation of stationary structures in the form of spatial solitons, $\Psi(x,y,z) = \psi(x,y) e^{i \beta z}$, where $\psi(x,y)$ is the soliton envelope, and $\beta$ is a soliton parameter, a nonlinear shift of the beam propagation constant. In order to analyze the vortex-like structures in a periodic potential, we present the field envelope in the form, $\psi(x,y) = |\psi(x,y)|\exp[i \varphi(x,y)]$, and assume that the accumulation of the phase $\varphi$ around a singular point (at $\psi=0$) is $2\pi M$, where the integer $M$ is a topological charge of the phase singularity. We consider spatially localized structures in the form of {\em vortex-like bright solitons} with the envelopes decaying at infinity. Such structures may exist when the soliton eigenvalue $\beta$ is inside a gap of the linear Floquet-Bloch spectrum of the periodic structure (\cite{Mingaleev:2001-5474:PRL, Ostrovskaya:2003-160407:PRL}). More importantly, a self-induced waveguide created by the vortex soliton is double-degenerated, and it supports simultaneously two modes, $|\psi(x,y)| \cos \varphi$ and $|\psi(x,y)| \sin \varphi$, for the same value of $\beta$. For symmetric vortex-like configurations, i.e. those possessing a 90$^\circ$ rotational symmetry, this is always the case.

The profiles of stable symmetric vortex solitons, mentioned above, resemble closely a ring-like structure of the soliton clusters in homogeneous media, see Sect.~\rsect{cluster}. Using the reduced Hamiltonian approach developed by~\cite{Desyatnikov:2002-53901:PRL} in homogeneous media,  \cite{Alexander:2004-63901:PRL} have generalized it to construct the discrete vortex solitons as a superposition of a finite number of the fundamental (no nodes) solitons, similar to Eq.~\reqt{ans}. Here, in contrast to the case of a homogeneous medium, the positions of individual solitons are fixed by the lattice potential, provided the lattice is sufficiently strong. This approximation is valid when the overlapping integrals between solitons with numbers $n$ and $m$ are the small parameter, $c_{n \ne m} \ll 1$. It was rigorously demonstrated by~\cite{MacKay:1994-1623:NLN, Aubry:1997-201:PD} that under such conditions the soliton amplitudes are slightly perturbed due to their interaction, and one can seek stationary solutions using the perturbation approach. In the first order a simple constraint for the soliton phases $\alpha_n$ have been obtained, cf. Eq.~\reqt{phase},
\begin{equation}
\leqt{balance}
  \sum_{m=0}^{N-1} c_{n m} \sin( \alpha_m - \alpha_n ) = 0 .
\end{equation}
In the sum \reqt{balance}, each term defines the energy flow between the solitons with numbers $n$ and $m$, so that the equations \reqt{balance} introduce a condition for {\em a balance of energy flows} which is a necessary condition for stable propagation of a soliton cluster and the vortex-soliton formation. These conditions are satisfied trivially when all the solitons are in- or out-of-phase. The nontrivial solutions of Eqs.~\reqt{balance} correspond to the vortex-like soliton clusters have been analyzed only for symmetric configurations (\cite{Eilbeck:1985-318:PD, Eilbeck:2003-44:Proc}), and even then some important solutions have been missed. Using this approach, \cite{Alexander:2004-63901:PRL} introduced different novel types of \textit{asymmetric vortex solitons}, some of them are shown in Fig.~\rpict{fig21}. Moreover, the existence properties of asymmetric vortex-like solutions are highly nontrivial, due to specific properties of the coupling coefficients calculated for realistic periodic structures. Note also that the symmetric modes can be represented as the angular Bloch modes (\cite{Ferrando:nlin.PS/0411059:ARXIV}).

\pict[0.708]{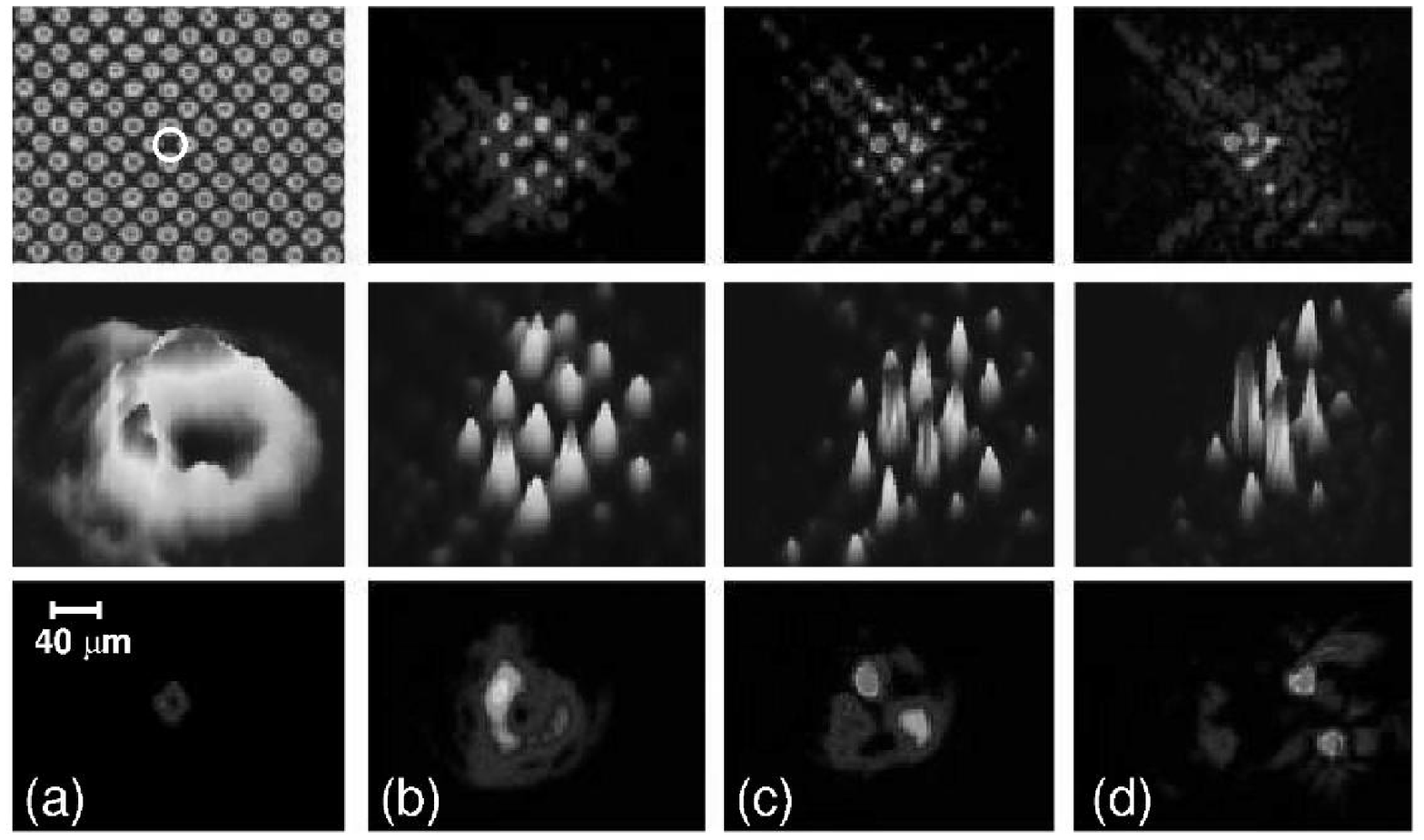}{fig22}{Experimental observation of an optical vortex propagating with (top panel) and without (bottom panel) an optically-induced lattice. Middle panel - three dimensional representation. (a) input intensity pattern of the lattice (top) and input (middle) and linear output vortex beam (bottom). (b-d) intensity patterns of the vortex beam at crystal output with a bias field of 600, 1200, and 3000~V/cm, respectively. The bright ring in the lattice pattern (a) indicates the location of the input vortex. After~\cite{Neshev:2004-123903:PRL}.}

In a number of recent publications, authors develop the concept of higher-order lattice solitons, such as dipole solitons (\cite{Yang:2004-1662:OL}), ``quasivortices'' (\cite{Kevrekidis:2004-56612:PRE}), or even periodic soliton
trains  (\cite{Chen:2004-143902:PRL, Kartashov:2004-2831:OE}). These multi-humped states can be constructed as the bound states of individual lowest order lattice solitons (\cite{Kevrekidis:2001-9615:JPA}). Though the simplest stationary solutions with in-phase solitons also exist, usually the stable configuration requires the edge-type $\pi$-phase dislocations between neighboring sites (\cite{Kartashov:2004-1918:OL}). The approach developed by~\cite{Alexander:2004-63901:PRL} includes into this rich family the higher order solitons with nested screw phase dislocations and nonzero angular momentum.

To provide evidence for the existence of the  discrete vortex
solitons two separate groups performed independent experimental
investigations (\cite{Neshev:2004-123903:PRL,
Fleischer:2004-123904:PRL}). Both relied on optical induction to
create a nonlinear lattice in a photosensitive (photorefractive)
material (\cite{Efremidis:2002-46602:PRE}). In this technique
ordinarily polarized light is periodically modulated (by
interference or by imaging a mask) to induce a 2D array of
waveguides in an anisotropic photorefractive crystal. A separate
probe beam of extraordinary polarization acquires a vortex
structure by passing through a phase mask and is then launched
into the array. The degree of nonlinearity is controlled by
applying a voltage across the \textit{c}-axis of the crystal
(photorefractive screening nonlinearity) and controlling the
intensity of the probe beam.

Typical experimental results are summarized in Figs.~\rpict{fig22}. A two-dimensional square lattice was first created, with its principal axes oriented in the diagonal directions shown in the top panel of Fig.~\rpict{fig22}(a). The resulting periodic structure acts as a square array of optically induced waveguides for the probe beam. The vortex beam, shown in the bottom panel of Fig.~\rpict{fig22}(a), was then launched straight into the middle of the lattice ``cell" of four waveguides, as indicated by a bright ring in the lattice pattern. Due to the coupling between closely spaced waveguides of the lattice, the vortex beam exhibits discrete diffraction when the nonlinearity is low [Fig.~\rpict{fig22}(b)], whereas it forms a discrete vortex soliton at an appropriate level of higher nonlinearity [see Fig.~\rpict{fig22}(c,d), the top-middle panel]. As predicted, the observed discrete diffraction and discrete self-trapping of the vortex beam in the photonic lattice is remarkably different from that in a homogeneous medium [see Fig.~\rpict{fig22}(bottom row)].

In addition, we mention that the concept of lattice solitons has
been recently explored theoretically in lattices induced by Bessel
beams \cite{Kartashov:2004-93904:PRL, Kartashov:2004-444:JOB}. In
this case the optically induced potential posesses a cylindrical
symmetry and support stable soliton complexes in the form of
ring-shaped multipoles and necklaces
(\cite{Kartashov:2004-65602:PRE}). Remarkably, such lattices allow
stable ring vortex soliton to exist even in self-defocusing medium
(\cite{Kartashov:2005-00000:PRL}).

Further generalization of the concept of lattice vortex solitons
include higher-band vortex solitons (\cite{Manela:2004-2049:OL}),
made of two components from different bands, composite vortices in
Bessel lattices (\cite{Kartashov:2005:JOSB}), and in
conventional honeycomb lattices made in quadratic nonlinear media
(\cite{Xu:2005-000:PRE}). We also notice recent investigations of
higher-order antiguiding modes (\cite{Yan:2004-104:OE}) and
optical vortices (\cite{Ferrando:2004-817:OE}) in photonic crystal
fibers which feature interesting similarities with higher-order
solitons and vortices in optically-induced lattices.

\section{Links to vortices in other fields}

Many diverse concepts in physics, ranging from the vortex clusters
in quantum dots (\cite{Saarikoski:2004-116802:PRL}) to the data
vortex switch architecture in the networks of optical waveguides
(\cite{Yang:2002-1242:JLT}), may find their analogies with the
physics of optical vortices. The concepts of linear singular
optics can be expanded to the femtosecond regime
(\cite{Bezuhanov:2004-1942:OL}) as well as to apply in the
entirely new wavelength domain, e.g., the X-ray regime
(\cite{Peele:2002-1752:OL, Peele:2003-2315:OE,
Turner:2004-2960:OE}). The fundamental physical concept of phase
singularities finds many promising applications, such as
extensively studied manipulation of micro-objects by optical
tweezers and spanners (\cite{Simpson:1996-2485:JMO,
Friese:1998-348:NAT, Gahagan:1999-533:JOSB, Koumura:1999-152:NAT,
Paterson:2001-912:SCI, MacDonald:2002-1101:SCI,
Grier:2003-810:NAT}). Moreover, in many cases, the physics of
optical vortices is useful for getting a deeper insight into the
novel phenomena described by similar nonlinear models. Here, we
mention only three of such rapidly growing fields that include
vortex states in dissipative systems, vortices in Bose-Einstein
condensates, and the application of singular optical beams to
quantum information. Each of these topics is rather extended and
diverse, and deserves a separate overview.

\subsection{Vortices in dissipative optical systems}
\lsect{dissipative}

Above, we described the vortex solitons and related phenomena in
optical systems with the help of the conservative NLS-type
equations. The NLS equation can be linked to a more general
dissipative Ginzburg-Landau (GL) model as its conservative limit.
The theoretical approach of~\cite{Ginzburg:1950-1064:ZETF} was
introduced to describe the phenomena of superconductivity
(\cite{Cyrot:1973-103:RPP}). \cite{Abrikosov:1957-1174:JETP}
developed further this approach for the type II superconductors,
more common in nature, and showed that the flux penetrates the
superconductor in the form of a regular array of flux tubes or
vortices (Abrikosov vortices). In two dimensions, the
\textit{complex} GL equation (CGL) admits extensively studied
quantized vortices (\cite{Cross:1993-851:RMP,
Pismen:1999:VorticesNonlinear}), the stationary limit of the
latter equation is also known as the stationary Gross-Pitaevskii equation. A
large number of publications is devoted to the study of vortices
in boson condensates (such as superconductors and superfluids),
described by the CGL equation (see, e.g.,
\cite{Ovchinnikov:1997-199:PDE, Ovchinnikov:1998-1277:NLN,
Ovchinnikov:1998-1295:NLN, Ovchinnikov:2002-153:rar}, and
references therein). In the context of nonlinear optics, the
latter model corresponds to the stationary NLS with
self-defocusing type nonlinearity, discussed above in
Sect.~\rsect{dark}. It is interesting to note that related
\textit{integrable} complex sine-Gordon model admits the
continuous families of the non-radially symmetric dark vortex
solitons (\cite{Barashenkov:2002-2121:NLN}), while their existence
remains an open question for the nonintegrable CGL equation
(\cite{Ovchinnikov:2000:preprint}).

The analogy between superfluids and laser optics was recognized as
early as 1970 (\cite{Graham:1970-31:rar}). In particular, the
vortex solutions to laser equations were found
by~\cite{Coullet:1989-403:OC, Tamm:1990-1034:JOSB} and intensively
studied latter, both theoretically and experimentally
(\cite{Weiss:1999-151:APB}). The concept of vortices in lasers is
connected to the \textit{dissipative optical solitons} (DOS), or
auto-solitons. This kind of solitons was initially predicted for
wide-aperture nonlinear interferometers excited by external
radiation (\cite{Rozanov:1988-1399:OPSR}) and for laser systems
with saturable absorbers (\cite{Rozanov:1992-1394:OPSR}). If the
material relaxation time is much smaller than that for the field
in optical resonator (the so-called class-A laser), the master
Maxwell-Bloch equations can be reduced to a CGL equation
(\cite{Mandel:1997:TheoreticalProblems})
\begin{equation}
\leqt{CGL} \frac{\partial E}{\partial \zeta} = (\delta+i) \Delta_d
E+ f\left(|E|^2\right) E+E_i.
\end{equation}
Here the evolution variable $\zeta$ stands for time, in the
resonator schemes, or the propagation coordinate $z$, in a bulk
medium, $\delta$ is the effective diffusion coefficient, and the
diffraction operator $\Delta_d$ is acting in $d=1,2,3$
``transverse'' coordinates. The parameter $E_i$ represents an
external driving plane-wave field, and the nonlinearity
$f\left(|E|^2\right)$ is a complex function, so that the
conservative limit Eq.~\reqt{eq} is given by $\delta=E_i=0$ and
$f\left(|E|^2\right)=iF\left(|E|^2\right)$.

\pict[0.708]{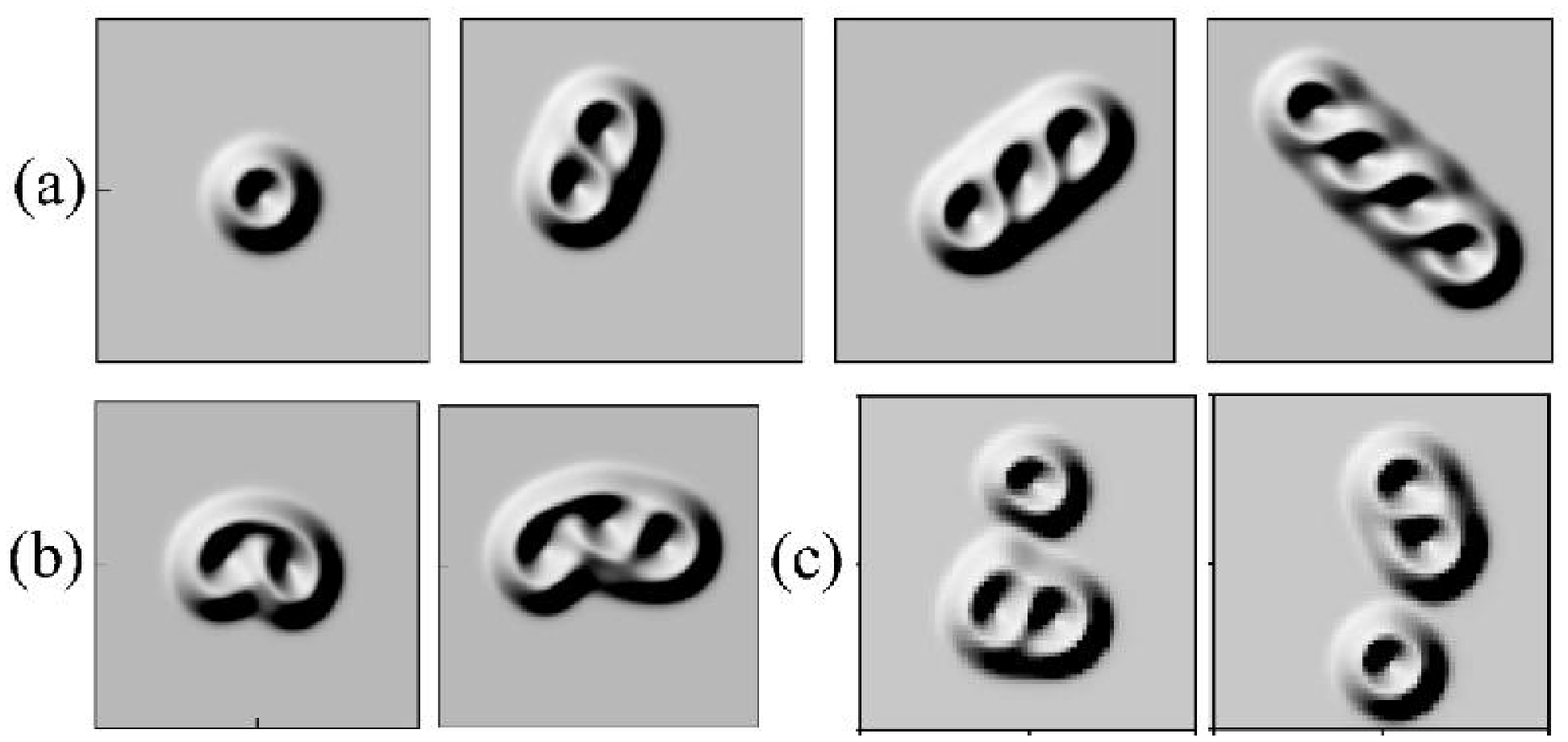}{fig23}{(a,b) The instantaneous transverse
distributions of intensity of laser radiation for the regime of
rotating chains with (a) different number of single-charged
dislocations with $m=+1$, and (b) the opposite topological charges
of the links $m=-1,+1$ (on the left) and $m=-1,+1,+1$ (right).
After~\cite{Rozanov:2003-843:OPS}. (c) Interaction between laser
vortices. The instantaneous transverse intensity distributions of
laser radiation for the regime of three solitons with topological
charges $m =-1$ during the evolution at dimensionless time $t=0$
and $t=1250$ (\cite{Rozanov:2004-88:OPS}).}

The wide-aperture interferometers, filled with passive or active
nonlinear media (\textit{optical cavities}) and excited by
external radiation $E_i$, exhibit optical bistability
(\cite{Gibbs:1985:OpticalBistability}). Due to the \textit{spatial
hysteresis} (\cite{Rozanov:2002:SpatialHysteresis}) there possible
domain walls connecting two stable and otherwise spatially
homogeneous transmitted waves. Domain walls are usually identified
with switching waves, and they represent building blocks for DOS's
--- the localized solutions emerging as bound states of switching
waves. This implies the discreetness of the DOS spectrum in
contrast to the continuous ``families'' of conservative solitons.
The external radiation $E_i$ determines the frequency and the
phase of DOS's, or ``cavity solitons''; they exist on the nonzero
background and have oscillating tales. Several DOSs may interact
(\cite{Afanasjev:1997-6020:PRE, Ramazza:2002-66204:PRE,
Schapers:2000-748:PRL, Schapers:2003-227:IQE, Tlidi:2003-216:IQE})
and form bound states or clusters of cavity solitons
(\cite{Vladimirov:2002-46606:PRE}).  In contrast to the
conservative model discussed in Sect.~\rsect{cluster}, the
clusters may appear due to the oscillating effective interaction
potential (\cite{Malomed:1991-6954:PRA, Malomed:1998-7928:PRE})
with a discrete set of the equilibrium distances between cavity
solitons. A large number of publications is devoted to the study
of cavity solitons and their link to spontaneous pattern
formation, as also discussed in the recent review papers
(\cite{Arecchi:1999-1:PRP, Rozanov:2000-462:UFN,
Firth:2002-54:rar, Peschel:2003-51:IQE, Lugiato:2003-193:IQE}),
and reflected in the comprehensive list of references prepared
by~\cite{Mandel:2004-R60:JOB}.

Interesting applications of vortices such as the pump beams in externally driven cavities was suggested for degenerate optical parametric oscillator, as discussed by~\cite{Oppo:2001-66209:PRE}, and in the vertical-cavity surface-emitting lasers. In the former case, the stable domain walls appear as being trapped in the beam,
while in the latter case the cavity solitons perform a uniform rotary motion along the crater of a doughnut-shaped holding beam (\cite{Barland:2003:EPN}). Micro-cavities offer novel possibilities for the cavity soliton generation and control (\cite{Barland:2002-699:NAT, Debernardi:2003-109:IQE, Maggipinto:2003-206:IQE, Vahala:2003-839:NAT}), and the examples include the spontaneous generation of the ``optical vortex crystals'' (\cite{Scheuer:1999-230:SCI}).

In lasers, the phase of the field is free ($E_i=0$ in Eq.~\reqt{CGL}) and the topological solitons are possible (\cite{Firth:2002-54:rar}). The comprehensive overview of the theoretical studies and the experimental generation of laser vortices can be found in~\cite{Weiss:1999-151:APB, Weiss:2003:Vortices}. In lasers with saturable absorbers, i.e. when nonlinearity in Eq.~\reqt{CGL} is given by $f\left(|E|^2\right) = -1+g_0/(1+|E|^2)-a_0/(1+b|E|^2)$, novel types of solitons may appear, such as transversely asymmetric and rotating structures without phase dislocations and radially symmetric vortices with higher-order topological charges
(\cite{Rozanov:1995-868:OPSR, Fedorov:2003-197:IQE}). Furthermore, bright dissipative vortex solitons can form strongly coupled ``vortex clusters'', as shown in Fig.~\rpict{fig23}(a,b), as well as weakly coupled bound states (\cite{Rozanov:2004-427:JETP}). The latter exhibit spontaneous rotation during the evolution, see Fig.~\rpict{fig23}(c).

Stabilization of dissipative vortex solitons in the cubic-quintic CGL and new types of radially symmetric solitons, such as erupting, flat-top, and composite vortices, were reported recently by~\cite{Crasovan:2001-59:PLA, Crasovan:2001-16605:PRE}. In the same model, the existence of stable clusters of dissipative solitons rotating around a central vortex core was predicted by~\cite{Skryabin:2002-44101:PRL}. This extends the results presented in Sect.~\rsect{cluster} to the case of dissipative CGL systems.

Another example of the dissipative optical systems supporting topological spatial solitons is given by the nonlinear interferometer formed by a liquid crystal light valve with a feedback. The so-called ``triangular solitons'' with the rich structure of phase singularities were found by~\cite{Ramazza:2004-s266:JOA, Bortolozzo:nlin.PS/0407036:ARXIV} in this system.

The vectorial generalization of the CGL model and corresponding patterns of phase dislocations studied by~\cite{Hernandez-Garcia:2000-744:PRL, Hoyuelos:2003-176:PD}.
Similar polarization patterns and vectorial defects were observed in numerical simulations of the three-wave optical parametric oscillators by~\cite{Santagiustina:2002-36610:PRE}.

Finally, we mention the three-dimensional generalization of cavity solitons, namely the ``bubbles with a dark skin'' and 3D Turing structures in sinchronously pumped degenerate, and 3D vortex rings in non-degenerate optical parametric oscillators (\cite{Weiss:1999-151:APB}).

\subsection{Vortices in matter waves}
\lsect{BEC}

The concept of optical vortices can take us away from optics itself and emphasize the relevance of optical solitons and optical vortices to other fields of nonlinear physics. In particular, the study of vortices is an important research topic in the rapidly developing field of coherent matter waves and nonlinear atom
optics. In particular, vortices appear in the nonlinear dynamics of the Bose--Einstein condensates and they provide a close link between self-focusing of light in nonlinear optics and the nonlinear dynamics of matter waves.

The phenomenon known as Bose--Einstein condensation (BEC) was actually predicted in 1924 for systems whose particles obey the Bose statistics and whose total particle number is conserved. It was shown that there exists a critical temperature below which a finite fraction of all particles condenses into the same quantum state. Since 1995, the BEC phenomenon has been observed using several different types of atoms, confined by a magnetic trap and cooled down to extremely low temperatures (\cite{Anderson:1995-198:SCI, Bradley:1995-1687:PRL,
Davis:1995-3969:PRL, Fried:1998-3811:PRL}).

From a mathematical point of view, the dynamics of BEC wave function can be described by an effective mean-field equation known as the Gross--Pitaevskii (GP) equation (\cite{Dalfovo:1999-463:RMP}). This is a classical nonlinear equation that takes into account the effects of particle interaction through an effective mean field. As a matter of fact, the complete theoretical description of a BEC requires a quantum many-body approach (\cite{Dalfovo:1999-463:RMP}). The many-body Hamiltonian describing $N$ interacting bosons is expressed through the boson field operators $\hat{\Phi}({\bf r})$ and $\hat{\Phi}^{\dagger}({\bf r})$ that, respectively, annihilate and create a particle at the position ${\bf r}$. A mean-field approach is commonly used for the interacting systems to overcome the problem of solving exactly the full many-body Schr{\"o}dinger equation. Apart from the convenience of avoiding heavy numerical work, mean-field theories allow one to understand the behavior of a system in terms of a set of parameters that have a clear physical meaning. Actually, most of the experimental results show that the mean-field approach is very effective in providing both qualitative and quantitative predictions for the static and dynamic properties of the trapped ultracold gases.

Because of the similarities between the GP equation in the BEC theory and the NLS equation in nonlinear optics, many of the phenomena predicted and observed in nonlinear optics are expected to occur for the BEC macroscopic quantum states, even though the underlying physics can be quite dissimilar. In particular, this includes the dynamics of BEC vortices (\cite{Williams:1999-568:NAT, Garcia-Ripoll:2000-4264:PRL}) recently reviewed by~\cite{Fetter:2001-r135:JPCM}.

Historically, quantum vortices in trapped atomic gases were first observed in 1999, using two-component condensates (\cite{Matthews:1999-2498:PRL}). The possibility of trapping more than one BEC component arises from the hyperfine atomic structure. Atoms in internal states with different total angular momentum may coexist in the BEC fraction, and it is possible to induce transitions between their different states. To form a vortex soliton, a phase gradient was imprinted in one of the BEC components, which caused it to rotate. The system was stabilized
at a configuration in which the nonrotating component was localized at the center of the trap acting as an effective potential on the rotating component, which resided in the outer region.

\pict[0.708]{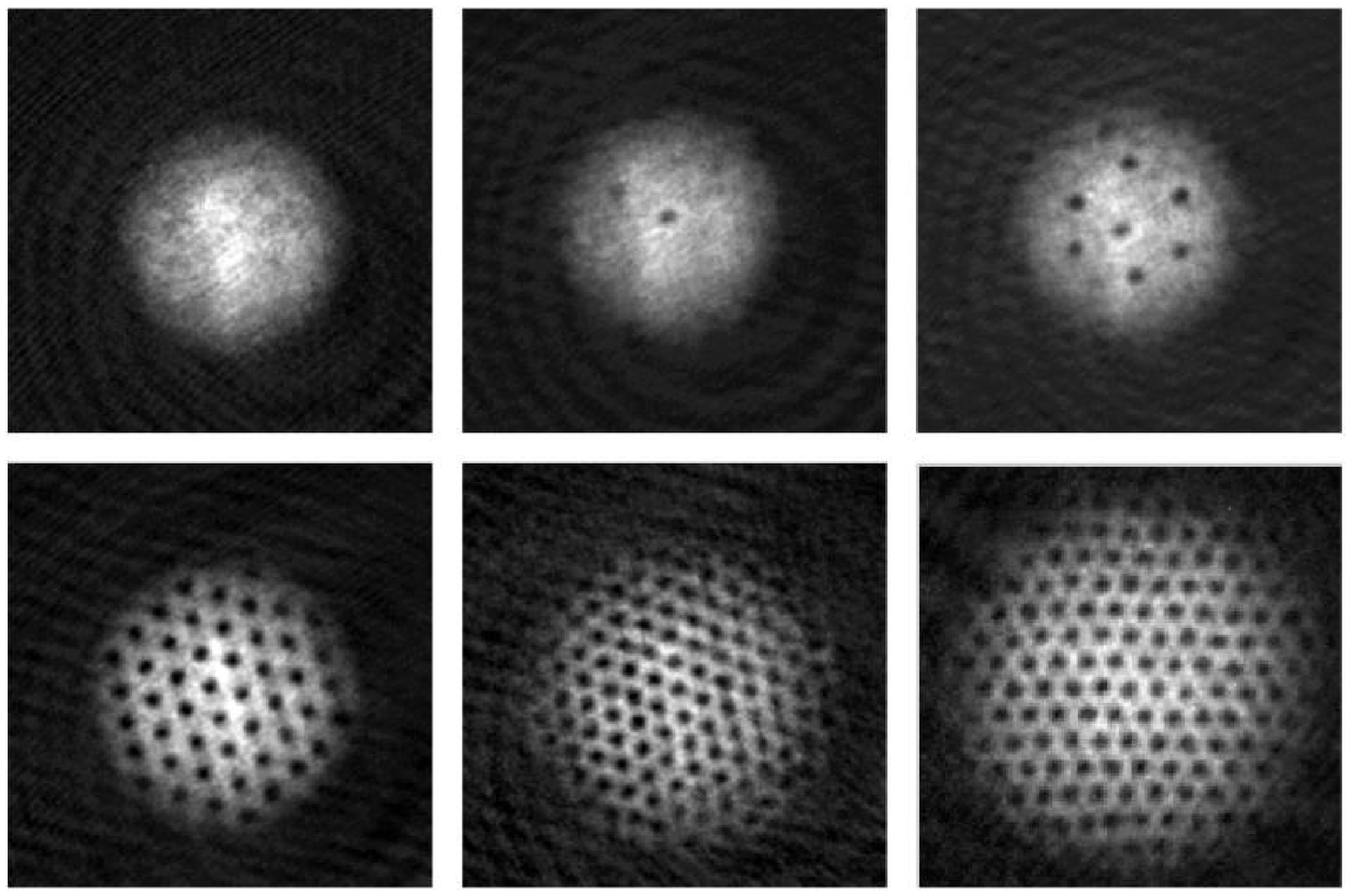}{fig24}{Generation of vortices and vortex lattices in a  rotating $^{87}$Rb condensate. (Courtesy P. Engels).}

The main properties of a two-component BEC can be described using
a system of two incoherently coupled GP equations, similar to the
two coupled NLS equations that describe optical vector solitons,
except for the presence of a trapping potential that prevents the
condensate with repulsive interaction from spreading. The
two-component GP equation has solutions with remarkable properties
(\cite{Garcia-Ripoll:2000-4264:PRL}). In a situation where the two
components overlap considerably, the creation of a vortex in just
one of the components is not dynamically stable. The reason is
that the angular momentum can be transferred after some time from
one component to the other one, initiating a cyclic process. If
one only monitors the density profile of the two atomic species,
it may seem that the vortex disappears and eventually reappears in
a periodic manner.

If a nearly two-dimensional trap is made to rotate, the situation
changes qualitatively. In the rotating frame the Coriolis force
manifests itself through an additional term in the
Hamiltonian, $-\Omega L_z$, where $L_z$ is a
component of the angular momentum operator ${\bf L}$ and $\Omega$
is the angular frequency (\cite{Fetter:2001-r135:JPCM}). This
additional force produces the centrifugal barrier proportional to
$L_z^2z$, where $L_z = m \hbar$ is the angular momentum per
particle. Thus, it is always energetically costly to have a high
angular momentum. However, for nonzero values of $\Omega$, it may
be energetically favorable to develop small positive values of
$L_z$. If $\Omega$ is sufficiently large, solutions with $L_z$
greater than $\hbar$ may have the lowest energy; a vortex is
created in this situation.

For $\Omega$ greater than a critical frequency $\Omega_c=0.22 \omega_0$, where $\omega_0$ is the trap frequency, the lowest energy corresponds to the state with $L_z = \hbar$. As the frequency increases, states with increasingly high angular
momentum become the effective ground state in the rotating frame, thereby creating vortices with a topological charge $m > 1$. Such configurations are unstable and decompose into several single-charge vortices (\cite{Butts:1999-327:NAT}). The
interaction between vortices is generally believed to be repulsive. Thus, at rotating frequencies high enough to generate many stable vortices, vortices tend to move apart and drift toward the borders of the condensate, where they would disappear if their existence were not favored by the rotation. The result is that, at very high angular frequencies, vortices tend to form a regular array, as also observed experimentally. Figure~\rpict{fig24} shows examples of vortices for a rotating Rb-vapor BEC. The array formation is akin to what has been long known for type II superconductors, where it is the presence of a magnetic field that forms a triangular vortex crystal---the so-called Abrikosov lattice (\cite{Abrikosov:1957-1174:JETP}).

In another experiment by~\cite{Madison:2000-806:PRL}, the formation of a regular vortex array was observed in a $^{87}$Rb BEC as the number of stable vortices raises from 0 to 4 by increasing the angular frequency. The formation of a triangular vortex lattice with as many as 130 vortices was observed in an
experiment where BEC was obtained using sodium atoms (\cite{Abo-Shaeer:2001-476:SCI}). The optimum configuration in terms of size and regularity is achieved after 500~ms. For times shorter than that, regular order is not completely established, and a blurry structure is formed because of the misalignment of some vortices with respect to the rotation axis. For times much
longer than 500~ms, inelastic collisions induce atom losses and a decrease in the number of vortices. The spin texture of various vortex-lattice states at higher rotation rates and in the presence of an external magnetic field has been presented recently by~\cite{Mizushima:2004-43613:PRA}, and intriguing reshaping from the triangular to a square lattice has been observed by~\cite{Schweikhard:2004-210403:PRL}.

Several experiments have studied the nucleation of vortices in a BEC stirred by a laser beam. In the experiment by~\cite{Raman:2001-210402:PRL}, vortices were generated in a BEC cloud stirred by a laser beam and observed with time-of-flight
absorption imaging. Depending on the stirrer size, either discrete resonances or a broad response was visible as the stir frequency was varied. Stirring beams that were small compared to the condensate size generated vortices below the critical rotation frequency for the nucleation of surface modes, suggesting a local mechanism of vortex generation. In addition, it was observed that the centrifugal distortion of the condensate induced by a rotating vortex lattice led to bending of the vortex lines.

Recent developments in the topic of vortices in BEC were summarized by~\cite{Kevrekidis:cond-mat/0501030:ARXIV}, including discrete vortices in periodic lattices. Corresponding continuous model is very similar to the optically induced photonic lattices, described in Sect.~\rsect{lattice}, and supports the ``gap vortex soliton'' identified by~\cite{Ostrovskaya:2004-19:OE}. However, there are examples of the phenomena which have no analogy in optics, for example the structural transitions of the lattice of vortices in the rapidly rotating periodic potential, reported by~\cite{Pu:cond-mat/0404750:ARXIV}. Similarly, even the properties of a single vortex soliton in BEC and self-defocusing Kerr media are very similar, the stable vortex dipoles in nonrotating BEC, introduced by~\cite{Crasovan:2003-63609:PRA}, can not exist in optical system without the external trapping potential, see Sect.~\rsect{dark}.

Different from singular vortices, the nonsingular topologically nontrivial states, or skyrmions, have attracted an attention since the creation of BEC spinors (\cite{Matthews:1999-2498:PRL}). Two-dimensional (2D) particlelike solitons of this kind are sometimes referred as the coreless vortices. Skyrmions can be created out of the ground state, in which all the spins are aligned, by reversing the average spin in a finite region of space (\cite{AlKhawaja:2001-918:NAT}). Than the Skyrmion is characterized by the winding number, the analog of a topological charge. The half-charge skyrmion has been successfully generated in experiment with three-component BEC by~\cite{Leanhardt:2003-140403:PRL}, and possible spin textures for anti- and ferromagnetic interactions in this system were compared by~\cite{Mueller:2004-33606:PRA}. The stability of 3D (\cite{AlKhawaja:2001-918:NAT}) skyrmions, composed of two coaxial tori, and 2D (\cite{Zhai:2003-43602:PRA}) single-charged skyrmions has been studied in two-component BEC. The general conclusion been that skyrmions can exist as a metastable state only, though the stabilizing mechanisms were suggested for single- and multiply-quantized skyrmions by~\cite{Savage:2003-10403:PRL, Ruostekoski:2004-41601:PRA}. Skyrmions have important applications in nuclear physics and quantum-Hall effect (\cite{Makhankov:1993:SkyrmeModel}), it is expected that observation of these structures in BEC would enable a direct comparison between theory and experiment.

\subsection{Optical vortices and quantum information}

During the recent years a new fascinating avenue for applications
of optical vortices in the area of quantum optics has been
identified. As discussed extensively throughout this chapter,
light beams with nested optical vortices carry orbital angular
momentum, a property that holds as well for the mode functions
that describe the photon quantum states, including states
corresponding to single photons or to entangled pairs. The quantum
angular momentum of light contains a spin and an orbital
contribution, and in general only the total angular momentum is an
observable quantity (\cite{Cohen:1989:Photons}). However, within
the paraxial regime, both contributions can be measured and
manipulated separately (\cite{VanEnk:1994-963:JMO,
VanEnk:1994-497:EPL, Simpson:1997-52:OL, Barnett:2002-S7:JOB,
Neil:2002-53601:PRL, Leach:2004-13601:PRL}). The spin contribution
is described by a two-dimensional state, thus can be employed to
generate \textit{qubits}, whereas the orbital contribution can
generate multi-dimensional quantum entangled states, or
\textit{qudits}, with an arbitrarily large number of entanglement
dimensions. While the spin angular momentum is a workhorse of
quantum optics and quantum information
(\cite{Bouwmeester:2000:PhysicsQI}), only recently the orbital
angular momentum has been added to the toolkit
(\cite{Arnaut:2000-286:PRL,Mair:2001-313:NAT,
Molina-Terriza:2002-13601:PRL}).

\cite{Allen:1992-8185:PRA} showed a decade ago that paraxial
Laguerre-Gaussian laser beams, with a nested vortex, carry a
well-defined orbital angular momentum associated to their spiral
wave fronts (\cite{Allen:1999-291:ProgressOptics}). The formal
analogy between paraxial optics and quantum mechanics implies that
such modes are the eigenmodes of the quantum mechanical angular
momentum operator. The Laguerre-Gaussian modes form a complete
Hilbert set and can thus be used to represent the quantum photon
states within the paraxial regime of light propagation. The
quantum angular momentum number carried by the photon is then
represented by the topological charge, or winding number $m$, of
the corresponding mode, and each mode carries an orbital angular
momentum of $m\hbar$ per photon. Multi-dimensional vector photon
states can be constructed with controllable projections into modes
with well-defined winding numbers, thus providing
higher-dimensional alphabets \cite{Molina-Terriza:2002-13601:PRL}.
In particular, mode functions in the form of vortex-pancakes allow
the manipulation, including the addition and removal, of specific
projections of the vector states. It is worth stressing that beams
without nested vortices, or alternatively with a complex
topological structure, can also carry orbital angular momentum. A
beautiful illustrative example was shown by
\cite{Santamato:2002-871:OE} in the classical regime, by studying
the optical angular momentum transfer to transparent particles
using light beams carrying zero average angular momentum. The
concept applies as well to the quantum regime in a variety of
beams shapes and geometries. However, because Laguerre-Gaussian
modes carrying nested optical vortices are eigenstates of the
quantum orbital angular momentum operator, vortices play  a
central role in the area.

\pict[0.708]{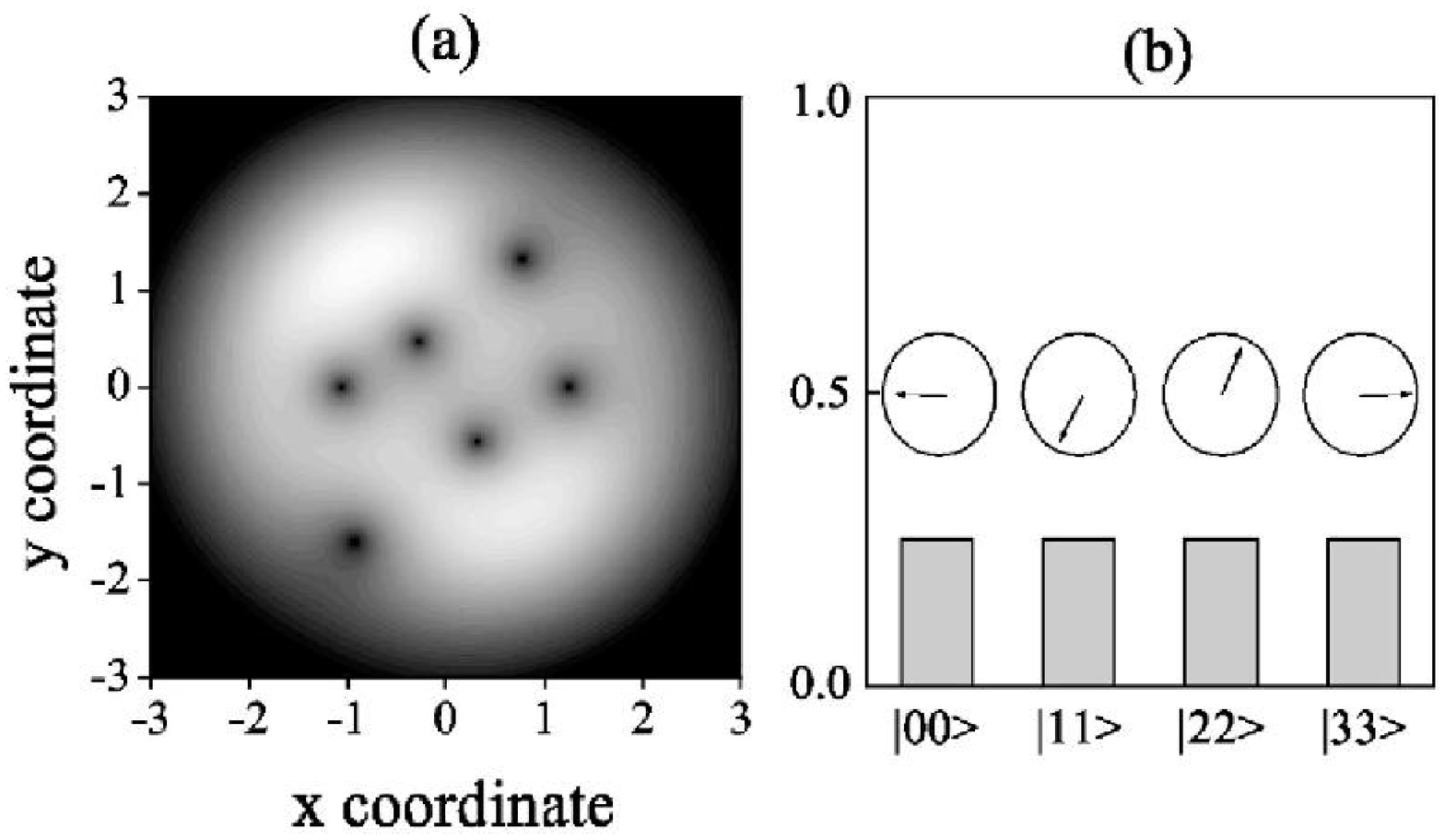}{fig25}{Intensity profile of the pump beam
that generates the maximally entangled qu-quart. (a) Intensity
profile, (b) the generated quantum state. In (a), the transverse
coordinates are normalized to the beam width. In (b), the bars
show the weight distribution and the clocks show the phase
distribution of the quantum state.
After~\cite{Torres:2003-52313:PRA}.}

The generation of quantum states entangled in orbital angular
momentum relies in the process of spontaneous parametric down
conversion (\cite{Klyshko:1967-23:JETPL, Arnaut:2000-286:PRL,
Mair:2001-313:NAT, Franke-Arnold:2002-33823:PRA,
Padgett:2002-777:JMO}). The generated entangled two-photon states
can be prepared in desired states by making use of transverse
engineering of quasi-phase-matched
geometries,\cite{Torres:2004-376:OL} or by appropriate tailoring
of the spatial characteristics of the pump beam. The latter can be
accomplished by a variety ways, including by pumping the nonlinear
crystal with several nested optical vortices
(\cite{Torres:2003-52313:PRA}). An illustrative example of the
potential engineering of quantum entangled states with pump beams
with nested vortex pancake is shown in Fig.~\rpict{fig25}. The
characterization of the entangled photon pairs in terms of
eigenstates of the orbital angular momentum operator yields the
concept of \textit{quantum spiral bandwidth}, which was found to
depend on the shape of the beam that pumps the down-converting
crystal, and on the material properties and length on the crystal
(\cite{Torres:2003-50301:PRA}). All these schemes are awaiting
experimental demonstration, albeit the related concept of
entanglement concentration put forward by
\cite{Torres:2003-52313:PRA} has been observed experimentally
using an alternative approach, as mentioned below.

A fundamental question that arises is the conservation of OAM in
the process of photon down-conversion (\cite{Arnaut:2000-286:PRL,
Eliel:2001-5208:PRL, Visser:2002-33814:PRA,
Barbosa:2002-53801:PRA, Caetano:2002-41801:PRA}). In collinear
down-conversion, the two-photon entangled state constituted by the
signal and idler photons is described by a transverse mode
function that is globally paraxial. Therefore, the orbital angular
momentum of all the involved photons is a well-defined quantity
that in the absence of momentum transfer between light and matter
must be conserved, a feature that within the experimental accuracy
is consistent with the observations by~\cite{Mair:2001-313:NAT},
in the quasi-collinear geometry used. In non-collinear geometries
the relation between the orbital angular momentum and the
vorticity of the mode function is not necessarily given by simple
algebraic rules, as  most clearly illustrated in highly
non-collinear settings (\cite{Molina-Terriza:2003-155:OC,
Torres:2004-1939:OL}).

The quantum applications of optical vortices holds promise for
exciting developments in the near future, in particular to the
proof-of-principle demonstration of quantum algorithms and to
explore fundamental quantum features in higher-dimensional Hilbert
spaces. Significant advances along this direction have been
already achieved during the last years after the observation of
OAM entanglement by~\cite{Mair:2001-313:NAT}. Important steps
include the development of interferometric schemes that might be
used to sort out single photons according to their OAM
(\cite{Vasnetsov:2001-464:QE, Leach:2002-257901:PRL}), the
generation of qutrits encoded in OAM and their use to observe
violation of Bell inequalities in three-dimensional Hilbert spaces
(\cite{Vaziri:2002-240401:PRL}), the demonstration of
concentration of higher-dimensional entanglement
(\cite{Vaziri:2003-227902:PRL}), the triggered production,
transmission and reconstruction of qutrits for different quantum
communication protocols (\cite{Molina-Terriza:2004-167903:PRL}),
the use of qutrits for quantum bit commitment
(\cite{Langford:2004-53601:PRL}), the proposal of innovative
set-ups to efficiently measure high-dimensional entanglement
(\cite{Oemrawsingh:2004-217901:PRL}), and the demonstration of
coin-tossing algorithms based in qutrits
(\cite{Molina-Terriza:2004-000:PRL}). Vortices are also being used
to explore fundamental quantum features, like the uncertainty
principle for angular position and angular momentum
(\cite{Franke-Arnold:2004-103:NJP}), or the effects induced by the
quantum vacuum on the perfect zero of a classical field (the
vortex core) (\cite{Berry:2004-S178:JOA}). Much more is expected
to come soon, as this area of research is at its infancy.

\section{Concluding remarks}

\pict[0.708]{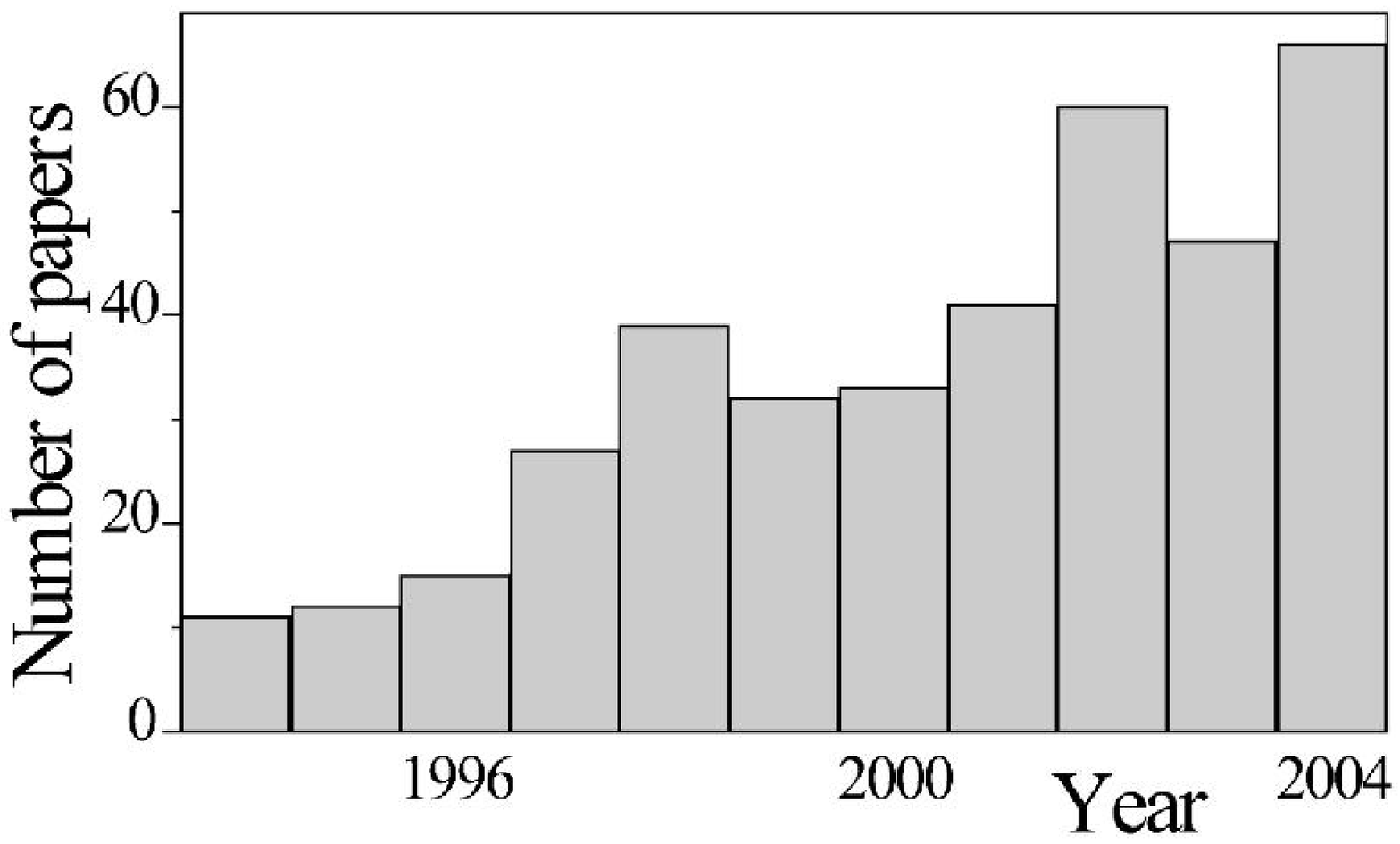}{fig26}{The number of papers cited in this
review vs. the publication year.}

We have presented a comprehensive overview of exciting research in
the field of {\em nonlinear singular optics} that studies the
propagation of optical vortices and optical beams carrying an
angular momentum in nonlinear media. Understanding and controlling
the properties of optical vortices could lead to applications in
the near future, ranging from optical communications and data
storage to the trapping, control, and manipulation of particles
and cold atoms. Indeed, optical vortices provide an efficient way
to control light by creating reconfigurable waveguides in bulk
media. The study of phase singularities in optical parametric
processes not only suggests novel directions of fundamental
research in optics but also provides links to other branches of
physics. For example, the recent discovery of a rich variety of
exotic topological defects in unconventional superfluids (such as
$^3$He-A) and superconductors points to the likelihood that deep
analogies exist between vortices in complex superfluids and
singularities in light waves.

\section{Acknowledgements} \lsect{acknowledgements}

We thank many of our colleagues for collaboration and useful discussions, most especially T. Alexander, S. Carrasco, Z. Chen, L. Crasovan, C. Denz, Y. Kartashov, W. Krolikowski, B. Luther-Davies, B.A. Malomed, D. Mihalache, S. Minardi, G. Molina-Terriza, D. Neshev, E. Ostrovskaya, D. Petrov, M.F. Shih, M. Soskin, A. Sukhorukov, G. Swartzlander, and J.P. Torres for sharing their knowledge with us and for their constant support and collaboration. ASD acknowledges the hospitality of ICFO during his visit to Barcelona.

This work was produced with the assistance of the Australian Research Council (ARC); the Center for Ultra-high bandwidth Devices for Optical Systems (CUDOS) is an ARC Center of Excellence. LT acknowledges support by the Generalitat de Catalunya and by the Spanish Government.

\begin{small}

\end{small}
\end{document}